%% file: pr.tex
\documentstyle[epsfig]{myelstart}
\newcommand{\ie}{\it i.e. \rm}
\newcommand{\etal}{\it et al \rm}

\newcommand{\tr}{\mathop{\rm tr}}

\begin{document}
\begin{frontmatter}
\title{Recent mathematical developments in the Skyrme model}

\author{T. Gisiger and M.B. Paranjape}
\address{Groupe de Physique des Particules, Universit\'e de Montr\'eal, C.P. 
6128, succ. centre-ville, Montr\'eal, Qu\'ebec, Canada, H3C 3J7}

\begin{abstract}
In this review we present a pedagogical introduction to recent, more
mathematical developments in the Skyrme model. Our aim is to render these
advances accessible to mainstream nuclear and particle physicists. We
start with the static sector and elaborate on geometrical aspects of the
definition of the model. Then we review the instanton method which yields an
analytical approximation to the minimum energy configuration in any sector of
fixed baryon number, as well as an approximation to the surfaces which join
together all the low energy critical points. We present some explicit results
for $B=2$. We then describe the work done on the multibaryon minima using 
rational maps, on the topology of the
configuration space and the possible implications of Morse theory. Next we turn
to recent work on the dynamics of Skyrmions. We focus exclusively on the low
energy interaction, specifically the gradient flow method put forward by
Manton. We illustrate the method with some expository toy models. We end this
review with a presentation of our own work on the semi-classical quantization 
of nucleon states and low energy nucleon-nucleon scattering.
\end{abstract}
\end{frontmatter}

\section{Introduction}
\input {introduction}

\section{The Skyrme model: Lagrangian and notation}
\input {section2}

\section{Study of the statics of the Skyrme model}
\input {section3}

\subsection{Geometry of the Skyrme model}
\input {section31}

\subsection{The instanton method}
\input {section32}

\subsection{Rational maps and multi-baryon number states of the 
Skyrme model}
\input {section33}

\subsection{Search for a sphaleron in the Skyrme model: Morse theory}
\input {section34}

\section{Dynamics of the Skyrme model: Soliton-Soliton scattering}
\input {section41}

\subsection{Application of Manton's method to the $B=2$ 
sector of the Skyrme model}
\input {section42}

\subsection{Low energy nucleon-nucleon scattering}
\input {section43}

\subsubsection{Lagrangian of the Skyrmion-Skyrmion system}
\input {section431}

\subsubsection{Quantization of the Skyrmion spin/isospin states}
\input {section432}

\subsubsection{Nucleon-nucleon scattering}
\input {section433}

\ack
We thank G.E. Brown for inviting us to write this paper and R.B. MacKenzie for
his useful comments on the manuscript. This
work supported in part by NSERC of Canada and FCAR of Qu\'ebec.

\end{document}

%% file: introduction.tex
The Skyrme model\cite{Skyrme} 
was first proposed by T.H.R. Skyrme\footnote{For an 
interesting compilation 
of the life, work and influence in physics of T.H.R. Skyrme, see {\em 
Selected Papers, with commentary, of Tony Hilton Royle Skyrme\/} (World 
Scientific Series in 20th Century Physics--Vol. 3), G.E. Brown editor} 
in the sixties, as a revolutionary idea 
for incorporating baryons in the non-linear sigma model description of the 
low-energy interactions of pions. This sigma model consists of a unitary 
matrix valued field $U(\vec x,t)$ of dimension $2\times 2$ or $3\times 3$ 
depending on the number of light quark flavours that are considered. The 
dynamics is described by the Lagrange density
\begin{equation}
{\cal L} = - {f_\pi^2\over 4} 
\tr (U^\dagger \partial_\mu U U^\dagger \partial^\mu U)
\end{equation}
where $f_\pi^2$ is the pion decay constant. Skyrme noted the existence of 
topologically non-trivial field configurations of finite energy. These were 
however, unstable against collapse, which can be adduced by simple 
application of scaling arguments. 
Skyrme then added a higher derivative term to the Lagrange density 
rendering these 
configurations stable. This term is now called the Skyrme term and the Skyrme 
Lagrange density ${\cal L}_{\mathrm{sk}}$ is given by
\begin{equation}
{\cal L}_{\mathrm{sk}} =
-{f_\pi^2 \over 4}\; \tr(U^\dagger \partial_\mu U U^\dagger \partial^\mu U)
+ {1\over 32 e^2} \;
\tr( [U^\dagger \partial_\mu U,U^\dagger \partial_\nu U]^2)
\end{equation}
where $e$ is a new, dimensionless coupling constant. Since each derivative 
corresponds to a momentum, this term is clearly of higher order in the 
low-energy (momentum) approximation. Skyrme proposed the interpretation of 
these 
topological {\em solitons} (stable, localized, finite-energy solutions of the 
classical equations of motion) as the nucleons and identified the topological 
winding number of the soliton with the baryon number.
The technology of 
quantum field theory in the sixties was not sufficiently advanced to treat 
solitons and it took almost twenty years before the ideas of 
Skyrme were 
revived by Balachandran {\it et al} \cite{Balachandran} and 
Witten\cite{Witten}, and vindicated with surprising accord with 
experiment\cite{Adkins}. 
Witten\cite{Witten} described another topological density which should be 
added in the effective action, the celebrated 
Wess-Zumino-Novikov-Witten\cite{Witten,Wess-Zumino,Novikov} term 
\begin{equation}
\Gamma_{WZNW} = -{{\mathrm{i}} N\over 240 \pi^2}
\int_{\mathrm{D_5}} {\d}^5 x \epsilon^{\mu\nu\alpha\beta\gamma}
\tr [ 
U^\dagger\partial_\mu U\,
U^\dagger \partial_\nu U\,
U^\dagger \partial_\alpha U\, 
U^\dagger \partial_\beta U\,
U^\dagger \partial_\gamma U
]
\end{equation}
where $U\in \mathrm{SU}(3)$, $N \in \Zset$ ($D_5$ is in fact a 5-dimensional 
manifold with only its boundary giving the usual 4-dimensional space-time). 
Witten\cite{Witten} showed its relation to the underlying microscopic 
theory of 
the strong interactions QCD, with the number of colours giving the quantized 
coefficient $N$ in $\Gamma_{WZNW}$. Since 
these seminal papers there has been an enormous amount of work relating the 
Skyrme model to phenomena in nuclear and particle physics, for instance 
targeting the spectrum of excitations of baryons\cite{Excitations}, 
the inclusion of strange degrees of freedom in the model\cite{Sriram}, the 
nucleon-nucleon potential\cite{Jackson-Jackson-Pasquier,Vinh}, 
scattering $\pi-N$ states\cite{Mattis}, high density baryon matter as a Skyrme 
crystal\cite{Cristal} and the nucleon-anti-nucleon 
annihilation
\cite{VNNbar-AP,VNNbar-Num,VWWW-NNbar,NNbar-Sim,NNbar-IC,Amado-et-cie} 
to name a few.
We will not consider these developments in detail here and refer the 
interested reader to the literature and to the many excellent reviews on the 
subject\cite{Brown,Walhout,Rho,Sanyuk}. 

Concurrently, there were certain mathematical advances in the Skyrme model 
which were not strongly based on making any contact with phenomenology, but 
moreover to understand the mathematical content of the model. These concerned 
two main areas, that of the exact nature of the minima or critical points of 
the static energy functional and secondly a description of the scattering of 
the corresponding solitons. It turns out in a certain approximation, that of 
low energy scattering, these two aspects are not unrelated. There were 
also some auxiliary mathematical and physical results, concerning 
geometrical insights into the model. Our review will primarily focus on these 
relatively recent mathematical developments. We start with the static sector 
and introduce the Skyrme model on a general Riemannian manifold. Then we 
present the methods using instantons and rational maps to obtain useful 
Skyrme configurations, followed by a short description of Morse theory and its 
application in the model. Next we move to the dynamics where we treat the 
gradient flow method put forward for studying soliton scattering. We terminate 
with it applications to the baryon number 2 sector of the model. 
Most of the advances 
which we will consider were made by N.S. Manton, among others (our 
original contributions to this subject are secondary).
Since these advances generally use the language and formalism of differential 
geometry, the reader should be familiar with these notions (any standard
course on tensor analysis/differential geometry/general relativity 
should be adequate\cite{Flanders,Nakahara,Nash,Eguchi}).  

We start the study of the statics of the model in section 3 
with the geometrical aspects of the model
as first discussed by Manton and Ruback\cite{Manton-Ruback} and then by
Manton\cite{Mantongeo} and Loss\cite{Loss}. 
We show in detail how
the model can be understood as a theory of elasticity in curved space. 
This fruitful approach allows us to explain, for example, 
why the Skyrmion (the lowest 
energy solution of the model with baryon number 1) does not saturate the 
Bogomolnyi bound in ordinary space, but does so for a space with great 
curvature. This situation is connected to 
chiral symmetry restoration and deconfinement. In section 
3.1 we first review the theory of non-linear elasticity, both in flat and 
curved space, as well as some tools of differential geometry. We then move on 
to show how those tools can allow us to 
connect elasticity and the Skyrme model. 
This shows how it is possible to formally interpret a classical 
field theory as an elasticity theory.  

Section 3.2 describes the instanton method put forward by Atiyah 
and Manton\cite{Atiyah-Manton-1} to approximate critical points 
of the Skyrme energy functional. 
Indeed, the Skyrme model has always been plagued (as T.H.R. Skyrme realized 
already in the sixties) by the absence of analytical solutions: to obtain 
solutions of the model, one always has to solve partial 
differential equations numerically, 
except in the simplest cases like the $B=1$ Skyrmion, where 
one has only to solve an ordinary differential equation numerically. 
The instanton 
method permits one to obtain analytical expressions (in the simplest 
cases) or at least limit the difficulties to solving ordinary differential 
equations. The tradeoff is that in this method (which is by nature approximate)
the error introduced is hard to estimate, except by comparing 
the resulting approximation with the full ``exact'' numerical solution of the 
problem. The method seems nevertheless to work reasonably well, and has been 
helpful in investigating the bound states in the sector $B=2$, 3 and 4, and 
Skyrme crystals as well\cite{Atiyah-Manton-2,Leese-Manton,Manton-Sutcliffe}.

The next subsection 3.3 presents the use of 
rational maps\cite{Houghton} in finding approximations
to minima with particular symmetries for sectors comprising of a range of
baryon numbers. Rational maps are mappings of the complex plane to itself
comprised of ratios of polynomials. The choice of the coefficients of the 
polynomials can encode the mappings with complicated symmetries, especially 
when viewed via stereographic projection as mappings of $S^2\rightarrow S^2$.
The former $S^2$ corresponds to the angular degrees of freedom about a Skyrme
configuration while the latter corresponds to a selected $S^2$ in the group
${\mathrm{SU}}(2)$ obtained via the Hopf projection as will be explained
in subsection 3.3. 

We give in subsection 3.4 a 
short introduction to Morse theory and how it could be used to find new 
solutions to the equations of motion\cite{Isler}.  
Morse theory relates the existence of critical points of a function defined 
on a manifold to non-trivial topological aspects of the manifold. Here the 
function in question is the energy functional defined on the manifold of the 
space of all static field configurations. Although the method has not borne
fruit in the Skyrme model, it has already been useful in the analysis leading 
to the sphaleron of the standard electroweak model\cite{Mantonsph}. 

In the next section (section 4), we present some recent 
developments in the 
study of the dynamics of the model, namely Skyrmion-Skyrmion and 
nucleon-nucleon scattering. 

We start by presenting in section 4.1 Manton's method\cite{Mantongfc} for 
truncating the degrees of freedom of a system, thereby possibly
rendering it tractable. This formalism was published by Manton after his work 
on BPS monopole (topological solitons of the massless $\mathrm{SU}(2)$ Higgs 
model) 
scattering\cite{Gibbons}. It is a much more general method in the sense 
that it 
is applicable not only to soliton problems (namely scattering thereof) but 
to many systems with a great number of degrees of freedom  
under the right circumstances. Even 
though the underlying ideas are intuitively simple, the formalism needs 
to be expressed using differential geometry. This is why in section 4.2 
we introduce this method using a number of simple examples to help 
us emphasize the ideas behind the method and to get a feel for how it works. 
Results obtained using this method are then compared with ``exact'' numerical 
results. In section 4.3 we present the application of the method to the case
$B=2$ in the Skyrme model.

We close this review with a discussion of our work on 
Skyrmion-Skyrmion\cite{Gisiger1} and 
nucleon-nucleon\cite{Gisiger2} scattering using Manton's method but in a 
semi-classical, rather than purely quantum mechanical perspective.

%% file: section2.tex
The Skyrme model including mass is described by the Lagrange density,
\begin{eqnarray}
{\cal L}_{\mathrm{sk}} =
-{f_\pi^2 \over 4}\; \tr(U^\dagger \partial_\mu U U^\dagger \partial^\mu U)
&+& {1\over 32 e^2}\;
\tr( [U^\dagger \partial_\mu U,U^\dagger \partial_\nu U]^2)
\nonumber
\\
&-& {1\over 2} m_\pi^2 \tr(U + U^\dagger)\label{eq:lsk}
\end{eqnarray}
where $U(x)$ is a unitary matrix valued field. The Lagrangian is also often 
written in terms of $F_\pi= 2 f_\pi$ but this is only a matter of convention. 
The quantum fluctuations of 
$U(x)$ represent the low energy mesons made up of quark-anti-quark 
pairs. In the chiral limit, all light quarks $(u,d,s)$ are massless and 
degenerate and the corresponding flavour symmetry dictates that 
$U(x)$ is a $3\times 3$ matrix.
In this review, however, we will be more focused on the fact that $U(x)$ 
is a unitary matrix; the simplest example of this is a $2\times 2$ matrix, 
the only case that we will 
consider here. Phenomenologically, this means we 
consider the explicit breaking of the $\mathrm{SU}(3)_{\mathrm{f}}$ symmetry 
to be large. 
The Skyrme Lagrangian (\ref{eq:lsk}) 
corresponds to the first two terms of a systematic
expansion in derivatives of the effective Lagrangian describing low energy
interactions of pions plus the mass term. 
It should be derivable from QCD hence 
$f_\pi$, $e$ and $m_\pi$, the pion mass, are
in principle calculable parameters. These calculations are actually unfeasable
and we take $f_\pi$, $e$ and $m_\pi$ from phenomenological fits. We then  
find $f_\pi$ to be in the range of 130--190 MeV and $e\simeq 5$. 
In this article, we will always take $m_\pi\rightarrow 0$ for simplicity.
 
The energy functional, coming from the static part 
of the Lagrangian, can be written in a more 
elegant way by adopting a convenient choice of units (so-called natural 
units)\cite{Manton-Ruback}:
\begin{equation}
E_{\mathrm{sk}} = \int {\d}^3 \vec x\,\Biggl[
-{1 \over 2}\; \tr(U^\dagger \partial_{\mathrm{i}} U 
U^\dagger \partial_{\mathrm{i}} U)
- {1\over 16}\;
\tr( [U^\dagger \partial_{\mathrm{i}} U,
U^\dagger \partial_{\mathrm{j}} U]^2)\Biggr],
\label{eq:lsknu}
\end{equation}
where the unit of energy is now $f_\pi/4 e\simeq 6\;\mathrm{MeV}$ and the unit 
of length $2/e f_\pi\simeq 0.6\;\mathrm{fm}$.

The baryons arise as topological solitonic solutions of the
equations of motion\cite{Skyrme}. These topological solitons
correspond to non-trivial mappings
of $\Rset^3$ plus the point at infinity into $\mathrm{SU}(2)$:
\begin{equation}
U(\vec x): \Rset^3 + \infty \to \mathrm{SU}(2) = S^3.
\end{equation}
But topologically
\begin{equation}
\Rset^3 + \infty = S^3;
\end{equation}
thus the homotopy classes of mappings
\begin{equation}
U(\vec x): S^3 \to S^3,
\end{equation}
which define $\Pi_3(S^3) = \Zset$,
characterize the space of configurations. The topological charge of each 
sector is given by
\begin{equation} 
N = {1 \over 24 \pi^2} \int{d^3 \vec x\, \epsilon^{\mathrm{ijk}} \;
\tr(U^\dagger \partial_{\mathrm{i}} U\, 
U^\dagger \partial_j U\, 
U^\dagger \partial_{\mathrm{k}} U)},
\label{eq:nb}
\end{equation}
which is an integer and is identified with the baryon number
\cite{Skyrme,Witten}. 

\begin{figure}
\centering
\mbox{\epsfig{figure=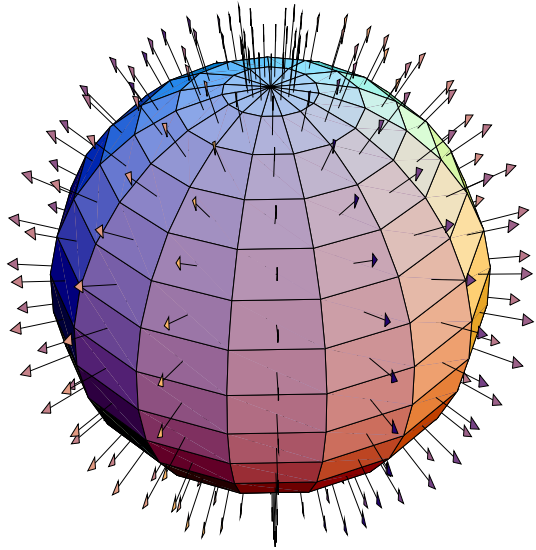,width=3.0truein}}
\caption{Artistic impression (by T. Gisiger) of the field of the Skyrmion 
or hedgehog.}
\label{fig:fig0}
\end{figure}

The solution with baryon number $N=1$ with lowest energy is the 
Skyrmion. It is parametrized presumably (since this has still not been 
rigorously, mathematically proven) by the following expression
\begin{equation}
U(\vec x) = {\mathrm e}^{{\mathrm i} \hat r\cdot\vec\tau f(r)}
\label{eq:usk}
\end{equation}
where $\vec\tau$ are the Pauli matrices, $\hat r$ is the unit position vector 
and $r$ is its length. The field $U(\vec x)$ therefore points radially, and is
sometimes called ``hedgehog'' (see figure $\ref{fig:fig0}$).
$f(r)$ is a function to be obtained from the equations 
of motion of the theory by replacing the ansatz (\ref{eq:usk}) in the static 
equations of the theory:
\begin{equation}
\partial_{\mathrm{k}} \Biggl( {f_\pi^2\over 2} \partial_{\mathrm{k}} U 
U^\dagger
+ {1\over 8 e^2 } \biggl[\partial_{\mathrm{i}}U U^\dagger,
\biggl[ \partial_{\mathrm{k}}U U^\dagger,
\partial_{\mathrm{i}}U U^\dagger\biggr]\biggr]\Biggr) = 0\label{eq:eqg}
\end{equation}
which give for (\ref{eq:usk})
\begin{equation}
( {\rho\over 4} + 2 \sin^2{F} ) F'' + {1\over 2} \rho F' + \sin{2f} F'^2 
- {1\over 4} \sin{2 F} - {\sin^2{F} \sin{2 F}\over \rho^2}
= 0\label{eq:eqsk}
\end{equation}
where $\rho \equiv 2 e f_\pi r$ and $f(r)= F(\rho)$.
This equation has to be solved numerically using the boundary conditions 
imposed on the ansatz by the requirements that it have 
baryon number unity and finite energy, namely 
$f(0) = \pi$ and $f(+\infty)=0$.  
The asympotic behaviour of $f(r)$ for large values of $r$ is obtainable from 
(\ref{eq:eqsk}):
\begin{equation}
f(r) \rightarrow {\kappa \over r^2}.
\label{eq:fsk}
\end{equation}
Using numerical integration
we find that $\kappa\simeq 2.16/e^2 f_\pi^2$
\cite{Jackson-Jackson-Pasquier,Schroers}
(The equation for $f(r)$ can equally well be obtained by replacing the ansatz 
(\ref{eq:usk}) in the energy functional (\ref{eq:lsknu}), 
extremizing relative to 
$f(r)$, and integrating over $\theta$ and $\phi$). 

Using the definitions
\begin{equation}
{\cal R}^{\mathrm{a}}_\mu(U)=-{{\mathrm i} \over 2}\; \tr[\tau^{\mathrm{a}} 
\partial_\mu U U^\dagger]
\label{eq:1fru}
\end{equation}
\begin{equation}
{\cal L}^{\mathrm{a}}_\mu(U)=-{{\mathrm i} \over 2}\; \tr[\tau^{\mathrm{a}} 
U^\dagger \partial_\mu U]
\label{eq:1flu}
\end{equation}
\begin{equation}
D_{\mathrm{ab}}(U) = {1\over 2}\;\tr[\tau^{\mathrm{a}} U \tau^{\mathrm{b}}
U^\dagger]
\label{eq:du}
\end{equation}
the Skyrme Lagrange density can be written in the following fashion
\begin{eqnarray}
{\cal L}_{\mathrm{sk}}={f_\pi^2\over 2} {\cal R}(U)_\mu\cdot{\cal R}(U)^\mu
- {1\over{4 e^2}}\biggl [&&{\cal R}(U)_\mu\cdot{\cal R}(U)^\mu\;
{\cal R}(U)_\nu\cdot{\cal R}(U)^\nu\nonumber
\\
&-&
{\cal R}(U)_\mu\cdot{\cal R}(U)^\nu\;{\cal R}(U)_\nu\cdot{\cal
R}(U)^\mu\biggr]
\label{eq:lsk1f}
\end{eqnarray}
where the ``$\cdot$'' represents, for example, 
${\cal R}_\mu\cdot{\cal R}^\mu = 
{\cal R}^{\mathrm{a}}_\mu {\cal R}^{\mathrm{a}\mu}$.
We can separate this into a kinetic energy ${\cal T}$ which is the part
quadratic in time derivatives and a potential energy ${\cal V}$ without any
time derivatives:
\begin{equation}
{\cal T}={f_\pi^2\over 2} {\cal R}_0\cdot{\cal R}_0 -
{1\over{2 e^2}}\biggl [{\cal R}_0\cdot{\cal R}_{\mathrm{i}}\;
{\cal R}_0\cdot{\cal R}_{\mathrm{i}} -
{\cal R}_0\cdot{\cal R}_0\;{\cal R}_{\mathrm{i}}
\cdot{\cal R}_{\mathrm{i}}\biggr ]
\label{eq:t1f}
\end{equation}
\begin{equation}
{\cal V}=-{f_\pi^2\over 2} {\cal R}_{\mathrm{i}}\cdot{\cal R}_{\mathrm{i}} -
{1\over{4 e^2}}\biggl [{\cal R}_{\mathrm{i}}\cdot{\cal R}_{\mathrm{i}}\;
{\cal R}_{\mathrm{j}}\cdot{\cal R}_{\mathrm{j}} -
{\cal R}_{\mathrm{i}}\cdot{\cal R}_{\mathrm{j}}\;
{\cal R}_{\mathrm{j}}\cdot{\cal R}_{\mathrm{i}}\biggr ].
\label{eq:v1f}
\end{equation}
The parametrization of a Skyrmion with (iso)orientation defined by a time 
dependent $\mathrm{SU}(2)$ matrix $A(t)$ and position $\vec R(t)$ is
\begin{equation}
U(\vec x,t) = A(t)\, U(\vec x - \vec R(t)) \,A(t)^\dagger.
\label{eq:uskar}
\end{equation}
This gives a Skyrmion 6 degrees of freedom.
After replacing this ansatz in the Skyrme Lagrangian and integrating over all 
space, we find\cite{Adkins}:
\begin{equation}
L = -2 M + {1\over 2} M \dot{\vec R^{\;2}} + 2 \Lambda\; \bigl(
{\cal R}_0^{\mathrm{a}}(A)\, {\cal R}_0^{\mathrm{a}}(A)\bigr)
\label{eq:l1sk}
\end{equation}
where
\begin{eqnarray}
M &=&\; 4\pi \int_0^{+\infty} r^2dr\nonumber
\\
&\times& { \biggl\{ {1\over 8} f_\pi^2 \biggl[\biggl({\partial f\over \partial
r}\biggr)^2\!\!+ 2\,{\sin^2 f\over r^2}\biggr]+{1\over 2 e^2}{\sin^2 f\over
r^2}
\biggl [{\sin^2 f\over r^2} + 2\biggl({\partial f\over \partial
r}\biggr)^2\biggr] \biggr\} }\label{eq:msk}
\end{eqnarray}
is the mass of a Skyrmion and (see for instance \cite{Walhout})
\begin{equation}
\Lambda = (ef_\pi )^3\int{r^2 dr \sin^2 f\biggl[ 1+ {4\over (ef_\pi)^2 }
\biggl(f'^2+ {\sin^2 f\over r^2}\biggr)\biggr] }
\label{eq:misk}
\end{equation}
is its moment of inertia. This roughly gives the Skyrmion a mass of 850 
MeV (quite close to the nucleon mass), or in natural units of energy 
$\simeq 1.23\times 12 \pi^2\cite{Mantongeo}$. The moment of inertia is roughly 
equal to $(1/195)$ MeV$^{-1}$\cite{Walhout}.

%% file: section3.tex
We now start the first part of this review which is devoted to the study of 
the static sector of the model. Most of the material concerns geometric 
aspects of the model, but a few pages are devoted to the instanton method,
rational maps and Morse theory.

%% file: section31.tex
In this subsection we will first briefly review the theory of non-linear 
deformations of a body. Most of the material comes from Ogden\cite{Ogden} 
and readily generalizes to the case of a field theory. We then take up
the case of the Skyrme model defined on various spatial 
manifolds, presenting the findings of Manton\cite{Mantongeo} and 
Manton and Ruback\cite{Manton-Ruback}. 

\subsubsection{Non-linear deformation of a body}

\begin{figure}
\centering
\mbox{\epsfig{figure= 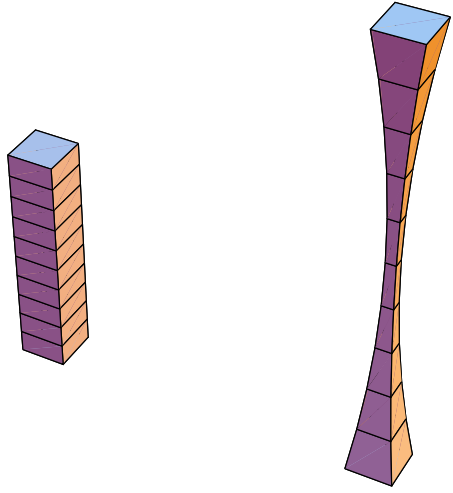,width=4.0truein}}
\caption{Example of deformation of a solid (left) before the deformation and 
(right) after a simple stretching along the $z$ axis.} 
\label{fig:fig1}
\end{figure}

Let us consider a body $B$ at rest, \ie, free of oscillations or 
interior motion: 
the action of exterior forces (like gravity) if any, and that of the
interior forces caused by the nature of the body (the interactions
between the atoms of the body, for instance) exactly cancel each other.
We will call this state of the body its initial configuration.
Let us now change the shape of the body by applying forces to it, until it
reaches a new static configuration. Then it is possible to characterize
this action by a function $\vec \chi$ which we now define. Let $\vec X$ 
be the initial position of a given point $P$ of the body, and $\vec x$ its
position in the final configuration. Then we can define the deformation
function $\vec \chi$ as mapping $\vec X$ to $\vec x$ for every point of the
body:
\begin{eqnarray}
\vec \chi&:&\Rset^3\rightarrow \Rset^3\nonumber
\\
& & \vec X\mapsto \vec x=\vec\chi(\vec X).\label{eq:xchiX}
\end{eqnarray}
For our needs, we will restrict ourselves to $\vec\chi$ being continuous
and twice differentiable.
In the case of a real dynamical system, 
$\vec\chi$ could depend on time but here we will only consider 
static configurations. See Figure $\ref{fig:fig1}$ for an 
example of deformation parametrized by a function $\vec{\chi}$.

To make things more concrete, let us introduce coordinate systems in the 
initial and final configurations. Using two different coordinate systems to 
describe the deformation of a body can seem like overkill but we follow this 
course for two reasons. The first is that it enables one to choose coordinate 
systems which best suit each configuration, simplifying the computations that 
follow (which are usually quite complex in real life problems). The second, 
more important reason is that it will make the jump to Manton's elasticity 
formalism in the frame of field theory easier and more natural. We 
note that we will only be working with bases of vectors which are locally 
orthogonal, and that most of the following equations are written in 
cartesian coordinates. All tensorial equations are readily generalized 
to arbitrary curvilinear systems.

Let $O$ be the origin of the system in the initial configuration and 
$\{\vec E_{\mathrm{m}}\}$ its base. 
Following the convention used by Manton\cite{Mantongeo}, we will use the 
indices $m, n, o, p$ to denote basis vectors. The position 
$\vec X$ of a point of the body is defined by
\begin{eqnarray}
\vec X &=& (\vec X\cdot\vec E_{\mathrm{m}})\; 
\vec E_{\mathrm{m}}\nonumber
\\
&\equiv& X^{\mathrm{m}} \vec E_{\mathrm{m}} 
\end{eqnarray}
where summation over $m$ is implicit. $\vec X$ then has 
coordinates $X^{\mathrm{m}}$ in this coordinate system.

Let $o$ and $\{\vec \xi_\mu\}$ be the origin and base of the coordinate 
system we
choose to describe the final configuration, respectively (We will use
the indices $\mu,\nu\,\rho,\sigma$ here). Then we express $\vec x$
in the final basis as
\begin{eqnarray}
\vec x &=& (\vec x\cdot\vec 
\xi_\mu)\; \vec \xi_\mu\nonumber
\\
&\equiv& x^\mu \vec \xi_\mu.
\end{eqnarray}
Then equation (\ref{eq:xchiX}) can be written in terms of the components
\begin{equation}
x^\mu = \chi^\mu(X^{\mathrm{m}}).\label{eq:xmchixmu}
\end{equation}

The deformation of the body is completely contained in the map $\vec\chi$.
However the map contains much redundancy which we will isolate next.
Under deformation an element of length ${\d}\vec X$ transforms 
according to
\begin{equation}
{\d} \vec x = {\partial \vec \chi\over \partial\vec X} {\d} \vec X
\label{eq:xddX}
\end{equation}
or in coordinates
\begin{equation}
{\d} x^\mu = {\partial \chi^\mu\over \partial X^{\mathrm{m}}} 
{\d} X^{\mathrm{m}}.
\end{equation}
Let us introduce the second order tensor
\begin{equation}
A = A_{\mu {\mathrm{m}}}\; \vec \xi_\mu \otimes \vec 
E_{\mathrm{m}}\label{eq:geo7}
\end{equation}
with
\begin{equation}
A_{\mu {\mathrm{m}}} = {\partial \chi^\mu \over 
\partial X^{\mathrm{m}}}\label{eq:geo8}
\end{equation}
which is called the deformation gradient 
(relative to the reference configuration). 
This is just the local Jacobian of the deformation defined by
(\ref{eq:xchiX}). 

Intuitively it is clear that $A$ completely represents the deformation, 
but as we shall 
demonstrate, it is also sensitive to the choices of the bases 
$\{\vec E_{\mathrm{m}}\}$ 
and $\{\vec \xi_\mu\}$ which have no physical content. For 
instance, a physical quantity such as the amount of energy 
stored in the body by the deformation should not depend on these choices.

Let us consider new bases $\{ \vec E'_{\mathrm{m}}\}$ and $\{ \vec \xi'_\mu\}$ 
defined by
\begin{equation}
\vec E'_{\mathrm{m}} = \Omega_{\mathrm{mn}} \vec 
E_{\mathrm{n}}\label{eq:geo8p1}
\end{equation}
\begin{equation}
\vec \xi'_\mu = W_{\mu\nu} \vec \xi_\nu\label{eq:geo8p2}
\end{equation}
where $\Omega$ and $W$ are orthogonal matrices. The new bases correspond to 
a new choice of orientation for the bases in the initial and final 
configuration. The tensor $A$ expresses itself in each basis as 
follows:
\begin{eqnarray}
A & = & A_{\mu \mathrm{m}}\; \vec \xi_\mu \otimes \vec E_{\mathrm{m}}\nonumber
\\
& = & A'_{\nu \mathrm{n}}\; \vec \xi'_\nu \otimes \vec 
E'_{\mathrm{n}}\label{eq:geo9}
\end{eqnarray}
where
\begin{equation} 
A'_{\nu \mathrm{n}} = W_{\nu\mu} A_{\mu \mathrm{m}} 
(\Omega^{\mathrm{T}})_{\mathrm{mn}}\label{eq:geo10}
\end{equation} 
and $\Omega^{\mathrm{T}}$ is the transpose of $\Omega$.
Therefore $A$ depends strongly on the choice of the bases. 
In the normal theory of 
elasticity, $A$ is considered to be non singular: no 
${\d}\vec X$ can be deformed
into a line element ${\d}\vec x$ with 
zero length. Such an annihilation of a line 
element would imply an infinite force acting on the body, which is  
unrealistic. But in the next subsections, devoted to the study of the Skyrme 
model as a theory of elasticity, we will see that there is physical meaning 
to a singular or in fact zero tensor $A$.

A step toward a better description of the deformation is to use the 
right Cauchy-Green deformation tensor $D = A A^{\mathrm{T}}$. It is symmetric 
and positive definite by construction and under the change of coordinate axes 
mentioned above, $D \rightarrow D' = W D W^{\mathrm{T}}$ which is physically 
more sound but still sensitive to the choice of the coordinate system of the 
final configuration.

The right Cauchy-Green tensor is related in a simple way to the usual strain 
tensor of elasticity. To show this let us define the displacement 
vector $\vec u$: 
\begin{equation}
\vec u(\vec X) = \vec x - \vec X = \vec \chi(\vec X) - \vec X\label{eq:geo10p1}
\end{equation}
which measures how much a point moves during the deformation. It is usually 
assumed to be very small in the ordinary, \ie {\em linear\/}, theory of 
elasticity. 
Define $G$ the displacement gradient
\begin{equation}
G_{\mu \mathrm{m}} = (A - \mathrm{I})_{\mu \mathrm{m}} 
= {\partial u^\mu \over \partial X^{\mathrm{m}}}.
\end{equation}
Then one can define two tensors of strain:
\begin{eqnarray}
E &=& {1\over 2}(G + G^{\mathrm{T}} + G^{\mathrm{T}} G)\nonumber
\\
&=& {1\over 2} ( A A^{\mathrm{T}} - \mathrm{I})={1\over 2}(D-{\mathrm{I}})
\label{eq:defe}
\end{eqnarray}
and
\begin{eqnarray}
F &=& {1\over 2}(G + G^{\mathrm{T}} + G G^{\mathrm{T}})\nonumber
\\
&=& {1\over 2} ( A^{\mathrm{T}}A - \mathrm{I}).\label{eq:deff}
\end{eqnarray}

In component formulation this gives :
\begin{equation}
E_{\mu \mathrm{m}} = {1\over 2}\biggl( {\partial u^\mu\over \partial 
X^{\mathrm{m}}}+
{\partial u^{\mathrm{m}}\over \partial X^\mu}+ \sum_{\mathrm{n}=1}^3
 {\partial u^{\mathrm{n}}\over \partial X^\mu} {\partial u^{\mathrm{n}}\over 
\partial X^{\mathrm{m}}}\biggr)
\end{equation}
and 
\begin{equation}
F_{\mu \mathrm{m}} = {1\over 2}\biggl( {\partial u^\mu\over \partial 
X^{\mathrm{m}}}+
{\partial u^{\mathrm{m}}\over \partial X^\mu}+ \sum_{\mathrm{n}=1}^3
 {\partial u^\mu\over \partial X^{\mathrm{n}}} {\partial u^{\mathrm{m}}\over 
\partial X^{\mathrm{n}}}\biggr).
\end{equation}
$E$ is the ordinary strain tensor considered in Ogden\cite{Ogden} 
while the reader 
will recognize $F$ to be the strain tensor defined in the theory of elasticity,
see for example  
Landau\cite{Landau}. Notice that the physical content of $D$ and $E$ coincide
since they differ by a translation and a factor, as seen from 
equation $(\ref{eq:deff})$.

The non-redundant description of the deformation is furnished by 
functions of the matrix $D$ 
which are invariant under conjugation by orthogonal matrices. 
These functions are given by the secular (determinant) equation of the matrix 
$D$:
\begin{equation}
\det (D- \delta I) = \delta^3 + I_1(D) \delta^2 + I_2(D) \delta + I_3(D) =0.
\label{eq:eqsec}
\end{equation}
One finds
\begin{eqnarray} 
I_1 & = & \tr D\label{eq:defi1}
\\
I_2 & = & {1\over 2}[ (\tr D)^2 - \tr (D^2)]\label{eq:defi2}
\\
I_3 & = & \det D\label{eq:defi3}.
\end{eqnarray}
The set of eigenvalues of $D$ is invariant under orthogonal conjugation, as 
are the coefficients $I_{\mathrm{i}}$, which can be verified easily. The set 
of invariants 
$I_{\mathrm{i}}$ is complete since the eingenvalues of $D$ are 
uniquely determined by equation (\ref{eq:eqsec}). 

These eigenvalues can in fact be 
expressed in terms of the eigenvalues of a matrix obtained from $A$. This
can be seen via the following analysis. Any 
non-singular, real matrix $A$ admits a polar decomposition
\begin{eqnarray}
A &=& R U
\\
&=& V R
\end{eqnarray}
where $R$ is an orthogonal matrix and $U$ and $V$ are symmetric,
non-singular, real 
matrices\cite{Ogden,Reed-Simon}. Let $\vec u_{\mathrm{i}}$ and 
$\lambda_{\mathrm{i}}$ denote the eigenvectors and corresponding eigenvalues
of $U$ 
\begin{equation}
U \vec u_{\mathrm{i}} = \lambda_{\mathrm{i}} \vec u_{\mathrm{i}}
\qquad i = 1,\;2,\;3
\end{equation}
where the $\vec u_{\mathrm{i}}$ can be chosen orthonormal since $U$ is a 
symmetric, real (hermitian) matrix. Then
\begin{equation}
A \vec u_{\mathrm{i}} = R U \vec u_{\mathrm{i}} = 
\lambda_{\mathrm{i}} (R \vec u_{\mathrm{i}}) = 
V (R \vec u_{\mathrm{i}}),\label{eq:valprU}
\end{equation}
hence $R \vec u_{\mathrm{i}}$ is an eigenvector of $V$ with the same 
eigenvalue
$\lambda_{\mathrm{i}}$. 

Now
\begin{equation}
D = AA^{\mathrm{T}} = V^2 = R U^2 R^{\mathrm{T}},
\end{equation}
thus
\begin{equation}
D (R \vec u_{\mathrm{i}}) = \lambda_{\mathrm{i}}^2 (R \vec u_{\mathrm{i}})
\end{equation}
and the three invariants (\ref{eq:defi1}), (\ref{eq:defi2}) and 
(\ref{eq:defi3}) are
\begin{eqnarray} 
I_1 & = & \lambda_1^2 + \lambda_2^2 + \lambda_3^2\label{eq:defi1l}
\\
I_2 & = & \lambda_1^2 \lambda_2^2 + \lambda_2^2 \lambda_3^2 + \lambda_3^2 
\lambda_1^2\label{eq:defi2l}
\\
I_3 & = & \lambda_1^2 \lambda_2^2 \lambda_3^2\label{eq:defi3l}.
\end{eqnarray}
These invariants will be useful in writing the energy stored in the body due
to  
the deformation. The same applies to the Skyrme model, as we will show below.
We can give a physical interpretation of the three invariants in terms of 
stretching of a set of vectors and various associated geometrical quantities.

The simplest invariant is $I_3$ which is related to the change in the volume 
defined by three non-coplanar vectors ${\d}\vec X^{(1)}$, 
${\d}\vec X^{(2)}$ and ${\d}\vec X^{(3)}$, and their images 
${\d}\vec x^{(1)}$, 
${\d}\vec x^{(2)}$  and ${\d}\vec x^{(3)}$ under the deformation. By equation 
(\ref{eq:xddX}), we have
\begin{equation}
{\d}\vec x = A \cdot {\d}\vec X
\end{equation}
which gives
\begin{equation}
{\d} v = \det A\; {\d} V = \lambda_1 \lambda_2 \lambda_3 \;{\d}V
\end{equation}
where
\begin{eqnarray} 
{\d} V & = & {\d}\vec X^{(1)}\cdot {\d}\vec X^{(2)}\times {\d}\vec X^{(3)}
\\
{\d} v & = & {\d}\vec x^{(1)}\cdot {\d}\vec x^{(2)}\times {\d}\vec 
x^{(3)}
\end{eqnarray}
assuming that ${\d}\vec x^{(\mathrm{i})}$ and ${\d}\vec X^{(\mathrm{i})}$ form 
right handed 
triads. So the third invariant is just the square of the change of a 
volume under the deformation. 

Let us define the
stretch of a line element of the body as measured by the ratio between 
the length of the initial line element and that of its image under the
deformation. From (\ref{eq:xddX}) and (\ref{eq:geo8}) we have
\begin{equation}
\hat m |{\d}\vec x| = A \hat M |{\d}\vec X|\label{eq:dxAdX}
\end{equation}
where $\hat m$, $\hat M$ are unit vectors along the direction of ${\d}\vec x$ 
and ${\d}\vec X$ and $|\dots|$ indicates the length. Taking the norm of both
sides of $(\ref{eq:dxAdX})$ we 
define
\begin{equation}
\lambda(\hat M) \equiv {|{\d}\vec x|\over |{\d}\vec X|} 
= \sqrt{\hat M\cdot A^{\mathrm{T}} A \cdot \hat M}
\end{equation}
which gives the stretch in the direction $\hat M$ at 
$\vec X$. Taking $\hat M$ to be colinear to the $n$-th eigenvector 
$\hat u_{\mathrm{n}}$ of $A^{\mathrm{T}}A$ with eigenvalue 
$\lambda_{\mathrm{n}}^2$, we get
\begin{equation}
\lambda^2(\hat u_{\mathrm{n}}) = \lambda_{\mathrm{n}}^2.
\end{equation}
The first invariant $I_1$ is then just the sum of the squares of the stretch 
along the three eigenvectors of $U$. These eigenvectors correspond exactly to 
the usual principal directions of strain. Indeed from 
(\ref{eq:deff}) we see that 
the 
matrix $F$ corresponds to the usual definition of the strain tensor. However
\begin{equation}
F = {1\over 2} (U^2-1)
\end{equation}
hence 
\begin{equation}
F \vec u_{\mathrm{i}} = {1\over 2} (\lambda_{\mathrm{i}}^2 - 1)
\vec u_{\mathrm{i}}
\end{equation}
thus showing that 
the $\vec u_{\mathrm{i}}$ correspond to the principal directions of 
strain. 

The second invariant is the most subtle and is related to the change 
under the deformation of the 
area elements defined by an orthonormal triad. An 
orthonormal triad $\hat v_{\mathrm{i}}$ defines the three area elements
\begin{equation}
{1\over 2} \epsilon_{\mathrm{ijk}} \hat v_{\mathrm{j}}\times
\hat v_{\mathrm{k}}\qquad i,j,k = 1,2,3.
\end{equation}
Under the deformation these are transformed to
\begin{equation}
{1\over 2} \epsilon_{\mathrm{ijk}} (A\hat v_{\mathrm{j}})\times
(A\hat v_{\mathrm{k}}).
\end{equation}
Computing the norm and summing over $i$ yields
\begin{eqnarray}
&&{1\over 4} \epsilon_{\mathrm{ijk}} 
\epsilon_{\mathrm{ilm}} 
(A\hat v_{\mathrm{j}} \times A\hat v_{\mathrm{k}})
\cdot
(A\hat v_{\mathrm{l}} \times A\hat v_{\mathrm{m}}) 
\\
&=&
{1\over 2}
(A\hat v_{\mathrm{j}} \times A\hat v_{\mathrm{k}})
\cdot
(A\hat v_{\mathrm{j}} \times A\hat v_{\mathrm{k}}) 
\\
&=& {1\over 2}\biggl[
(A\hat v_{\mathrm{j}}\cdot A\hat v_{\mathrm{j}})
(A\hat v_{\mathrm{k}}\cdot A\hat v_{\mathrm{k}})  
-
(A\hat v_{\mathrm{j}}\cdot A\hat v_{\mathrm{k}})
(A\hat v_{\mathrm{j}}\cdot A\hat v_{\mathrm{k}})
\biggr]
\end{eqnarray}
With
\begin{equation}
\hat v_{\mathrm{i}} = v^{\mathrm{k}}_{\mathrm{i}} \vec u_{\mathrm{k}}
\end{equation}
where $v_{\mathrm{i}}^{\mathrm{k}}$ are the elements of an orthogonal
matrix (since $\hat v_{\mathrm{i}}$ are orthonormal), we obtain
\begin{equation}
A \hat v_{\mathrm{i}} = v^{\mathrm{k}}_{\mathrm{i}} A \vec u_{\mathrm{k}}
= v^{\mathrm{k}}_{\mathrm{i}} \lambda_{\mathrm{k}} R \vec u_{\mathrm{k}}
\end{equation}
so
\begin{eqnarray}
A \hat v_{\mathrm{j}}\cdot A \hat v_{\mathrm{k}} &=&
v^{\mathrm{l}}_{\mathrm{j}} v^{\mathrm{m}}_{\mathrm{k}}
\lambda_{\mathrm{l}} \lambda_{\mathrm{m}} 
(R \vec u_{\mathrm{l}}\cdot R \vec u_{\mathrm{m}})
\\
&=& v^{\mathrm{l}}_{\mathrm{j}} v^{\mathrm{l}}_{\mathrm{k}}
\lambda_{\mathrm{l}}^2
\end{eqnarray}
Hence the squared norm is now equal to
\begin{eqnarray}
&& {1\over 2} \biggl(
(v^{\mathrm{l}}_{\mathrm{j}} v^{\mathrm{l}}_{\mathrm{j}}\lambda_{\mathrm{l}}^2)
(v^{\mathrm{m}}_{\mathrm{k}} v^{\mathrm{m}}_{\mathrm{k}}\lambda_{\mathrm{m}}^2)
-
(v^{\mathrm{l}}_{\mathrm{j}} v^{\mathrm{l}}_{\mathrm{k}}\lambda_{\mathrm{l}}^2)
(v^{\mathrm{m}}_{\mathrm{j}} v^{\mathrm{m}}_{\mathrm{k}}\lambda_{\mathrm{m}}^2)
\biggr) 
\\
&=& {1\over 2} \biggl(
\sum_{\mathrm{l}} \lambda_{\mathrm{l}}^2
\sum_{\mathrm{m}} \lambda_{\mathrm{m}}^2
-
\sum_{\mathrm{l}} \lambda_{\mathrm{l}}^4
\biggr)
\\
&=& \lambda_1^2  \lambda_2^2 + \lambda_2^2  \lambda_3^2 + \lambda_3^2  
\lambda_1^2
\\
&=& I_2.  
\end{eqnarray}

\subsubsection{Geometrical framework for the Skyrme model}
A field corresponds to a mapping $\pi$ from the manifold $S$ of ordinary 
space to the manifold $\Sigma$ of the target space.
A field theory further specifies the dynamics obeyed by the field 
via the Euler-Lagrange equations obtained from the Lagrangian of the theory.
We will limit ourselves to the case where both $S$ and $\Sigma$ are 3-spheres, 
the former being a 3-sphere of radius $L$, while the latter is the 3-sphere 
of the $SU(2)$ group corresponding to isospin. The case of the ordinary 
Skyrmion defined on $\Rset^3$ is obtained by taking $L\rightarrow+\infty$. 

This modest generalization of the Skyrme model allows for a non-trivial 
application of the geometrical formalism of non-linear elasticity theory which 
we have just elaborated. Varying the radius $L$ allows us to cover the cases 
from extreme to zero curvature.

The map $\pi$ describes the Skyrme field, whether it is a group of waves with 
zero baryon number, a Skyrmion or a heavy nucleus of high baryon number. Even 
if $\pi$  bears similarity with the deformation map $\vec \chi$ of the above 
elasticity theory, it is different in a fundamental way: it maps a curved 
space onto another curved space. $\vec \chi$ only maps a set of points in 
$\Rset^3$ onto another set of points in $\Rset^3$. Apart from this fundamental 
difference, there are many similarities between the treatment of elastic 
bodies and field theories.

The initial spatial manifold with a given metric comes equipped with a 
tangent space at each point. The tangent space $T_{\mathrm{p}}(S)$ 
of $S$ at the point 
$p$ has a natural basis $\{\partial/\partial p^{\mathrm{i}}\}$. Although
linearly independent, these vectors are not necessarily
orthonormal. By a linear 
transformation, we can construct an orthonormal basis 
$\{\hat e_{\mathrm{m}}\}$ ($m=1$, 2, 3) of $T_{\mathrm{p}}(S)$ given by 
\begin{equation}
\hat e_{\mathrm{m}} = e^{\mathrm{i}}_{\mathrm{m}} 
{\partial\over\partial p^{\mathrm{i}}}
\end{equation}
where the coefficient of the linear transformation are called the dreibein. 
Doing the same construction at every point $p$ of $S$ defines an orthonormal 
frame at each point of $S$. We will follow the same convention as Manton and
use indices $i,j,k,l$ with reference to the coordinate basis and $m,n,o,p$ 
with reference to the orthonormal basis. 
If the reader has difficulty with these notions, we recommend the 
references already mentioned\cite{Flanders,Nakahara,Nash,Eguchi}.  

With coordinates $\pi^\alpha$ on $\Sigma$ and its given metric 
$\tau^{\alpha\beta}$ ($\alpha,\beta,
\delta,\gamma$ coordinate indices and $\mu,\nu,\rho,\sigma$ orthonormal basis 
indices), consider the image of the orthonormal frame field 
$\{\hat e_{\mathrm{m}}\}$
under $\pi$. 
We let $\pi^\alpha=\pi^\alpha(p^{\mathrm{i}})$ be the coordinates of the 
image of $p^{\mathrm{i}}$ for efficiency of notation.
According to the transformation law of a vector
\begin{equation}
e^{\mathrm{i}}_{\mathrm{m}} \rightarrow e^{\mathrm{i}}_{\mathrm{m}} 
{\partial \pi^\alpha\over \partial p^{\mathrm{i}}}.
\end{equation}
It is evident that the lengths and directions of the orthonormal triad 
are changed:
it is generally no longer orthonormal. As long as the Jacobian of the 
transformation
\begin{equation}
J^\alpha_{\mathrm{i}} = {\partial \pi^\alpha\over \partial p^{\mathrm{i}}}
\end{equation}
is non-singular, the image triad defines a basis of the image tangent space. 
This is the generic case but not at all the relevant one in many physical
situations, as we will see below.

The (inverse) metric of space $S$ is given by
\begin{equation}
t^{\mathrm{ij}} = \delta^{\mathrm{mn}} e^{\mathrm{i}}_{\mathrm{m}} 
e^{\mathrm{j}}_{\mathrm{n}}
\end{equation}
where $\delta^{\mathrm{mn}}$ is the usual Kronecker delta. Its image 
under $\pi$ is
\begin{equation}
t'^{\alpha\beta} = \delta^{\mathrm{mn}} e^{\mathrm{i}}_{\mathrm{m}} 
{\partial \pi^\alpha\over \partial p^{\mathrm{i}}}
e^{\mathrm{j}}_{\mathrm{n}} {\partial \pi^\beta\over \partial 
p^{\mathrm{j}}}.
\end{equation}
As we will see later the degree to which $t'^{\alpha\beta}$ differs from the 
intrinsic metric $\tau^{\alpha\beta}$ already existant on $\Sigma$, is a 
measure of the lack of isometricity of the map $\pi$ and the general energy 
functional for Skyrme type models measures this non-isometricity (an isometry 
is a map which preserves the metric \ie distances are left unchanged 
by the map). Intuitively the mapping $\pi$ produces a strain and the energy 
functional is a measure of the energy attributed to this strain. If
\begin{equation}
t'^{\alpha\beta} = \tau^{\alpha\beta}
\end{equation}
at every point of $\Sigma$ then the map is an isometry.

There are in fact four ways of testing whether or not the map is an isometry, 
in the event $\pi$ is invertible. In the following paragraph we will make 
precise 
the notions of push-forward, pull-back of the inverse metrics and metrics 
respectively of the spaces $S$ and $\Sigma$.

\begin{figure}
\centering
\mbox{\epsfig{figure= 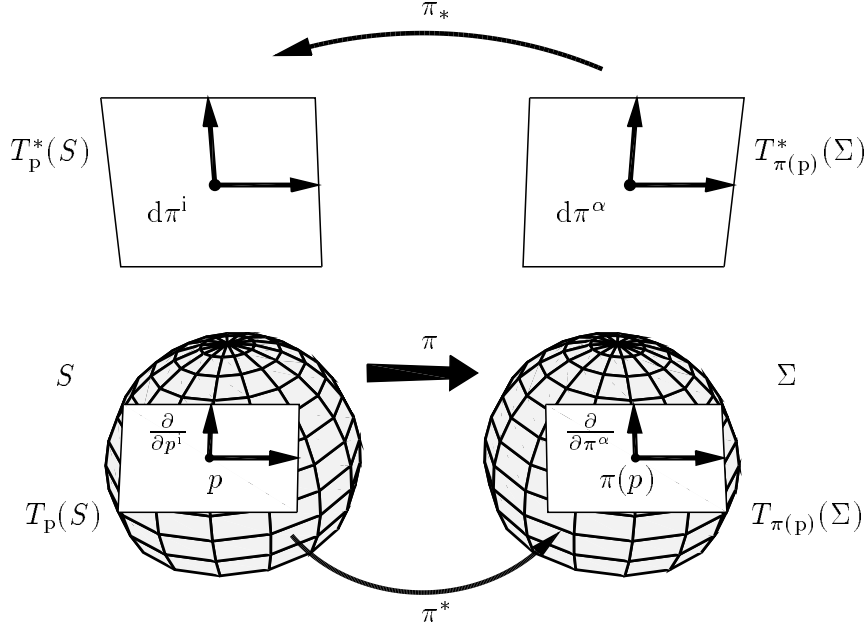,height=4.0in}}
\caption{Diagram of the push-forward and pull-back for the mapping 
$\pi:S\rightarrow\Sigma$. The same construction is possible for the inverse 
mapping $\pi^{-1}:\Sigma\rightarrow S$.}\label{fig:fig2}
\end{figure}

Any map $\pi$ between manifolds $S$ and $\Sigma$ defines a map $\pi^*$ called 
the {\em push-forward\/} between the corresponding tangent spaces and 
$\pi_*$ called the {\em pull-back\/} between their dual spaces (see figure 
$\ref{fig:fig2}$). We remind the reader that the dual 
space, sometimes called the space of differential forms, is simply the space 
of real valued {\em linear\/} functions of the tangent space. 
This means if $v\in 
T_{\mathrm{p}}$ and $\omega\in T^*_{\mathrm{p}}$ ($T^*_{\mathrm{p}}$ is just
notation for the dual 
space), $\omega$ is a 
linear function taking $v$ to $\Rset$, and we write 
$\langle\omega,v\rangle\in \Rset$ which is 
also called the contraction of $\omega$ with $v$. 
The mapping 
$\pi$ is defined as follows: 
\begin{eqnarray}
\pi: & S &\rightarrow \Sigma\nonumber
\\
& p &\mapsto \pi(p).
\end{eqnarray}
The {\em push-forward\/} $\pi^*$ is defined by
\begin{equation}
\pi^*:T_{\mathrm{p}}(S)\rightarrow T_{\pi(\mathrm{p})}(\Sigma)
\end{equation}
where $T_{\mathrm{p}}(S)$ is the tangent space at the point $p$ to 
the manifold $S$ 
while $T_{\pi(\mathrm{p})}(\Sigma)$ is the tangent space at the point 
$\pi(p)$ to the 
manifold $\Sigma$. A vector $v\in T_{\mathrm{p}}(S)$ with components 
$v^{\mathrm{i}}$ with respect 
to the coordinate basis $\{\partial/\partial p^{\mathrm{i}}\}$
\begin{equation}
v = v^{\mathrm{i}} {\partial\over \partial p^{\mathrm{i}}}
\end{equation}
is mapped to the vector $\pi^*(v)$ on $\Sigma$ given in terms of the 
coordinate basis $\{\partial/\partial\pi^\alpha\}$,
\begin{equation}
\pi^*(v) = v^{\mathrm{i}} {\partial \pi^\alpha\over \partial p^{\mathrm{i}}}\; 
{\partial\over \partial \pi^\alpha}
\end{equation}
hence the components transform as
\begin{equation}
v^{\mathrm{i}} \rightarrow {\partial \pi^\alpha\over \partial 
p^{\mathrm{i}}}\; v^{\mathrm{i}}.
\end{equation}
The {\em pull-back\/} $\pi_*$ is defined as follows:
\begin{equation}
\pi_*: T_{\pi(\mathrm{p})}^*(\Sigma) \rightarrow T^*_{\mathrm{p}}(S)
\end{equation}
where $T^*_{\pi(\mathrm{p})}(\Sigma)$ 
is the dual space to $T_{\pi(\mathrm{p})}(\Sigma)$ 
\ie the space of differential 1-forms, while $T^*_{\mathrm{p}}(S)$ is dual to 
$T_{\mathrm{p}}(S)$. We remind the reader that the coordinate basis 
of the cotangent 
space is defined with respect to the coordinate basis of the tangent space by 
the condition 
$\langle {\d}\pi^\alpha,\partial/\partial\pi^\beta\rangle=
\delta^\alpha_\beta$. A 1-form $\omega \in T^*_{\pi(\mathrm{p})}(\Sigma)$ with 
components $\omega_\alpha$ in the dual coordinate basis $\{{\d}\pi^\alpha\}$,
\begin{equation}
\omega = \omega_\alpha\; {\d}\pi^\alpha
\end{equation}
is mapped to the 1-form $\pi_*(\omega)$ of $T^*_{\mathrm{p}}(S)$ 
with dual coordinate 
basis $\{{\d} p^{\mathrm{i}}\}$
\begin{equation}
\pi_*(\omega) = \omega_\alpha\; {\partial \pi^\alpha\over \partial 
p^{\mathrm{i}}}\; {\d} p^{\mathrm{i}}
\end{equation}
hence the components transform as
\begin{equation}
\omega_\alpha\rightarrow \omega_\alpha\; {\partial \pi^\alpha\over 
\partial p^{\mathrm{i}}}.
\end{equation}
These transformation rules generalize tensorially on tensor products of the 
tangent and dual tangent spaces. Evidently, if $\pi$ is invertible then
\begin{equation}
\pi^{-1}: \Sigma \rightarrow S
\end{equation}
defines a push-forward and a pull-back in the opposite sense. We now
come to the point of computing images 
of the metric and the inverse metric under $\pi$ and under its inverse (if it 
exists). 

Starting with the metric $t$ on $S$ and $\tau$ on $\Sigma$ we have
\begin{eqnarray}
t &=& t_{\mathrm{ij}}\; {\d} p^{\mathrm{i}}\otimes {\d} p^{\mathrm{j}}
\\
\tau &=& \tau_{\alpha\beta}\; {\d}\pi^\alpha\otimes {\d}\pi^\beta
\end{eqnarray}
the corresponding pull-backs via $\pi\equiv\pi(p)$ for $\tau$ and
$\pi^{-1}\equiv p(\pi)$ for $t$ are
\begin{equation}
\tau' = \tau_{\alpha\beta} {\partial \pi^\alpha\over\partial p^{\mathrm{i}}} 
{\partial\pi^\beta\over \partial p^{\mathrm{j}}}\; {\d} p^{\mathrm{i}}
\otimes {\d} p^{\mathrm{j}}
\; \in  T^*_{\mathrm{p}}(S)\otimes T_{\mathrm{p}}^*(S)
\end{equation}
and
\begin{equation}
t' = t_{\mathrm{ij}} {\partial p^{\mathrm{i}}\over \partial\pi^\alpha} 
{\partial p^{\mathrm{j}}\over \partial\pi^\beta}\;
{\d}\pi^\alpha\otimes {\d}\pi^\beta\; \in T_{\pi(\mathrm{p})}^*(\Sigma) 
\otimes T_{\pi(\mathrm{p})}^*(\Sigma).
\end{equation}
Also for the inverse metrics 
\begin{equation}
t^{-1} =t^{\mathrm{ij}}\; {\partial\over \partial p^{\mathrm{i}}} 
\otimes{\partial\over\partial p^{\mathrm{j}}}
\end{equation}
\begin{equation}
\tau^{-1} = \tau^{\alpha\beta}\; {\partial\over\partial\pi^\alpha}\otimes
{\partial\over\partial\pi^\beta}
\end{equation}
we have the corresponding push-forwards via $\pi$ for $t^{-1}$ and 
$\pi^{-1}$ for $\tau^{-1}$,
\begin{equation}
(t^{-1})'  =  t^{\mathrm{ij}} {\partial\pi^\alpha\over\partial p^{\mathrm{i}}}
{\partial\pi^\beta\over\partial p^{\mathrm{j}}}\; 
{\partial\over\partial\pi^\alpha}\otimes
{\partial\over\partial\pi^\beta}
\;\in  T_{\pi(\mathrm{p})}(\Sigma) \otimes T_{\pi(\mathrm{p})}(\Sigma)
\end{equation}
and
\begin{equation}
(\tau^{-1})'  =  \tau^{\mu\nu} {\partial p^{\mathrm{m}}\over\partial\pi^\mu}
{\partial p^{\mathrm{n}}\over\partial \pi^\nu}\; {\partial\over \partial
p^{\mathrm{m}}}\otimes {\partial\over\partial p^{\mathrm{n}}}
\;\in T_{\mathrm{p}}(S) \otimes T_{\mathrm{p}}(S).
\end{equation}
Now the condition that the map is an isometry is given by any of the 
following statements:
\begin{eqnarray}
t' = \tau\qquad & \Leftrightarrow & \qquad 
t_{\mathrm{ij}} {\partial p^{\mathrm{i}}\over \partial\pi^\alpha} 
{\partial p^{\mathrm{j}}\over \partial\pi^\beta} = \tau_{\alpha\beta}
\label{eq:tptau}
\\
\tau' = t\qquad & \Leftrightarrow & \qquad
\tau_{\alpha\beta} {\partial \pi^\alpha\over\partial p^{\mathrm{i}}} 
{\partial\pi^\beta\over \partial p^{\mathrm{j}}} = t_{\mathrm{ij}}
\label{eq:taupt}
\\
(t^{-1})' = \tau^{-1}\qquad & \Leftrightarrow & \qquad
t^{\mathrm{ij}} {\partial\pi^\alpha\over\partial p^{\mathrm{i}}} 
{\partial\pi^\beta\over\partial p^{\mathrm{j}}} = \tau^{\alpha\beta}
\label{eq:tm1ptaum1}
\\
(\tau^{-1})'=t^{-1}\qquad & \Leftrightarrow & \qquad
\tau^{\alpha\beta} {\partial p^{\mathrm{i}}\over\partial\pi^\alpha} 
{\partial p^{\mathrm{j}}\over\partial \pi^\beta} = t^{\mathrm{ij}}
\label{eq:taum1ptm1}
\end{eqnarray}
If any one of these equations is true, they are all true. Equations 
(\ref{eq:tptau}) and (\ref{eq:taum1ptm1}) require that $\pi$ is invertible to 
make sense. As we will see, it is not necessary for $\pi$ to be invertible, 
hence (\ref{eq:taupt}) and (\ref{eq:tm1ptaum1}) are more fundamental.
They are all algebraically identical when $\pi$ is invertible. 
We take (\ref{eq:taupt}) as the 
fundamental relation imposing (locally at the point $p$) isometry, since 
(\ref{eq:taupt}) does not require the inverse mapping to exist.
(\ref{eq:tm1ptaum1}) is equally suitable.

We give an explicit example of the preceding formalism with the 
$\mathrm{SU}(2)$ 
Skyrme field defined on $S^3$. We take $S$, the initial spatial manifold, 
to be a topological and metrical $S^3$ of radius $L$. The target manifold is 
the manifold of the group $\mathrm{SU}(2)$ which also happens to be an $S^3$. 
Group 
manifolds come equipped with a natural metric, the so-called Haar measure, 
and this gives a natural radius of one to the target 3-sphere. 

With the Cartesian coordinates $X^1,X^2,X^3,X^4$ on $\Rset^4$, we define a 
3-sphere of radius $L$ embedded in $\Rset^4$ by the constraint
\begin{equation}
\sum_{\mathrm{i}=1}^{3} X^{\mathrm{i}} X^{\mathrm{i}} +(X^4)^2 = L^2.
\end{equation}
With the relation
\begin{equation}
x^{\mathrm{i}} = {2 L\over L - X^4}\; X^{\mathrm{i}}\label{eq:chcoordSR}
\end{equation}
where $i=1,2,3$ we effect the stereographic projection to $\Rset^3$ and obtain
the following metric 
\begin{equation}
t = {1\over (r^2/4 L^2 +1)^2} \sum^3_{{\mathrm{i}}=1} {\d} x^{\mathrm{i}}
\otimes {\d} x^{\mathrm{i}}
\end{equation}
where $r^2 = \sum^3_{\mathrm{i}=1}x^{\mathrm{i}} x^{\mathrm{i}}$. Thus we see 
that a stereographic projection
is simply a conformal transformation of flat space
(\ie the metric only changes by 
an overall, space dependent, scaling).  

The natural metric on the target manifold is best expressed in terms of the 
left-invariant 1-forms. These are a natural basis of the 
co-tangent space of $\mathrm{SU}(2)$. If $\pi^\alpha$ ($\alpha=1,2,3$) is any 
set of local coordinates on $\mathrm{SU}(2)$, they are defined by the
following generalisation of $(\ref{eq:1flu})$:
\begin{equation}
{\cal L}^\mu = -{{\mathrm i} \over 2} \tr \biggl(\tau^\mu U^\dagger(\pi) 
{\partial \over \partial \pi^\alpha} U(\pi) \biggr) d\pi^\alpha
\end{equation}
where $\tau^\mu$ are the Pauli matrices and $U$ taken for convenience to 
be in the fundamental representation of ${\mathrm SU}(2)$
(our notation is consistent for this subsection, later we will revert to the 
original notation of section 2).
These are left 
invariants since under the transformation
\begin{equation}
U\rightarrow V\; U
\end{equation}
where $V$ is a constant element of ${\mathrm SU}(2)$, ${\cal L}^\mu$ are 
invariant. (One can also define right invariant 1-forms by exchanging the 
role of $U$ and $U^\dagger$ as in $(\ref{eq:1fru})$). 

For example let us take the coordinates $\{\vec\pi\}$ defined by
\begin{equation}
U = \sqrt{1-\vec\pi^2} + {\mathrm i}\, \vec \pi\cdot\vec\tau = 
\pi_0 + {\mathrm i}\,\vec\pi\cdot\vec\tau.
\end{equation}
Note that $U$ is a function of $\pi^1,\pi^2,\pi^3$ and as such, is not a 
covariant expression, in the tensorial sense. Indeed, $\pi^\alpha$ are also 
just coordinates and do not transform tensorially either. Hence the expression 
for ${\cal L}^\mu$ in the specific coordinates chosen does not appear as a 
tensorial expression.  We find:
\begin{equation}
{\cal L}^\mu = {1\over\sqrt{1-\vec\pi^2}} \sum_{\alpha=1}^3 \biggl(
\delta^{\mu\beta}+\pi^\mu\pi^\beta - \delta^{\mu\beta} \vec\pi^2 +
\sum_{\beta=1}^3 \epsilon^{\mu\gamma\beta} \pi^\gamma\biggr)\; {\d}\pi^\beta
\label{eq:lmupi}
\end{equation}
which are the well known Maurer-Cartan forms\cite{Eguchi} 
written in this coordinate system.

This set of 1-forms is natural since, first of all, at the identity where 
$\vec\pi=\vec 0$, 
\begin{equation}
{\cal L}^\mu = {\d}\pi^\mu.
\end{equation}
Secondly we can obtain ${\cal L}^\mu$ at any other point in the group via 
the pull-back 
of an appropriate mapping defined using the group multiplication. Consider a 
general element $V_0$ in the group with corresponding coordinates 
$\vec\pi_{\mathrm{V}_0}$. 
The mapping of a neighborhood ${\cal V}_{\mathrm{V}_0}$ of 
$\vec\pi_{\mathrm{V}_0}$ to a 
neighborhood
${\cal V}_{\mathrm{I}}$ of the identity $\mathrm{I}$ is furnished by multiplication by 
$V_0^\dagger$. If a general element of ${\cal V}_{\mathrm{V}_0}$ is 
noted by $V$ with 
coordinates $\vec\pi_{\mathrm{V}}$, and a general element of 
${\cal V}_{\mathrm{I}}$ is noted by $U$ with 
coordinates $\vec\pi$ the map
\begin{eqnarray}
V_0^\dagger:& & {\cal V}_{\mathrm{V}_0}\rightarrow {\cal V}_{\mathrm{I}}
\nonumber
\\
& & V \mapsto U = V_0^\dagger V = (\sqrt{1-\vec\pi_{\mathrm{V}_0}^2} -
{\mathrm i}\, \vec\pi_{\mathrm{V}_0}\cdot\vec\tau) 
(\sqrt{1-\vec\pi_{\mathrm{V}}^2} -
{\mathrm i}\, \vec\pi_{\mathrm{V}}\cdot\vec\tau) 
\nonumber
\\
& & \qquad\qquad\qquad\;\;\equiv \sqrt{1-\vec\pi(\vec\pi_{\mathrm{V}_0})^2} 
+ {\mathrm i}\, \vec\pi(\vec\pi_{\mathrm{V}_0})\cdot\vec\tau
\end{eqnarray}
which gives 
\begin{equation}
\vec\pi(\vec\pi_{\mathrm{V}}) = - \sqrt{1-\vec\pi_{\mathrm{V}}^2}\; 
\vec\pi_{\mathrm{V}_0} +
\sqrt{1-\vec\pi_{\mathrm{V}_0}^2}\; \vec\pi_{\mathrm{V}} + 
\vec\pi_{\mathrm{V}_0}\times\vec\pi_{\mathrm{V}}.
\end{equation}
This induces the pull-back of ${\d}\pi^\mu = 
({\cal L}^\mu_\alpha|_{\vec\pi=\vec 0}){\d}\pi^\alpha$,
clearly where ${\cal L}^\mu_\alpha|_{\vec\pi=\vec 0} = \delta^\mu_\alpha$,
\begin{eqnarray}
{V^\dagger_0}_*({\cal L}^\mu_\alpha\biggl|_{\vec\pi=\vec 0}
\!\!\!\!\!\!\!\!{\d}\pi^\alpha) 
&=& {\cal L}^\mu_\alpha\biggl|_{\vec\pi=\vec 0} 
\;{\partial\pi^\alpha\over\partial\pi_{\mathrm{V}}^\beta}
\Biggl|_{\vec\pi_{\mathrm{V}} = \vec\pi_{\mathrm{V}_0}} \; 
{\d}\pi_{\mathrm{V}}^\beta
\nonumber
\\
&=& {\partial\pi^\mu\over\partial\pi_{\mathrm{V}}^\beta}
\Biggl|_{\vec\pi_{\mathrm{V}} = \vec\pi_{\mathrm{V}_0}} \; {\d}\pi^\beta
\nonumber
\\
&=& {1\over \sqrt{1-\vec\pi_{\mathrm{V}_0}^2}}
\biggl( \delta^{\mu\beta} + (\pi_{\mathrm{V}_0}^\mu\pi_{\mathrm{V}_0}^\beta - 
\vec\pi_{\mathrm{V}_0}^2 \delta^{\mu\beta} ) 
+ \epsilon^{\mu\gamma\beta} \pi_{\mathrm{V}_0}^\gamma \biggr) 
{\d}\pi_{\mathrm{V}}^\beta
\end{eqnarray}
which is exactly as we had found before in equation $(\ref{eq:lmupi})$, with 
$\vec\pi=\vec\pi_{\mathrm{V}_0}$. 

The metric on $SU(2)$ is given by
\begin{equation}
\tau = \delta_{\mu\nu}\, {\cal L}^\mu \otimes {\cal L}^\nu
\end{equation}
which clearly indicates the orthonormality of ${\cal L}^\alpha$, knowing that 
${\cal L}^\alpha_\mu$ are invertible as matrices at each point in the group. 
A short calculation then shows that this implies the metric
\begin{equation}
\tau=-\tr \biggl( U^\dagger(\pi) {\partial U(\pi)\over \partial \pi^\mu}
U^\dagger(\pi) {\partial U(\pi)\over \partial \pi^\nu} \biggr)\;
{\d}\pi^\mu\otimes {\d}\pi^\nu.
\end{equation}

Now we return to our setting where $\pi^\mu$ the coordinates on the group 
3-sphere are functions of $x^{\mathrm{i}}$ the coordinates on the spatial 
3-sphere, since 
we consider a mapping (that we call $\pi$) between these two spaces. The 
pull-back of the metric to the spatial $S^3$ via the mapping is given by
\begin{eqnarray}
\tau' &=& \pi_*(\tau) = \tau_{\alpha\beta}\;
{\partial\pi^\alpha\over\partial x^{\mathrm{i}}}
{\partial\pi^\beta\over\partial x^{\mathrm{j}}}\; 
{\d} x^{\mathrm{i}}\otimes {\d} x^{\mathrm{j}}
\nonumber
\\
&=& -\tr \biggl( U^\dagger(\pi) {\partial U(\pi)\over\partial \pi^\alpha}
 U^\dagger(\pi) {\partial U(\pi)\over\partial \pi^\beta}\biggr)
{\partial\pi^\alpha\over\partial x^{\mathrm{i}}} 
{\partial\pi^\beta\over\partial x^{\mathrm{j}}} \; 
{\d} x^{\mathrm{i}}\otimes {\d} x^{\mathrm{j}}
\nonumber
\\
&=& -\tr \biggl( U^\dagger {\partial U\over\partial x^{\mathrm{i}}}
U^\dagger {\partial U \over\partial x^{\mathrm{j}}}\biggr)\; 
{\d} x^{\mathrm{i}}\otimes {\d} x^{\mathrm{j}}
\nonumber
\\
&=& {2\over 1 - \vec\pi^2} {\partial \pi^\alpha\over\partial x^{\mathrm{i}}}
\biggl( \delta^{\alpha\beta} + 
(\pi^\alpha\pi^\beta - \vec\pi^2 \delta^{\alpha\beta}) \biggr) 
{\partial \pi^\beta\over\partial x^{\mathrm{j}}}\; {\d} x^{\mathrm{i}}
\otimes {\d} x^{\mathrm{j}}
\end{eqnarray}
where the last line is relevant to the coordinate system chosen on the group. 

The kinetic term of the Skyrme Lagrangian is obtained by contracting $\tau'$ 
with the inverse metric on the spatial manifold
\begin{equation}
t^{-1} = (1+r^2/4 L^2)^2 \delta^{\mathrm{ij}} {\partial\over\partial 
x^{\mathrm{i}}}\otimes {\partial\over\partial x^{\mathrm{j}}},
\end{equation}
\begin{eqnarray}
\langle\tau',t^{-1}\rangle & = & t^{\mathrm{ij}} 
{\partial\pi^\alpha\over\partial p^{\mathrm{i}}}
{\partial \pi^\beta\over \partial p^{\mathrm{j}}} \tau_{\alpha\beta}
\nonumber
\\
& = & - (1+r^2/4 L^2)^2 \sum^{3}_{\mathrm{i}=1} \tr \biggl( U^\dagger 
{\partial\over\partial x^{\mathrm{i}}} U\; U^\dagger 
{\partial\over\partial x^{\mathrm{i}}} U
\biggr).
\end{eqnarray}
Including the volume measure $\sqrt{g} = 1/(1+r^2/4 L^2)^3$
\begin{equation}
\sqrt{g} \langle\tau',t^{-1}\rangle = - {1\over 1+r^2/4 L^2}
\sum^{3}_{\mathrm{i}=1} \tr \biggl( U^\dagger
{\partial\over\partial x^{\mathrm{i}}} U\; 
U^\dagger {\partial\over\partial x^{\mathrm{i}}} U \biggr)
\end{equation}
which has the correct limit as $L\rightarrow +\infty$.

\subsubsection{Non-linear elasticity theory on a curved space and the Skyrme 
model}

Now we make the connection with the non-linear elasticity theory that we have
treated previously, except generalized to a curved space. Hence we suppose 
that $X^{\mathrm{i}}$ are coordinates on a curved space and that the particle 
$p$ of the body at the point $X_{\mathrm{p}}^{\mathrm{i}}$ is mapped to the 
point $x^{\mathrm{i}}_{\mathrm{p}} = 
\chi^{\mathrm{i}}(X^{\mathrm{j}}_{\mathrm{p}})$. The metric at the 
initial point is $t_{\mathrm{ij}}(X^{\mathrm{k}}_{\mathrm{p}})$ while at the 
image point it is $\tau_{\mathrm{ij}}(x^{\mathrm{k}}_{\mathrm{p}})
\equiv t_{\mathrm{ij}}(x^{\mathrm{k}})$. 
The initial triad is $e^i_{\mathrm{m}}(X^{\mathrm{k}}_{\mathrm{p}})$ while 
the final triad is 
$\xi^{\mathrm{i}}_{\mathrm{m}}(x^{\mathrm{k}}_{\mathrm{p}})\equiv 
e^{\mathrm{i}}_{\mathrm{m}}(x^{\mathrm{k}}_{\mathrm{p}})$. 
The initial orthonormal triad is mapped to the final triad in the following 
way:
\begin{equation}
\hat e_{\mathrm{m}}=e^{\mathrm{i}}_{\mathrm{m}}(X^{\mathrm{k}}) 
{\partial\over\partial X^{\mathrm{i}}} \rightarrow 
e^{\mathrm{i}}_{\mathrm{m}}(X^{\mathrm{k}}) 
{\partial\chi^\alpha\over\partial X^{\mathrm{i}}}
{\partial\over\partial x^\alpha}
= \e^{\mathrm{i}}_{\mathrm{m}}(X^{\mathrm{k}}) 
{\partial\chi^\alpha\over\partial X^{\mathrm{i}}}
\xi_\alpha^\mu(x^{\mathrm{k}}) \hat\xi_\mu(x^{\mathrm{k}})
\end{equation}
where $\xi_\alpha^\mu(x^{\mathrm{k}})$ is the inverse 
dreibein and $\hat\xi_\mu \equiv \xi^\beta_\mu\; \partial/\partial x^\beta$. 
Hence the analog of the tensor $A_{\mu \mathrm{m}}$ is
\begin{equation}
\tilde A_{\mathrm{m}}^\mu = e^{\mathrm{j}}_{\mathrm{m}}
(X^{\mathrm{k}}_{\mathrm{p}}) {\partial\chi^\alpha\over\partial 
X^{\mathrm{j}}}_{\mathrm{p}} \xi^\mu_\alpha(x^{\mathrm{k}}_{\mathrm{p}})
\end{equation}
which is the local Jacobian of the deformation. 

The right-Cauchy-Green deformation tensor is
\begin{equation}
D_{\mathrm{mm'}} = \tilde A_{\mathrm{m}}^{\mathrm{n}} 
\tilde A_{\mathrm{m'}}^{\mathrm{n}}
\end{equation}
and the ordinary strain tensor is given by
\begin{equation}
E = {1\over 2} (D - I).
\end{equation}   
The invariants are defined in the same way as before. 

Now we go to the completely general situation where we are mapping between two
different curved spaces. The deformation matrix now generalizes to
\begin{equation}
J_{\mathrm{m}}^\mu = e^{\mathrm{i}}_{\mathrm{m}} 
{\partial\pi^\alpha\over\partial p^{\mathrm{i}}} \zeta_\alpha^\mu
\label{eq:jmum}
\end{equation}
where the inverse dreibein $\zeta_\alpha^\mu$ is defined by the 
orthonormal basis $\hat \zeta_\mu 
\equiv \zeta^\alpha_\mu\; \partial/\partial\pi^\alpha$ 
in the space tangent to $\Sigma$,
while the equivalent strain tensor is
\begin{eqnarray}
D_{\mathrm{mn}} & = & \sum_{\mu=1}^3 J_{\mathrm{m}}^\mu J_{\mathrm{n}}^\mu
\nonumber
\\
& = & \sum_{\mu=1}^3 e^{\mathrm{i}}_{\mathrm{m}} 
{\partial\pi^\alpha\over\partial p^{\mathrm{i}}} 
\zeta_\alpha^\mu e^{\mathrm{j}}_{\mathrm{n}} 
{\partial\pi^\beta\over\partial p^{\mathrm{j}}} \zeta_\beta^\mu
\nonumber
\\
& = & e^{\mathrm{i}}_{\mathrm{m}} 
{\partial\pi^\alpha\over\partial p^{\mathrm{i}}} 
{\partial\pi^\beta\over\partial p^{\mathrm{j}}} 
e^{\mathrm{j}}_{\mathrm{n}} \tau_{\alpha\beta}.
\end{eqnarray}
Finally, taking the first invariant of $D$ by tracing over $m$ and $n$ gives
\begin{eqnarray}
\tr D & = & \sum_{\mathrm{m}=1}^3 
e_{\mathrm{m}}^{\mathrm{i}}
{\partial\pi^\alpha\over\partial p^{\mathrm{i}}} e^{\mathrm{j}}_{\mathrm{m}}
{\partial\pi^\beta\over\partial p^{\mathrm{j}}} \tau_{\alpha\beta}
\nonumber
\\
& = & t^{\mathrm{ij}} {\partial\pi^\alpha\over\partial p^{\mathrm{i}}} 
{\partial\pi^\beta\over\partial p^{\mathrm{j}}} \tau_{\alpha\beta}
\nonumber
\\
& = & -(1+r^2/4 L^2)^2 \sum_{\mathrm{i}=1}^3 \tr
\biggl ( U^\dagger {\partial\over\partial x^{\mathrm{i}}} U \;
U^\dagger {\partial\over\partial x^{\mathrm{i}}} U
\biggr)\label{eq:trd}
\\
& = & \lambda_1^2 + \lambda_2^2 + \lambda_3^2\label{eq:trdl}
\end{eqnarray}
Equation $(\ref{eq:trd})$ is clearly the usual kinetic term of the Skyrme 
model Lagrangian (in the limit where $L\rightarrow +\infty$).
 
The Skyrme term is obtained from the curvature tensor defined on the group 
manifold, which is pulled back to the space manifold and then contracted 
twice with the inverse metric there. 
The curvature tensor is most efficiently defined via the machinery of the 
exterior algebra and the spin connection. The spin connection is a 1-form 
\begin{equation}
\omega^\nu_\lambda = \omega_{\alpha\lambda}^\nu {\d}\pi^\alpha \equiv
\omega_{\mu\lambda}^\nu {\cal L}^\mu,\label{eq:spincon}
\end{equation}
which satisfies the structure equation
\begin{equation}
{\d} {\cal L}^\mu + \omega^\mu_\lambda \wedge {\cal L}^\lambda = 0
\label{eq:eqstr}
\end{equation}
and the ``metricity'' condition
\begin{equation}
\omega_{\nu\lambda} = - \omega_{\lambda\nu}
\end{equation}
where $\wedge$ is the wedge or exterior product (which is simply the 
antisymmetrized tensor product of the forms in question), and 
$\omega_{\nu\lambda}$ is the spin connection of equation 
$(\ref{eq:spincon})$ with  
index lowered by $\delta_{\mu\nu}$. The conditions 
$(\ref{eq:spincon})$ and $(\ref{eq:eqstr})$ are exactly 
equivalent to the conditions in the usual formulation of differential 
geometry that there is no torsion (the Christoffel symbol is symmetric in its 
lower two indices) and the metric is covariantly conserved (metricity).

The curvature is then given by the 2-form
\begin{eqnarray}
R^\mu_\nu & = & R^\mu_{\nu\lambda\rho}\; {\cal L}^\lambda\wedge{\cal L}^\rho 
\\
& = & {\d}\omega^\mu_\nu + \omega^\mu_\sigma \wedge \omega^\sigma_\rho.
\end{eqnarray}
The spin connection on the group manifold of the target space 
${\mathrm SU}(2)$ is well known\cite{Eguchi} and is given by
\begin{equation}
\omega_{\mu\nu} = \epsilon_{\mu\nu\lambda} {\cal L}^\lambda.
\end{equation}
Then using the relation
\begin{equation}
{\d}{\cal L}^\lambda = \epsilon^\lambda_{\mu\nu}\; {\cal L}^\mu\wedge
{\cal L}^\nu
\end{equation}
which is a little tedious to verify, a short calculation shows
\begin{equation}
R^\mu_{\nu\lambda\rho} = \epsilon^\mu_{\;\nu\sigma}\; 
\epsilon^\sigma_{\;\lambda\rho}.\label{eq:defRS3}
\end{equation}
The pull-back of the tensor $R_{\mu\nu\lambda\rho}$ (pull-backs are only 
defined for co-tangent space tensors) is then
\begin{equation}
R'_{\mathrm{ijkl}} = {\partial\pi^\alpha\over\partial p^{\mathrm{i}}}
{\partial\pi^\beta\over\partial p^{\mathrm{j}}}
{\partial\pi^\gamma\over\partial p^{\mathrm{k}}}
{\partial\pi^\delta\over\partial p^{\mathrm{l}}}
{\cal L}_\alpha^\mu {\cal L}_\beta^\nu {\cal L}_\gamma^\lambda 
{\cal L}_\delta^\rho\; R_{\mu\nu\lambda\rho}.\label{eq:defRpS3}
\end{equation}
We remark that the derivatives $\partial\pi^\alpha/\partial p^{\mathrm{i}}$ 
etc. serve 
only to change the variables from group manifold coordinates $\pi^\alpha$ to 
spatial coordinates $p^{\mathrm{i}}$, while the Maurer-Cartan forms 
contain the essential 
structure. Application of the following Fierz identities
\begin{eqnarray}
\tau^{\mathrm{a}}_{\mathrm{ij}} \tau^{\mathrm{b}}_{\mathrm{kl}} 
\epsilon^{\mathrm{abe}} 
& = & -{1\over 2} ( \tau^{\mathrm{e}}_{\mathrm{kj}} \delta_{\mathrm{il}} + 
\tau^{\mathrm{e}}_{\mathrm{il}} \delta_{\mathrm{kj}})
\\ 
\tau^{\mathrm{a}}_{\mathrm{ij}} \tau^{\mathrm{a}}_{\mathrm{kl}} & = & 
-{1\over 2} \tau^{\mathrm{a}}_{\mathrm{il}} \tau^{\mathrm{a}}_{\mathrm{kj}} +
{3\over 2} \delta_{\mathrm{il}} \delta_{\mathrm{jk}}
\end{eqnarray}
yields
\begin{equation}
R'_{\mathrm{ijkl}} = 2\, \tr \biggr( 
\bigl [ U^\dagger \partial_{\mathrm{i}} U, 
U^\dagger \partial_{\mathrm{j}} U \bigr]
\bigl [ U^\dagger \partial_{\mathrm{k}} U, 
U^\dagger \partial_{\mathrm{l}} U \bigr]
\biggr)
\end{equation}
and contracting twice with the inverse metric on space yields
\begin{eqnarray}
R' & = & t^{\mathrm{ik}} t^{jl} R'_{\mathrm{ijkl}}
\nonumber
\\
& = & (1 + r^2/4 L^2)^4 \delta^{\mathrm{ik}} \delta^{\mathrm{jl}} 
R'_{\mathrm{ijkl}}
\nonumber
\\
& = & (1 + r^2/4 L^2)^4 \sum_{\mathrm{i,j=1}}^3 \tr \biggl( 
\bigl[U^\dagger\partial_{\mathrm{i}} U,U^\dagger 
\partial_{\mathrm{j}} U \bigr]^2
\biggr)
\end{eqnarray}
which is obviously the Skyrme term. 

A more geometric and generally valid interpretation is obtained by (for any 
manifold $S$ and $\Sigma$) considering the squared norm of the pull-back of 
the area element defined by two dual basis vectors in the target 
space\cite{Mantongeo}. The area element defined by the two dual basis 
vectors is
\begin{eqnarray}
A^{\mu\nu} & = & \hat\zeta^\mu\wedge\hat\zeta^\nu = 
\zeta^\mu_\alpha \zeta^\nu_\beta\; 
{\d}\pi^\alpha \wedge {\d}\pi^\beta 
\nonumber
\\
& = & {1\over 2} \biggl( \zeta^\mu_\alpha \zeta^\nu_\beta -
\zeta^\mu_\beta \zeta^\nu_\alpha \biggr)\; {\d}\pi^\alpha\wedge {\d}\pi^\beta.
\end{eqnarray}
Its pull-back is given by
\begin{eqnarray}
A'^{\mu\nu} & = & A'^{\mu\nu}_{\mathrm{ij}}\, {\d}p^{\mathrm{i}}\wedge 
{\d}p^{\mathrm{j}}
\nonumber
\\
& = & {1\over 2} \biggl( \zeta^\mu_\alpha \zeta^\nu_\beta -
\zeta^\mu_\beta \zeta^\nu_\alpha \biggr) {\partial\pi^\alpha\over\partial 
p^{\mathrm{i}}}
{\partial\pi^\beta\over\partial p^{\mathrm{j}}}\; {\d} p^{\mathrm{i}}
\wedge {\d} p^{\mathrm{j}}.
\end{eqnarray}
Its squared norm is
\begin{equation}
|A'^{\mu\nu}|^2 = {1\over 4} t^{\mathrm{ik}} t^{\mathrm{jl}}
\biggl( 
\zeta^\mu_\alpha \zeta^\nu_\beta - \zeta^\mu_\beta \zeta^\nu_\alpha 
\biggr) 
\biggl(
\zeta^\mu_\gamma \zeta^\nu_\delta - \zeta^\mu_\delta \zeta^\nu_\gamma
\biggr)
{\partial\pi^\alpha\over\partial p^{\mathrm{i}}}
{\partial\pi^\beta\over\partial p^{\mathrm{j}}} 
{\partial\pi^\gamma\over\partial p^{\mathrm{k}}}
{\partial\pi^\delta\over\partial p^{\mathrm{l}}}
\end{equation}
and expressing $t^{\mathrm{ij}}= e^{\mathrm{i}}_{\mathrm{m}} 
e^{\mathrm{j}}_{\mathrm{n}} 
\delta^{mn}$, summing over $\mu$ and $\nu$ 
and using the definition of the deformation matrix $J_{\mathrm{m}}^\mu = 
e^{\mathrm{i}}_{\mathrm{m}} 
(\partial\pi^\alpha/ \partial p^{\mathrm{i}}) \zeta_\alpha^\mu$ (see equation
$(\ref{eq:jmum})$ gives
\begin{eqnarray}
\sum_{\mu,\nu} |A'^{\mu\nu}|^2 & = & \bigl( \tr
\bigl[JJ^{\mathrm{T}}\bigr] 
\bigr)^2 - \tr \big[JJ^{\mathrm{T}}]^2
\nonumber
\\
& = & \bigl (\tr [D]\bigr)^2 - \tr [D]^2
\nonumber
\\
& = & \lambda_1^2 \lambda_2^2 + \lambda_2^2 \lambda_3^2 + 
\lambda_3^2 \lambda_1^2.
\end{eqnarray}
This expression is completely general, allowing for any spatial 
and target
manifold. Specializing again 
to the case of $S^3$ and ${\mathrm SU}(2)$ we obtain
\begin{equation}
A^{\mu\nu} = \epsilon^{\mu\nu\sigma} A_\sigma
\end{equation}
where
\begin{equation}
A_\sigma = \epsilon_{\sigma\mu\nu} {\cal L}^\mu_\alpha {\cal L}^\nu_\beta\;
{\d}\pi^\alpha\wedge {\d}\pi^\beta.
\end{equation}
Then
\begin{equation}
A'_\sigma = \epsilon_{\sigma\mu\nu} {\cal L}^\mu_\alpha {\cal L}^\nu_\beta
{\partial\pi^\alpha\over\partial p^{\mathrm{i}}} 
{\partial\pi^\beta\over\partial p^{\mathrm{j}}}\;
{\d} p^{\mathrm{i}}\wedge {\d} p^{\mathrm{j}}
\end{equation}
and 
\begin{eqnarray}
|A'^{\mu\nu}|^2 &=&  
\epsilon_{\sigma\mu\nu}   
\epsilon_{\sigma\lambda\rho}
t^{\mathrm{ik}} t^{\mathrm{jl}}
{\cal L}^\mu_\alpha {\cal L}^\nu_\beta
{\cal L}^\lambda_\gamma {\cal L}^\rho_\delta 
{\partial\pi^\alpha\over\partial p^{\mathrm{i}}} 
{\partial\pi^\beta\over\partial p^{\mathrm{j}}}
{\partial\pi^\gamma\over\partial p^{\mathrm{k}}} 
{\partial\pi^\delta\over\partial p^{\mathrm{l}}}\nonumber
\\
&=& R'
\end{eqnarray}
by $(\ref{eq:defRS3})$ and $(\ref{eq:defRpS3})$.
We easily verify that
\begin{equation}
R'_{\mathrm{ijkl}} = A'^{\mu\nu}_{\mathrm{ij}} A'^{\sigma\tau}_{\mathrm{kl}} 
\delta_{\mu\sigma} \delta_{\nu\tau}.
\end{equation}  
This expression for $R'$ and the Skyrme term 
is in fact identical to that given in Manton\cite{Mantongeo}, 
however there is a
slight formal difference. We have pulled-back the area elements from the target
space (${\mathrm SU}(2)$) to the spatial manifold $S^3$ and computed the sum 
of their squared norms there. Manton\cite{Mantongeo} takes the area elements 
in the spatial
manifold and pushes forward their dual area elements (tangent space tensors) to
the target space and computes their squared norm in the target space. This 
gives the same energy functional. 

To complete our treatment of this example of $S^3$ mapped to 
${\mathrm SU}(2)$, we show the interpretation of the third invariant.
Consider the integral coming from the third invariant $(\ref{eq:defi3})$
\begin{eqnarray}
\int_{\mathrm{S}} \sqrt{I_3} \; \sqrt{\det t}\; {\d}^3 p 
& = & \int_{\mathrm{S}} 
\sqrt{\det D}\; \sqrt{\det t}\; {\d}^3 p
\nonumber
\\
& = & \int_{\mathrm{S}} \lambda_1 \lambda_2 \lambda_3 \sqrt{\det t}\; {\d}^3 p
\nonumber
\\
& = & \int_{\mathrm{S}} \det J \sqrt{\det t}\; {\d}^3 p
\nonumber
\\
& = & \int_{\mathrm{S}} \det \biggl 
( e^{\mathrm{i}}_{\mathrm{m}} {\partial\pi^\alpha\over\partial p^{\mathrm{i}}} 
\zeta^\mu_\alpha \biggr)\sqrt{\det t}\; {\d}^3 p
\nonumber
\\
& = & \int_{\mathrm{S}} \det \biggl( {\partial\pi\over\partial p}\biggr) 
\det \zeta\; {\d}^3p
\nonumber
\\
& = & (\deg\pi) \int_\Sigma \sqrt{\det \tau}\; {\d}^3\pi
\end{eqnarray}
where the factor $\deg\pi$ counts the number of times that the mapping $\pi$
wraps the initial manifold over the target manifold. In reality the last 
equation is
only valid locally on $S$, the integral gives the volume of the region 
covered in
$\Sigma$. This volume must be counted with the appropriate sign depending on
whether the relative orientation is preserved. Between regions where the 
relative
orientation changes sign is a zero of at least one of the $\lambda$'s. This 
gives
rise to natural boundaries which should be considered since the sign does not
change within these regions. Then summing up the volumes of the regions of
$\Sigma$ with the corresponding sign gives exactly the degree of the mapping
$\pi$, \ie the number of complete covering of $\Sigma$ that the mapping $\pi$
provides, multiplied by the volume of $\Sigma$. We assume that $S$ is a 
manifold without a boundary hence the mapping $\pi$ must cover $\Sigma$ an 
integral number of times. 

From the previous equation, we have the integral
\begin{equation}
\deg\pi = {1\over {\rm Vol}\; \Sigma} \int_{\mathrm{S}} \det\biggl 
({\partial \pi\over
\partial p}\biggr) \det\zeta\;  {\d}^3 p
\end{equation}
which can be expressed in terms of $U(p)$ to re-obtain the usual form of 
the baryon number $(\ref{eq:nb})$
\begin{eqnarray}
\deg\pi & = & {1\over 2\pi^2} \int_{\mathrm{S}} \det ({\cal L}) \det 
\biggl ( {\partial\pi\over\partial p}\biggr)\; {\d}^3 p
\nonumber
\\
& = & {1\over 2\pi^2} \int_{\mathrm{S}} \det ({\cal L} \;
{\partial\pi\over\partial p}\biggr)\; {\d}^3 p
\nonumber
\\
& = & {1\over 2\pi^2} \biggl({-{\mathrm i}\over 2}\biggr)^3 \int_{\mathrm{S}} 
\det \biggl (
\tr \biggl[ \tau^\mu U^\dagger {\partial U\over \partial \pi^\alpha}
\biggr] {\partial\pi^\alpha\over\partial p^{\mathrm{i}}}\biggr )\; {\d}^3p
\nonumber
\\
& = & {{\mathrm i}\over 16 \pi^2} \int_{\mathrm{S}} \det \biggl( \tr \biggl[
\tau^\mu U^\dagger {\partial U\over \partial p^{\mathrm{i}}}\biggr]
\biggr) {\d}^3p
\nonumber
\\
& = & {{\mathrm i}\over 16 \pi^2} \int_{\mathrm{S}} \epsilon_{\mu\nu\lambda} 
\tr \biggl[ \tau^\mu U^\dagger {\partial U\over\partial 
p^{\mathrm{i}}}\biggr]
\tr \biggl[ \tau^\nu U^\dagger {\partial U\over\partial 
p^{\mathrm{j}}}\biggr]
\tr \biggl[ \tau^\lambda U^\dagger {\partial U\over\partial 
p^{\mathrm{k}}}\biggr]
\epsilon^{\mathrm{ijk}}\; {\d}^3p
\nonumber
\\
& = & {1\over 24 \pi^2} \int_{\mathrm{S}} \tr\bigl[
U^\dagger \partial_{\mathrm{i}} 
U\;U^\dagger \partial_{\mathrm{j}} U\;U^\dagger \partial_{\mathrm{k}} U
\bigr] \,\epsilon^{\mathrm{ijk}}\; {\d}^3p 
\end{eqnarray}  
which is the familiar form of the baryon number in the Skyrme model.

The Skyrme energy function, according to Manton's formalism\cite{Mantongeo}, 
can be expressed as
\begin{equation}
E =  \int_{\mathrm{S}} \sqrt{\det t}\; {\d}^3p\; \bigl(
\lambda^2_1 + \lambda^2_2 + \lambda^2_3 +
\lambda^2_1 \lambda^2_2 + \lambda^2_2 \lambda^2_3 +
\lambda^2_3 \lambda^2_1
\bigr)
\end{equation}
which is easily re-expressed as
\begin{eqnarray}
E = \int_{\mathrm{S}} \sqrt{\det t}\; {\d}^3p\; 
[(\lambda_1 & \pm & \lambda_2 \lambda_3)^2 +
(\lambda_2 \pm \lambda_3 \lambda_1)^2 +
(\lambda_3 \pm \lambda_1 \lambda_2)^2]
\nonumber
\\
& \mp & 6 \int_{\mathrm{S}} \sqrt{\det t}\; {\d}^3p\; \lambda_1 
\lambda_2 \lambda_3.
\end{eqnarray}
This shows a novel way of demonstrating the Bogomolnyi bound: evidently
\begin{equation}
E \ge 6 \biggl|\; \int_{\mathrm{S}} \lambda_1 \lambda_2 
\lambda_3\sqrt{\det t}\; {\d}^3p\biggr|.
\end{equation}
The equality is attained only if (for winding number $+1$)
\begin{equation}
\lambda_1 = \lambda_2 \lambda_3 \qquad\qquad
\lambda_2= \lambda_3 \lambda_1\qquad\qquad
\lambda_3= \lambda_1 \lambda_2.
\end{equation}
This system has only three distinct solutions:
\begin{equation}
(\lambda_1,\lambda_2,\lambda_3): (0,0,0), \quad(1,1,1), \quad(1,-1,-1) + 
2\; {\rm permutations}.
\end{equation}
The trivial solution corresponds to mapping $S$ to a single point in $\Sigma$ 
and
is the usual vacuum solution. The third set of solutions is equivalent to the
second solution after a rotation by $180^\circ$ in the cotangent space of the
target manifold about a fixed axis. The second solution implies that the map is
everywhere an isometry, \ie, the two $S^3$ have the same radius, 1. This
shows that for an infinite initial sphere, which corresponds to the case of
$\Rset^3$,
the Bogomolnyi bound is not saturated and, as is well known, the map is far 
from
the identity map. Manton and Ruback\cite{Manton-Ruback} and 
Manton\cite{Mantongeo} show that as the radius of the
initial 3-sphere decreases, the map attains the form of the identity for a 
radius of
$\sqrt{2}$. For more details and further applications we refer the reader to
the 
literature\cite{Manton-Ruback,Mantongeo,Loss,Jackson-Wirzba-Castillejo,Jackson-Manton-Wirzba}.

We close this section with a few general words on this formalism. First of 
all the
$\lambda$'s are not independent dynamical variables. Infinitesimal arbitrary
perturbations are allowed, however integrating to finite deformations is 
subject to
consistency conditions. For example there is no deformation of a given
configuration which can yield $\lambda_{\mathrm{i}}=1$ over a finite region, 
if the region is not
iso-metric to a part of the target manifold.
Actually a smooth mapping $\pi$ will always give rise to a smooth
set of $\lambda$'s. A configuration with a discontinuous set of
$\lambda$'s is not attainable even though the corresponding energy integral
is finite.

We have 
also made the intuitive paradigm that the Jacobian matrix is a measure of the
deformation, and hence of the energy. However
this is somewhat misleading since $\lambda_{\mathrm{i}}=0$ is clearly 
a very deformed situation all the same corresponding to zero energy density. 
For a physical
elastic body, $\lambda_{\mathrm{i}}=0$ is in fact
an infinite-energy deformation hence
the corresponding energy functional is not at all like the Skyrme energy
functional. Very schematically, the energy density of an elastic body is
\begin{equation}
\epsilon_{\mathrm{el. body}} \sim (D-I)^2
\end{equation}
while in the Skyrme model it is like
\begin{equation}
\epsilon_{\mathrm{Sk.}} \sim D + D^2.
\end{equation}
Thus the Skyrme ground state is around $D=0$ which is quite unlike the
case of the elastic body, where $D=1$.   

This completes our exposition of the interpretation of a field theory, 
specifically the Skyrme model, as a non-linear elasticity theory. In the 
next two subsections we will look in more detail at the Skyrme model and its 
static, low energy configurations. First we will elaborate on the instanton 
method for obtaining an analytical ansatz for the set of relevant low energy 
configurations, and second we will describe the use of rational maps to 
obtain reasonable ans\"atze for multi-baryonic minima.

%% file: section32.tex
The instanton method uses the known solutions of 4-dimensional Euclidian 
Yang-Mills theory called instantons\cite{Coleman-i} and their moduli spaces to 
obtain Skyrme field configurations. The relation between the two seems 
tenuous at first, however, the known global topology and symmetries of the 
instanton moduli space and its similarities to expected properties of 
low-energy Skyrme field configurations seems to point in that direction. 
Consider the case of $B=2$, here know for two widely separated Skyrmions, 
there are 12 independent degrees of freedom. We expect the relevant 
low-energy space of configurations to also have 12 dimensions. 
Manton\cite{Mantongfc}  proposed that this sub-manifold could be obtained as 
the union of all gradient flow curves linking together all the low-energy 
critical points. We will return to this subject in much detail in section 4.

This idea to obtain the 12 dimensional sub-manifold, ${\cal M}_{12}$ which 
should serve 
as the correct truncation of the full field theory description of the 
interactions and dynamics of two Skyrmions, from gradient flow curves 
although in principle sound is 
in practice only numerically, approximately implementable. A consideration
of the symmetries involved, led Atiyah and Manton\cite{Atiyah-Manton-1} 
to suggest that an analytical 
construction of a manifold, 
which might be a reasonable approximation to the true 
sub-manifold, could be obtained from certain instanton configurations and 
their holonomies. 

The observation consists of the following two steps. First of all, from any 
${\mathrm SU}(2)$ instanton configuration in $\Rset^4$, it is possible to 
obtain a unitary matrix valued field defined on $\Rset^3$ by
\begin{equation}
U(\vec x) = P \exp \Biggl\{- 
\int_{-\infty}^{+\infty} {\d}\tau\; A_0(\vec x, \tau)\Biggr\},\label{eq:holo}
\end{equation}
where $P$ denotes the path ordered integral.
We will show later, how the baryon number of $U(\vec x)$ is equal to the 
instanton number of $A_\mu(\vec x,\tau)$. Secondly it has been known that the 
space of configurations of two instantons interpolates continuously and 
smoothly from an axially symmetric, localized configuration to two, 
individual, ``spherically'' symmetric, well (infinitely) separated 
instanton configurations. It is a matter of verification that the 
corresponding Skyrmion holonomies interpolate smoothly between the toroidal 
lowest energy deuteron to two infinitely separated single Skyrmions. It is 
also possible to obtain configurations which correspond to the spherically 
symmetric dibaryon type configuration. 

The most vexxing problem is that the two instanton configurations are 16 
dimensional, that is they have 16 independent parameters. One of these 
corresponds to a global time translation, the integral over the time 
direction removes this degree of freedom, leaving 15. This is larger than the 
12 dimensional manifold which is being sought. It is not evident what is the 
proper way to reduce the number of parameters by three. Nominally one should 
re-implement the gradient flow method on this sub-manifold of the 
configuration space starting from the highest energy critical point of the 
Skyrme energy functional restricted to the sub-manifold. Such a calculation 
has not been effected, numerically it is just as difficult to work with a 
discretized version of the full problem rather than the one defined on the 
sub-manifold, hence there does not seem to be a compelling motivation to 
study the gradient flow here. The problem has been studied in detail for the 
case of the most attractive channel, which we will return to a little later. 
The instanton method for three and higher baryons is not very efficient. 

\subsubsection{Topological numbers}

First let us solidify the connection between baryon number and instanton 
number. An instanton configuration with instanton number $k$ is technically 
defined as a connection on an ${\mathrm{SU}(2)}$ principal bundle over the 
four-sphere $S^4$ with second Chern number $C_2 = k$\cite{Eguchi}. 
The relationship 
between a gauge field defined on $\Rset^4$ and $S^4$ is obtained via 
stereographic projection. The usual instanton configurations which satisfy 
the Yang-Mills equations of motion defined on $\Rset^4$ have non zero field 
strength $F_{\mu\nu}$ in a localized region of space-time, and achieve 
a pure gauge field type configuration towards Euclidian infinity. 
This means that on 
the manifold at infinity of $\Rset^4$, which is topologically $S^3$, an 
instanton configuration defines a smooth group element valued 
configuration $U(\vec x,\tau)|_{(\vec x,\tau)\rightarrow \infty}$ 
and the gauge field is given by
\begin{equation}
A_\mu = U\partial_\mu U^\dagger.
\end{equation}
The second Chern number corresponds to the integral
\begin{eqnarray}
k & = & {-1\over 32\pi^2}
\int {\d}^4 x\; \epsilon^{\mu\nu\lambda\tau} 
\tr (F_{\mu\nu} F_{\lambda\tau})\nonumber
\\
& = & {1\over 32\pi^2} \int {\d}^4 x\; \partial_\mu K^\mu\nonumber
\\
& = & {1\over 32\pi^2} \oint_{\infty} {\d}\sigma_\mu K^\mu\label{eq:intc2}
\end{eqnarray}
where
\begin{equation}
K^\mu = -2\, \epsilon^{\mu\nu\lambda\sigma}\, {\tr} (
A_\nu F_{\lambda\sigma} - {2\over 3} A_\nu A_\lambda A_\sigma)
\end{equation} 
and
\begin{equation}
F_{\mu\nu} = \partial_\mu A_\nu - \partial_\nu A_\mu +[A_\mu,A_\nu].
\end{equation}
Replacing $A_\mu = U \partial_\mu U^\dagger$ gives
\begin{equation}
K^\mu = {4\over 3}\,\epsilon^{\mu\nu\lambda\sigma}\, {\tr}
\biggl ( U\partial_\nu U^\dagger\,
U\partial_\lambda U^\dagger\,
U\partial_\sigma U^\dagger
\biggr),
\end{equation}
and $k$ is given by the integral 
\begin{equation}
k = W(U) = {1\over 24 \pi^2} \oint_{\infty} {\d} \sigma_\mu\;
\epsilon^{\mu\nu\lambda\sigma}\, 
{\tr} 
\biggl ( 
U\partial_\nu U^\dagger\,
U\partial_\lambda U^\dagger\,
U\partial_\sigma U^\dagger
\biggr).
\end{equation}
This is exactly the measure of the winding number of the mapping of 
$S^3\rightarrow
{\mathrm SU(2)}$ defined by the group element at infinity. These mappings 
define elements of the homotopy group
\begin{equation}
\Pi_3( {\mathrm{SU(2)}}) = \Zset,\label{eq:defpi3z}
\end{equation}
the integer corresponding to the Chern number. The configuration on $\Rset^4$ 
does not attain a constant value at infinity, hence it is difficult to 
interpret this field as corresponding to a field on $S^4$ after stereographic 
projection. (To be precise, the stereographic projection involved is actually 
the conformal mapping taking $S^4\rightarrow \Rset^4$. $g_{\mu\nu} = 1
/(|x|^2/4 R^2 + 1)^2\;\eta_{\mu\nu}$ is the conformal transformation taking 
$\Rset^4$ to $S^4$ with the coordinate transformation exactly as in equation
$(\ref{eq:chcoordSR})$. This mapping has the 
advantage of mapping solutions of the equations of motion on one manifold to
solutions on the other because the Lagrangian of Yang-Mills theories is 
conformally invariant\cite{JNR}.) However the configuration on $S^4$ is also 
somewhat 
subtle, $A_\mu$ is actually a connection on a non-trivial ${\mathrm{SU(2)}}$
principal bundle defined over the base manifold $S^4$. These bundles are 
specified by 
fixing the transition function which maps the fibre ${\mathrm{SU(2)}}\equiv 
S^3$ over the ``northern hemisphere'' of $S^4$ to the $S^3$ over the 
``southern hemisphere'' at the ``equator''. The equator of $S^4$ is simply an 
$S^3$. Hence the transition functions are tantamount to defining a group 
element over the equatorial $S^3$. This means that one defines a mapping of 
the equatorial $S^3$ to the group ${\mathrm{SU(2)}}\equiv S^3$. Such mappings 
fall into the disjoint homotopy classes labelled exactly as in 
$(\ref{eq:defpi3z})$. 
The second Chern number of the bundle $(\ref{eq:intc2})$
is exactly equal to the integer characterizing the homotopy class of the 
transition function. 

Geometrically there is no constraint on the size of the 
coordinate charts; one need not restrict oneself to equal hemispheres. 
There is no hindrance to extending the southern hemisphere to include the 
whole of $S^4$ 
except for one point, the north pole. Indeed, in this way 
we will extend the solution of the equations of motion to almost everywhere 
on $S^4$. The solution will be singular at the north pole however only for the
connection. The integral $(\ref{eq:intc2})$ 
will be an integral over the whole sphere of only the 
field strengths associated with the connection
and the density $\tr (F_{\mu\nu}\tilde F^{\mu\nu})$ is non-singular 
over the whole sphere. The result must still 
give the second Chern number. Now the conformal projection of 
this field configuration to $\Rset^4$ will give the field configuration that 
satisfies the equations of motion on $\Rset^4$ and furthermore whose integral 
corresponding to $(\ref{eq:intc2})$
is also $C_2(A_\mu)$ (since $(\ref{eq:intc2})$ is independent of the metric). 
The values achieved at the north pole by 
$A_\mu$ defined by the limiting value of the configuration along any path 
leading to the north pole are all equal modulo 
gauge transformations. The conformal transformation maps this field on
$S^4$ to a configuration on $\Rset^4$ which becomes a pure 
gauge configuration at infinity since the field 
strength at the north pole is diluted over the entire $S^3$ manifold at 
infinity. 

\begin{figure}
\centering
\mbox{\epsfig{figure= 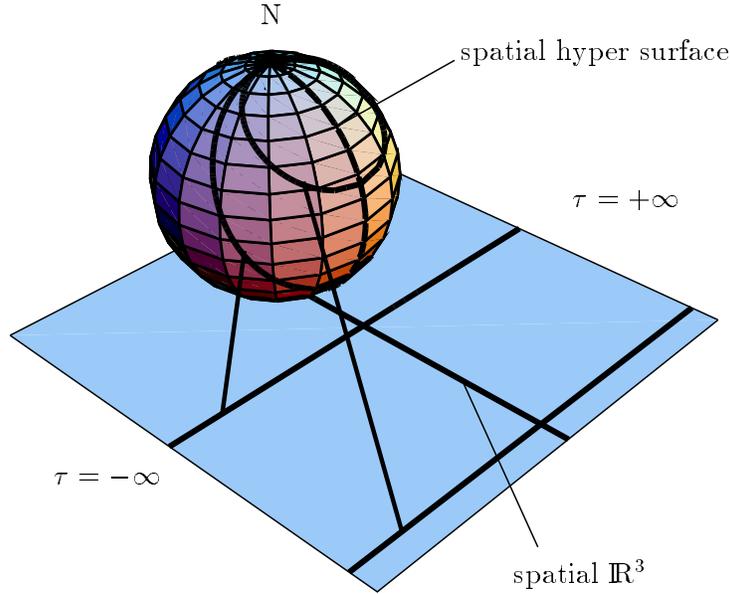,width=5.5truein}}
\caption{Diagram of the stereographic projection from $S^4$ to $\Rset^4$.}
\label{fig:fig3}
\end{figure}

The path ordered integral $(\ref{eq:holo})$ which defines $U(\vec x)$ starts
at $\tau=-\infty$ and follows a straight line to $\tau=+\infty$. This 
corresponds to a curve on $S^4$ which starts at the north pole, follows a 
particular path on the $S^4$ and returns back 
to the north pole (see figure $\ref{fig:fig3}$). 
The path on $S^4$ 
is simply given by the intersection of the $S^4$ with a 
2-plane, that which is defined by the line 
of integration in $\Rset^4$ and the north pole. This 
intersection
is actually just a circle. The set of such 
curves on $S^4$ leave the north pole 
on one side of a 3-dimensional hyper surface, circle around the $S^4$, and 
return to the north pole from the other side of the hypersurface. 
The hypersurface is in fact 
just a ``great'' 3-sphere, exactly like the equatorial $S^3$ except that it 
passes from north pole to south pole, and back. It is exactly the inverse 
(stereographic) projection of the spatial $\Rset^3$ (of the $\Rset^4$) 
onto the $S^4$. The curves 
leave the north pole, intersect this great $S^3$ exactly once and come back 
to the north pole in a symmetric fashion. Hence each curve on the $S^4$ 
defines a unitary matrix valued configuration on the great $S^3$ at the point 
where the curve intersects this great $S^3$. This configuration is by 
construction continuous. Consequently we manage to define a winding number 
$(\ref{eq:defpi3z})$.

The winding number is invariant under any continuous 
deformation which keeps a one 
to one relation between each curve and the points of the hypersurface. 
Envisage the following deformations. The lines of integration are well 
represented by 
the lines of forces emanating from and returning to an ideal pointlike 
``dipole'' 
situated at the north pole. The great sphere separates these 
lines of force at the north pole into outgoing lines on one side, and 
incoming lines on the other side. We simply imagine moving the two charges 
comprising the dipole apart, keeping the lines of integration the same 
as the lines of forces leaving the positive charge on one side and arriving 
at the negative charge on the other. Such a modification of the lines of 
integration will result in a homotopy of the original $U(\vec x)$, and hence 
will not change the homotopy type. Finally we will arrive at the 
situation where the two
charges occupy antipodal points of the four sphere (actually on the equator)
and the lines of force 
emanate symmetrically from one charge, cross the great 3-sphere and finally 
terminate on the opposite charge at the antipodal point. We make one further 
homotopy, we rigidly rotate the system of charges, lines of integration and the
great $S^3$ until they are vertical, such that the positive 
charge is at the south pole and the negative charge is at the north pole, and 
the usual equator now corresponds to the great $S^3$. Such a 
deformation requires a simultaneous redefinition of the stereographic 
projection, and a deformation of the 
$U(\vec x)$ (because the integration lines are changing) but it is  
clearly a continuous deformation keeping the homotopy type invariant. Finally
to be complete, we had started with closed line integrals 
originally (leaving and returning to the north pole) but now we have open line 
integrals starting at the south pole and terminating at the north pole. We 
can easily remedy this by adding one path to all of the others, starting at 
the north pole and descending down a fixed meridian to the south pole for all 
of the line integrals. This simply left-multiplies each $U(\vec x)$ by a 
constant unitary matrix, which again does not modify the homotopy type. 
Closed contour path ordered exponential integrals are gauge covariant hence 
we can return to the original description of the instanton on $S^4$
with more than two patches.
Now we have arrived at the starting point of the demonstration given 
in Manton and Atiyah\cite{Atiyah-Manton-2} where they show 
that the winding number of this 
configuration is the same as the instanton number (the second 
Chern number of the 
instanton bundle in question). This demonstration proceeds as follows.

\begin{figure}
\centering
\mbox{\epsfig{figure= 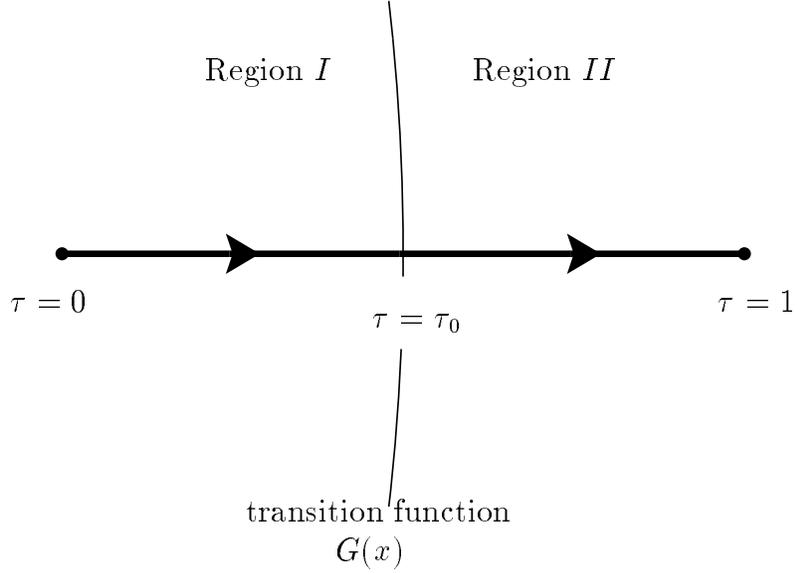,width=5.0truein}}
\caption{Diagram of the role of the transition function $G(x)$ between regions
$I$ and $II$.}
\label{fig:fig4}
\end{figure}

The first step is to use the gauge freedom to put the gauge field in the 
specific gauge where the component of the gauge field along the meridional 
directions vanishes.  This can be established in each patch
separately. Then the integral
\begin{equation}
P\exp\big(-\int A_\mu dx^\mu\big) =1
\end{equation}
since the inner product $A_\mu dx^\mu$ is zero along the meridional path. 
However the definition of the path ordered exponential, when the path of
integration crosses a boundary between patches is such that one must multiply 
the
contribution coming from the first leg of the path by the transition function
before continuing with the integral in the second patch (see figure 
$\ref{fig:fig4}$).
\begin{equation}
P\exp\big(-\int_0^1 A_\mu dx^\mu\big) = 
P\exp\big(-\int_{\tau_0}^1 A_\mu dx^\mu\big)
G(x)
P\exp\big(-\int_0^{\tau_0} A_\mu dx^\mu\big).
\end{equation}
Such an expansion is gauge covariant under simultaneous independent gauge
transformations in the patches $I$ and $II$ since the transition function is
defined to transform as
\begin{equation}
G(x)\rightarrow U_{\mathrm{II}} G(x) U^\dagger_{\mathrm{I}}
\end{equation}
while
\begin{eqnarray}
P\exp\big(-\int_0^{\tau_0} A_\mu dx^\mu\big)&\rightarrow&
U_{\mathrm{I}}(\tau_0) 
P\exp\big(-\int_0^{\tau_0} A_\mu dx^\mu\big)
U^\dagger_{\mathrm{I}}(0)
\\
P\exp\big(-\int_{\tau_0}^1 A_\mu dx^\mu\big)&\rightarrow& 
U_{\mathrm{II}}(1)
P\exp\big(-\int_{\tau_0}^1 A_\mu dx^\mu\big) 
U^\dagger_{\mathrm{II}}(\tau_0)
\end{eqnarray}
yielding
\begin{equation}
P\exp\big(-\int_0^1 A_\mu dx^\mu\big)\rightarrow 
U_{\mathrm{II}}(1)
P\exp\big(-\int_0^1 A_\mu dx^\mu\big)
U^\dagger_{\mathrm{I}} (0)
\end{equation}
Now since $A_\mu dx^\mu =0$ in our gauge, we obtain
\begin{equation}
U(\vec x) =U_0G(\vec x)
\end{equation}
where $G(\vec x)$ is the transition function at the equator.
Then, the baryon number
\begin{equation}
B = W(U(\vec x))=W(G(\vec x))=C_2(A_\mu )=k;
\end{equation}
hence the baryon number and the instanton number are identical.

\subsubsection{The sector $B=1$}

For $k=1$ the instanton profile is given
by
\begin{equation}
A_\mu = {\mathrm{i}} \bar\sigma_{\mu\nu} \partial_\nu \ln \rho
\end{equation}
where the definition of $\bar\sigma$ can be found in the article of Jackiw,
Nohl and Rebbi (JNR)\cite{JNR}, however the time 
component is explicitly
\begin{equation}
A_4={{\mathrm{i}}\over 2}{\nabla\rho\over\rho}\cdot\vec\tau
\end{equation}
with
\begin{equation}
\rho = 1+{\lambda\over (x_\mu -X_\mu )(x^\mu -X^\mu )}
= 1 + {\lambda\over |x-X|^2}.\label{eq:thinst}
\end{equation}
This is in the 't Hooft gauge\cite{tHooft}. 
The singularities in $\rho$ are gauge artefacts and hence do not contribute to
gauge covariant quantities such as the definition of the Skyrme field.  A local
gauge transformation moves the singularity to wherever we want, but of course
does not affect the $U(\vec x)$. We will not show this
here. The function $\rho$ $(\ref{eq:thinst})$ 
has 5 parameters, but three others are to be
added in because of global gauge transformations, which were 
factored out in the
definition of instantons.  This yields 8 parameters.  Evidently the 4 
translation
parameters $X^\mu$ are the center of mass coordinates, the spatial ones 
determine
the spatial center of mass of the corresponding Skyrmion, the temporal one is
absorbed by the integration in the time direction. Factoring out these
four leaves 4 parameters, 
$\lambda$
governing the overall scale, and three corresponding to global gauge
transformations. Hence
\begin{equation}
\rho = 1+{\lambda\over (r^2 +\tau^2)}
\end{equation}
where $r^2=\vec x\cdot\vec x$.  Then
\begin{equation}
A_4=-{i\over 2}{\lambda\over\rho}{2\vec x\cdot\vec\tau\over (r^2 +\tau^2)^2}
\end{equation}
and
\begin{eqnarray}
U(\vec x)&=& P\exp\Biggl(-\int_{-\infty}^\infty-{i\over 2}{\lambda\over\rho}
{2\vec x\cdot \vec\tau\over (r^2 +\tau^2)^2} d\tau\Biggr)\nonumber
\\
&=&\exp\biggl(i\lambda\vec
x\cdot\vec\tau\int_{-\infty}^\infty{d\tau\over (r^2+\tau^2)^2+\lambda
(r^2+\tau^2) }\biggr)\nonumber
\\
&=&e^{i\hat x\cdot\vec\tau\pi
\big( 1-(1+{\lambda\over r^2} )^{-{1\over 2}}\big)}
\end{eqnarray}
Thus
\begin{equation}
f(r)=\pi \biggl( 1-(1+{\lambda\over r^2} )^{-{1\over 2}}\biggr)
\end{equation}
which satisfies $f(0) = \pi$. This point is actually determined by the 
limit from
non-zero values of $r$, since the integral above is not well defined
for $r=0$.  The singularity is gauge dependent, as mentioned above, hence by
a local gauge transformation we can move the singularity away from  $r=0$
without affecting the value for the Skyrme field.  The Skyrme field so
obtained will be continuous at $r=0$, hence the value at $r=0$ can equally
well be defined as the value obtained from the limit of non-zero $r$.
Furthermore, $f(\infty ) = 0$.  For minimum energy one finds $\lambda =2.109$,
and the corresponding energy is $E=1.2432\times 12\pi^2$.  This exceeds the
numerically obtained minimum energy solution by only 1\%.

Adding in gauge transformations, \ie global iso-rotations simply combs the
Skyrmion profile without affecting the energy.  The  iso-rotation parameters 
are already evident in the JNR\cite{JNR} parametrization of the
instantons.  Here we have
\begin{equation}
\rho = {\lambda_1\over |x -X_1|^2 } +{\lambda_2\over |x -X_2|^2 }
\end{equation}
which has 10 parameters.  It is known that several of these are
local gauge artefacts, indeed, the instanton obtained is gauge equivalent to
the 't Hooft instanton $(\ref{eq:thinst})$ with
\begin{eqnarray}
X&=&{\lambda_1 X_2 +\lambda_2 X_1\over \lambda_1 +\lambda_2}
\\
\lambda &=& {\lambda_1\lambda_2\over (\lambda_1 +\lambda_2 )^2}
|X_1 -X_2|^2.
\end{eqnarray}

The JNR instanton generates, however, a Skyrme field differing from that 
obtained
from the 't Hooft instanton by a global gauge transformation given by
\begin{equation}
U_0 =
{(X_2 -X_1)^0\over |X_2 -X_1|}+
{\mathrm{i}} {(\vec X_2 -\vec X_1 )\over |X_2 -X_1|}\cdot\vec\tau.
\label{eq:gtinst}
\end{equation}
This completes the case $k=1(B=1)$.

\subsubsection{The sector $B=2$}

In this subsection we present the findings of Atiyah and 
Manton\cite{Atiyah-Manton-2}, where they relate the parameters (moduli) of 
$k=2$
instanton configurations introduced analytically by JNR and geometrically by 
Hartshorne\cite{Hartshorne}, to
the various parameters of the corresponding $B=2$ Skyrme configurations.  

For $k=2 (B=2)$, the JNR parametrization is
\begin{equation}
\rho = {\lambda_1\over |x-X_1|^2} + 
{\lambda_2\over |x-X_2|^2} +
{\lambda_3\over |x-X_3|^2}
\end{equation}
a 15 parameter solution.  It is clear that the overall scale of the
$\lambda$'s is never a
parameter, yielding 14 parameters.  In addition there is an explicit one
parameter family of local gauge transformations included in $\rho$, 
reducing the
number of true parameters to 13. We will return to this redundancy later. 
Integrating over Euclidean time to obtain the
Skyrme field reduces the number of parameters to 12.  Finally putting in the 3
iso-rotational degrees of freedom, as they are not included in the solution,
implies that the corresponding Skyrmion fields will have a total of 15
parameters.

In general, for higher $k$, there is no local gauge
transformation in the JNR expression for $\rho$, which thus has $5k+4$
parameters.  Integrating over time removes 1 but adding in three for 
global iso-rotations yields
in general $5k+6$ parameter Skyrme fields.  This is obviously not the full
complement of $6k$ that we expect for $B=k$ Skyrmions.  The full instanton
moduli space is actually supposed to be $8k$ dimensional.  These include 4
positions, 1 scale and 3 iso-rotations per instanton, not removing the 3 
overall
iso-rotation parameters.  Integrating to get the Skyrmions removes one
parameter, implying an $8k-1$ dimensional manifold.  This manifold would
correspond to $3$ positions, $3$ isorotations and 
$1$ scale per Skyrmion, and $k-1$
relative ``time" coordinates.  These time coordinates  serve simply to fix the
order of the individual Skyrmion fields in the corresponding product ansatz
type configuration, if the Skyrmions are well separated in these time
coordinates.  This ordering is of course quite irrelevant if the
Skyrmions are well separated spatially.  However the interpolation between
different orderings is quite important when they are spatially close together.

\begin{figure}
\centering
\mbox{\epsfig{figure= 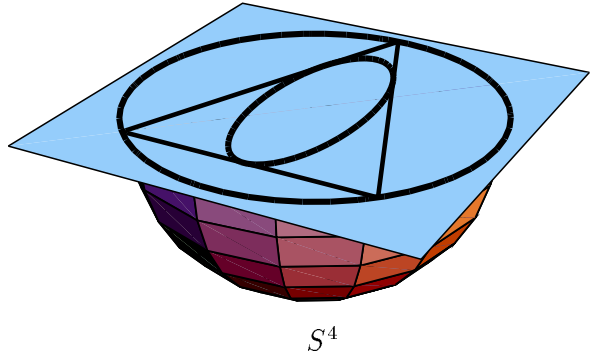,width=5.0truein}}
\caption{Diagram of the $S^4$ cut by a plane, exhibiting a triangle and 
ellipse.}
\label{fig:fig5}
\end{figure}

The full $8k$ dimensional manifold of instanton solutions is well
understood algebraically but not analytically.  
The largest manifold of analytically explicit
solutions corresponds to the $5k+4$ dimensional manifold of JNR.  For $k=2$
these have 13 parameters and for which there is an algebraic characterization
given by Hartshorne\cite{Hartshorne}. 
Hartshorne proves that there is a   1-1 correspondence
between the instanton solutions (on $S^4$) and a set of ellipses that are
interior to the $S^4$ (we imagine that the $S^4$ in question is embedded in
$\Rset^5$).  The ellipses lie in a 2-dimensional plane that intersects the
$S^4$ in a (coplanar) circle, subject to one condition.  The condition insists
that the ellipse can be circumscribed by a triangle whose vertices lie on the
circle (see figure $\ref{fig:fig5}$).
This condition is called the Poncelet condition, and he proved that if such a
triangle exists, then there is a one parameter family of such triangles, 
called a porism\cite{Atiyah-Manton-2}. 
The Poncelet condition is effectively one 
condition on the parameters
of the ellipse, for a fixed circle.  An ellipse is parametrized in general by 5
parameters: 2 give the semi-major and semi-minor axes, 2 fix the position of
the center and one fixes its orientation in the plane relative to a fixed set
of coordinate axes (rotations).  The Poncelet condition leaves 4 degrees of
freedom.  Now a 2-plane passing through the origin in $\Rset^5$ is specified 
by a
division of the $\Rset^5$ into the 2-dimensional space of the plane and the 
3-dimensional space orthogonal to the plane. 
The action of the orthogonal group
${\mathrm{O}}(5)$ generates all the different possibilities 
from any given initial one.
However, if the orthogonal group acts only in the 2-plane or in the 
3-dimensional orthogonal space, then we obtain nothing new.  
Hence the dimension
of the space of 2-planes in $\Rset^5$ passing through the origin is 
${\mathrm{dim}}\, {\mathrm{O}}(5) -
({\mathrm{dim}}\, {\mathrm{O}}(3) + 
{\mathrm{dim}}\, {\mathrm{O}}(2))= 10 - (3+1)=6$, where we have used 
that the dimension of
${\mathrm{O}}(N)$ is $N(N-1)/2$. Removing the condition that the plane 
passes through the
origin allows 3 translations, one for each independent orthogonal direction,
yielding 9 degrees of freedom.  Adding in the four degrees of freedom 
of the ellipse yields 13 in total.

Given the circle and ellipse, the corresponding family of triangles is neatly
described by a cubic equation.  Suppose that $s=\tan ({\theta\over 2})$ is a
variable along the circle ($\theta\in [-\pi ,\pi ]$ an angular variable around
the circle).  Then the vertices of the triangle correspond to $s_1,s_2,s_3$,
which are without loss of generality the roots of a cubic polynomial equation,
\begin{equation}
p_0 s^3 +p_1 s^2 +p_2 s +p_3 =0.\label{eq:eq1}
\end{equation}
Clearly there are some global constraints that they the
$p_{\mathrm{i}}$'s must satisfy so that the cubic $(\ref{eq:eq1})$
has three real roots, but
these constraints do not remove any degrees of freedom.  However, the 
$p_{\mathrm{i}}$'s are real,
their overall scale is irrelevant, and they are not all zero, hence they
define a ray in $\Rset^4$.  The space of all rays in 
$\Rset^4$ is called the real
projective space of dimension three, denoted $\Rset P_3$. 
The Poncelet condition is
expressed in this light by requiring that the coefficients, of the cubic
equations for all the triangles in the porism, must lie along a 
straight (projective)
line in $\Rset P_3$.  Hence if
\begin{equation}
q_0 s^3 +q_1 s^2 +q_2 s +q_3 =0 .
\end{equation}
is a cubic whose roots correspond to the vertices of another solution of the
Poncelet condition, then all solutions are obtained by the interpolation
\begin{equation}
(\mu p_0-\nu q_0)s^3 +(\mu p_1-\nu q_1)s^2 +(\mu p_2-\nu q_2)s +(\mu p_3-\nu
q_3)=0.\label{eq:ponccond}
\end{equation}
Now if for some $\bar\mu$ and $\bar\nu$ the coefficient $\bar\mu p_0+\bar\nu
q_0=0$, then the surviving quadratic equation gives two of the roots, while
the last root is pushed off to $s=\pm\infty$ (which is the same point for 
either
sign).  The requirement that the ellipse be interior to the circle implies that
the roots of $(\ref{eq:ponccond})$ 
are real and distinct for all $\mu$ and $\nu$.  This last
constraint precludes the possibility that also $\bar\mu p_1+\bar\nu
q_1=0$  since then we get one finite root but two roots get pushed off to
$s=\pm\infty$, which means that they are the same point on the circle.  This is
not permitted for an ellipse that is interior to the circle.

\begin{figure}
\centering
\mbox{\epsfig{figure= 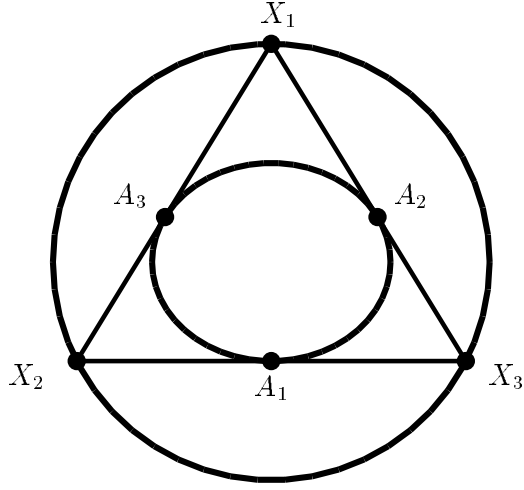,width=3.0truein}}
\caption{Diagram of a triangle showing the triangle $X_1 X_2 X_3$, and the
ellipse it defines, touching it at the points $A_1$, $A_2$ and $A_3$.}
\label{fig:fig6}
\end{figure}

The JNR parametrization corresponds to instantons in $\Rset^4$. 
However these are
related to those defined on $S^4$ by a conformal transformation.  This
corresponds to a specific stereographic projection of $S^4$  to $\Rset^4$. 
This projection takes circles to circles and triangles and ellipses in the
interior of the $S^4$ also to triangles and ellipses, respectively.  
The JNR parametrization
immediately gives us the ellipse and the circle. The points $X_{\mathrm{i}}$
determine a circle and the vertices of a triangle. Then we use
\begin{equation}
{\lambda_1\over\lambda_2}={X_1A_3\over A_3X_2}, \quad 
{\lambda_2\over\lambda_3}=
{X_2A_1\over A_1X_3}, \quad {\lambda_3\over\lambda_1}={X_3A_2\over A_2X_1}
\label{eq:lambXA}
\end{equation}
to determine the points $A_{\mathrm{i}}$, which are defined to be on the line
joining $X_{\mathrm{j}}$ with $X_{\mathrm{k}}$ (with $i,j,k$ all distinct, 
and where $X_1 A_3$ signifies the
distance between $X_1$ and $A_3$ along the line joining $X_1$ with $X_2$ for
example). The $\lambda_{\mathrm{i}}$'s and the $X_{\mathrm{i}}$'s are part of
the JNR parameters, the $A_{\mathrm{i}}$'s are uniquely determined by the 
ratios
of the $\lambda_{\mathrm{i}}$'s.  The $X_{\mathrm{i}}$'s determine the circle
uniquely, and the ellipse is determined also uniquely by the points
$A_{\mathrm{i}}$ along with the condition that the ellipse be circumscribed by
the triangle (see figure $\ref{fig:fig6}$).

\begin{figure}
\centering
\mbox{\epsfig{figure= 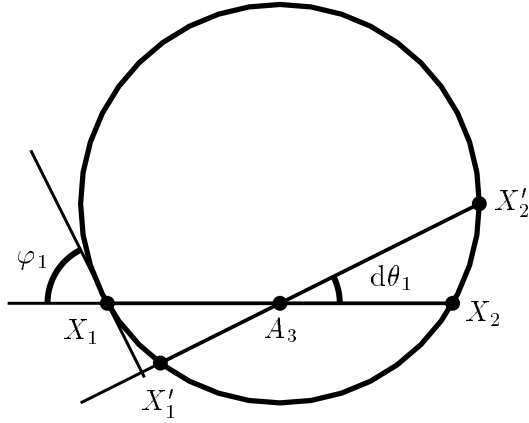,width=3.0truein}}
\caption{Diagram of the infinitesimal rotation of $X_1 X_2$ to $X_1' X_2'$.
Similar rotations occur to the other two sides of the triangle.}
\label{fig:fig7}
\end{figure}

The description of the instanton in terms of the triangle and ellipse allows
us to readily understand the symmetries of the instanton and hence the
resulting Skyrmion.  The  $\lambda_{\mathrm{i}}$'s seem to correspond to one
triangle in the porism, however they in fact afford an interpretation in terms
of an infinitesimal variation of the triangle within the porism.  If
$X^\prime_{\mathrm{i}}$ are the vertices of an infinitesimally close triangle
to the one determined by the $X_{\mathrm{i}}$'s, then the line
$X_1^\prime X_2^\prime$ is an infinitesimal rotation of the line $X_1X_2$ about
the point of tangency (see figure $\ref{fig:fig7}$). By elementary geometry
\begin{eqnarray}
X_1 X_1' &=& (X_1 A_3) {{\d} \theta_{\mathrm{1}}\over 
\sin{\varphi_{\mathrm{1}}}}
\\
X_2 X_2' &=& (A_3 X_2) {{\d} \theta_{\mathrm{1}}\over 
\sin{\varphi_{\mathrm{1}}}}
\end{eqnarray}
hence
\begin{equation}
{X_1X_1^\prime\over X_2 X_2^\prime} ={X_1A_3\over A_3X_2}
={\lambda_1\over\lambda_2} 
\end{equation}
using $(\ref{eq:lambXA})$.
Thus for an infinitesimal change of the triangle, the vertices move along the
circle by a distance proportional to the $\lambda_{\mathrm{i}}$'s.  This means
that the JNR data determines a 1 parameter family of cubic equations.  With
$X_{\mathrm{i}}X_{\mathrm{i}}^\prime ={\d}\theta_{\mathrm{i}}/
\sin \varphi_{\mathrm{i}}$, and $s_{\mathrm{i}}=
\tan{\theta_{\mathrm{i}}\over 2}$, then
\begin{equation}
{\d} s_{\mathrm{i}}={1\over 2} (1+s_{\mathrm{i}}^2){\d}
\theta_{\mathrm{i}}.\label{eq:eq2}
\end{equation}
The family of cubic equations
\begin{eqnarray}
\mu &&(s-s_1)(s-s_2)(s-s_3) +\nu \biggl(
\lambda_1(1+s^2_1)(s-s_2)(s-s_3) +\nonumber
\\
&&\lambda_2(1+s^2_2)(s-s_3)(s-s_1) + \lambda_3(1+s^2_3)(s-s_1)(s-s_2) 
\biggr)=0
\end{eqnarray}
has three real roots which correspond to the three angles
$\theta_{\mathrm{i}}$.  Indeed if $\nu =0$ the roots are $s_1, s_2$ and $s_3$.

For an infinitesimal $\nu$, $s_{\mathrm{i}} =s_{\mathrm{i}} +ds_{\mathrm{i}}$,
solving to first non-trivial order yields
\begin{eqnarray}
\mu\, ds_1ds_2ds_3 &+&d\nu\, \biggl(\lambda_1(1+s^2_1)\,ds_2ds_3 +\nonumber
\\
&&\lambda_2(1+s^2_2)\,ds_1ds_3 +
\lambda_3(1+s^2_3)\,ds_1ds_2 \biggr) =0
\end{eqnarray}
implying ${\d}s_{\mathrm{i}}={-3 {\d}\nu\over \mu} \lambda_{\mathrm{i}}
(1+s_{\mathrm{i}}^2)$ which corresponds to 
the desired variation $(\ref{eq:eq2})$.

Now we can address the case where the circle on $S^4$ passes through
the ``north pole", the point from which we do the stereographic
projection.  In this case the circle projects to a straight line in
$\Rset^4$ and the ellipse also projects to the same line.  The
projections of the porism of triangles gives a triplet of points
along the line, coming from the vertices.  These triples are again the
roots of a cubic equation as before, however the parameter is just an
affine parameter along the line.  If $X_{\mathrm{i}}$ and
$X_{\mathrm{i}}^\prime$ are two infinitesimally separated triplets, we
can define weights (up to an overall constant) by
\begin{equation}
{\lambda_1\over\lambda_2}={X_1X_1^\prime\over X_2X_2^\prime} \qquad
{\lambda_2\over\lambda_3}={X_2X_2^\prime\over X_3X_3^\prime} \qquad
{\lambda_3\over\lambda_1}={X_3X_3^\prime\over X_1X_1^\prime}
\label{eq:varinf}
\label{eq:eq3}
\end{equation}
then the JNR potential is as before.  Conversely, given the
$X_{\mathrm{i}}$ along a line in $\Rset^4$, we may invert the
stereographic projection and reconstruct the instanton and its
associated circle and ellipse on $S^4$ equally well. The inverse images of 
the $X_{\mathrm{i}}$'s determine the circle and the vertices of the triangle, 
the weights determine the ellipse.  The corresponding line of cubics is
defined directly for the affine parameter $s$ on the line in $\Rset^4$. An 
infinitesimal variation $(\ref{eq:varinf})$ implies the variation 
\begin{equation}
{\d}s_{\mathrm{i}} = \lambda_{\mathrm{i}} {\d}s.
\label{eq:dsdsi}
\end{equation}
Then the projective line of cubics is
\begin{eqnarray}
\mu(s-s_1)&&(s-s_2)(s-s_3)+\nu [\lambda_1(s-s_2)(s-s_3) +\nonumber
\\
\lambda_2
&&(s-s_3)(s-s_1) +\lambda_3(s-s_1)(s-s_2) ]=0
\end{eqnarray}
and an infinitesimal variation ${\d}\nu$ yields $ds_{\mathrm{i}} 
={-3 d\nu\over
\mu}\lambda_{\mathrm{i}}$ in concord with $(\ref{eq:dsdsi})$. 
The corresponding
points on $S^4$ projecting to these roots gives the porism of
triangles there.  

\begin{figure}
\centering
\mbox{\epsfig{figure= 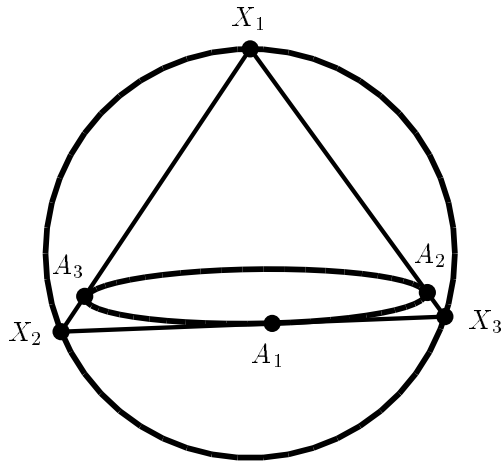,width=2.75truein}}
\caption{Triangle corresponding to a pair of well separated instantons or, 
correpondingly, to two well separated Skyrmions in the product ansatz.}
\label{fig:fig8}
\end{figure}

We will next consider two special cases which give rise to
interesting $B=2$ Skyrmions.  First consider the case of an ellipse
of very high eccentricity, such that it almost touches the surface of
the sphere $S^4$ at two points (see figure $\ref{fig:fig8}$). 
The instanton degenerates to two
$k=1$ instantons near these points.  Stereographic projection gives 2
well separated and localized instantons in $R^4$.  For simplicity  we
take $X_{\mathrm{i}}$ not be collinear.  With $\lambda_1=1$ (without
loss of generality) and  $\lambda_2, \lambda_3 <<1$
then for $x$ near $X_2$ the JNR potential is approximately
\begin{equation}
\rho = {1\over |x-X_1|^2} + {\lambda_2\over |x-X_2|^2} +
{\lambda_3\over |x-X_3|^2}\approx {1\over |x-X_1|^2} +
{\lambda_2\over |x-X_2|^2}\label{eq:JNRpot}
\end{equation}
which is exactly the JNR potential for a single instanton
centered at ${X_2+\lambda_2X_1\over (1+\lambda_2)}\approx X_2$ with
scale parameter $\lambda_2|X_2-X_1|^2$.  The scale being
proportional to $\lambda_2$, this allows for the possibility that the
instanton here has little overlap with the corresponding one located
at $X_3$.  The integration along the time lines gives two well
localized Skyrmions.  The minimum energy single Skyrmions are
obtained for $\lambda \approx 2.109$, which is not particularly
small.  However given that the scale size of the instantons is
quadratically dependent on the separation, it is always possible to
choose this so that the resulting single Skyrmions are well separated
and of minimal energy.

If $X_1$ is relatively near $X_2$ and it changes while keeping
everything else fixed, the Skyrmion at position $X_2$ will vary over
all possible orientations while the one at $X_3$ will remain
essentially unchanged. $(\ref{eq:JNRpot})$
implies a gauge transformation $(\ref{eq:gtinst})$
\begin{equation}
U =
{(X_2 -X_1)^0\over |X_2 -X_1|}+
{\mathrm{i}} {(\vec X_2 -\vec X_1 )\over |X_2 -X_1|}\cdot\vec\tau.
\end{equation}
which covers ${\mathrm{SO}}(3)$, hence all
possible orientations, twice as $X_1$ varies over a 3 sphere centered
on $X_2$.  Hence all possible relative orientations are permitted and
this case contains all well separated Skyrmions which gives rise to
the product ansatz.  

The second case that we will consider is the spherically symmetric
situation, which gives rise to hedgehog fields for $B=2$.  Consider
the potential
\begin{eqnarray}
\rho &=& {\lambda_1\over r^2+(\tau-\tau_1)^2} + {\lambda_2\over
r^2+(\tau-\tau_2)^2} + {\lambda_3\over r^2+(\tau-\tau_3)^2}\nonumber
\\
&=&{N(r,\tau)\over 
\prod_{{\mathrm{i}}=1}^3(r^2+(\tau-\tau_{\mathrm{i}})^2)}
\end{eqnarray}
with $r^2=\vec x\cdot\vec x$.  The corresponding JNR instantons are
situated along the time axis.  $N(r,\tau)$ is a quartic
polynomial in $\tau$ with $r$ dependent coefficients which is positive
since the denominators can never simultaneously vanish.  $N$ is
actually gauge invariant.  $N$ has $r$ dependent roots $\alpha (r),
\alpha^*(r), \beta (r), \beta^* (r)$, with the imaginary parts of $\alpha$
and $\beta$, $\Im (\alpha ), \Im (\beta)
> 0$.  Then
\begin{eqnarray}
\ln\rho = \ln (\tau-\alpha ) &+&\ln (\tau-\alpha^*) + \ln (\tau-\beta ) +\ln
(\tau-\beta^*) \nonumber
\\
&&-\sum_{{\mathrm{j}}=1}^3 \biggl( \ln{ (\tau-\tau_{\mathrm{j}} -
{\mathrm{i}}r)} + \ln {(\tau-\tau_{\mathrm{j}} + 
{\mathrm{i}}r)}\biggr)
\end{eqnarray}
which in turns yields
\begin{eqnarray}
{1\over \rho}{{\d}\rho\over {\d}r}= -{1\over (\tau-\alpha)}
{{\d}\alpha\over {\d}r}
&-&{1\over (\tau-\alpha^*)}{{\d}\alpha^*\over {\d}r}
-{1\over (\tau-\beta)}{{\d}\beta\over {\d}r}
-{1\over (\tau-\beta^*)}{{\d}\beta^*\over {\d}r} \nonumber 
\\
&+&{\mathrm{i}}\sum_{{\mathrm{j}}=1}^3
\biggl( {1\over \tau-\tau_{\mathrm{j}} - {\mathrm{i}}r} 
- {1\over \tau-\tau_{\mathrm{j}} + {\mathrm{i}}r}\biggr).
\end{eqnarray}
Then
\begin{equation}
A_4={{\mathrm{i}} \over 2}{1\over \rho}{{\d}\rho\over {\d}r} 
\hat r\cdot\vec\tau
\end{equation}
so the Skyrme field from the integration along time lines would
give
\begin{equation}
f(r)= -{1\over 2}\int_{-\infty}^\infty {1\over \rho}
{{\d}\rho\over {\d}r} {\d}\tau-\pi\label{eq:infr2}
\end{equation}
The $\pi$ comes from the difference between instantons on $S^4$ and $\Rset^4$,
the integration along time lines in $\Rset^4$ must be closed with a 
semi-circle at infinity to give a truly closed integration contour. With 
this addition the boundary condition $U|_\infty\rightarrow 1$ is satisfied.
The integral $(\ref{eq:infr2})$
may be computed using standard techniques of contour 
integration, yielding
\begin{eqnarray}
f(r) &=& \pi {\mathrm{i}}\, {{\d}\over {\d}r} (\alpha +\beta ) +2 \pi
\nonumber
\\
&=& \pi {\mathrm{i}}\, {{\d}\over {\d}r} \biggl(\Im (\alpha) 
+\Im (\beta) \biggr) +2 \pi
\end{eqnarray}
using  $\alpha +\alpha^* +\beta +\beta^*$ is independent of $r$.  The
roots of a quartic can be found in closed form, however this does not
elucidate the properties of the configuration.

The JNR potential has 5 parameters, $\tau_{\mathrm{i}}$ and the 2
independent ratios of the $\lambda_{\mathrm{i}}$.  $\tau_3$ can be
varied arbitrarily by a transformation moving along the line of
cubics (changing the porism of triangles), leaving 4 parameters.
We will call these $T_1$,
$T_2$, $\Lambda_1$ and $\Lambda_2$ in the equivalent 
't Hooft parametrization. (Take $\lambda_3=1$, $\tau_3=0$ and go to 
the 't Hooft parametrization.)
A rigid translation of the $T_{\mathrm{i}}$'s does nothing hence we take 
$T_1=-T$ and $T_2=T$, leaving 3 parameters. Finally imposing time reversal 
symmetry $\Lambda_1=\Lambda_2=\Lambda$, we get
\begin{equation}
N(r,\tau)=(r^2 +\tau^2)^2 +2(\Lambda +T^2)r^2 +2(\Lambda -T^2)\tau^2 + T^4
+2\Lambda T^2
\end{equation}
which implies that $\Im(\alpha) =\Im(\beta)$ and since this is just a 
quadratic equation in $\tau^2$
\begin{equation}
\biggl(\Im(\alpha)\biggr)^2 = 
{1\over 2} (R^2 + \Lambda - T^2 + 
( (r^2 + \Lambda +T^2) - \Lambda^2)^{1\over 2})
\end{equation}
and 
\begin{equation}
f(r) = 2\pi - 2\pi {{\d}\over {\d}r} \Im (\alpha) 
\end{equation}
As $T\rightarrow\infty$ we can find the form of $\alpha$, yielding
\begin{equation}
f(r) = 2\pi \biggl( 1 - (1+{\Lambda\over r^2} )^{-{1\over 2}}\biggr).
\end{equation}
This has energy  $E=1.855535\times 24\pi^2$ when $\Lambda =2.6211$.
Actually the true minimum occurs for $T^2\approx 84.6$ and $\Lambda =
2.6427$, with $E= 1.855529\times 24\pi^2$ which has been shown
numerically (see \cite{Atiyah-Manton-2}).  

The time centered 't Hooft potential
\begin{equation}
\rho = 1 +{\Lambda + \Delta\over r^2 +(\tau-T)^2} + {\Lambda - \Delta\over
r^2 +(\tau+T)^2}
\end{equation}
gives an approximate potential
\begin{equation}
f(r) = 2\pi -\pi \Biggl(1 +{\Lambda + \Delta\over r^2 }\Biggr)^{-{1\over 2}}-
\pi \Biggl(1 +{\Lambda - \Delta\over r^2 }\Biggr)^{-{1\over 2}}
\end{equation}
which is the product of two $B=1$ hedgehogs of scale parameters
$\Lambda + \Delta$ and $\Lambda - \Delta $.  The minimum occurs  at
$\Delta = 0$, $\Lambda = 2.6211$ with $E=1.855536\times 24\pi^2$.
The energy is symmetric in $\Delta $ so it is a reasonable conjecture
that the minimum actually occurs at $\Delta = 0$.

The spherically symmetric $B=2$ hedgehog  is stable against
perturbations preserving that symmetry.  Its unstable modes violates
this symmetry and span a 6 dimensional vector space.  Under 
${\mathrm{O}}(3)$
rotations these decompose into two 3-dimensional irreducible
sub-spaces, ${\mathrm{O}}(3)$ is vectorial for one and axial for the other.

In terms of the product ansatz we lower the energy if two coincident
Skyrmions are displaced -- along  3 independent axes (axial mode) or
rotated relatively in iso-space about 3 independent axes (vectorial
mode). There are actually 3 modes which increase the energy while
preserving the hedgehog form, the 3 independent parameters found above,
$\Lambda_1$, $\Lambda_2$ and $T$.  
There are 6 zero modes corresponding to
translations and global rotations (equivalently iso-rotations) and 
including the 6 modes which decrease the energy -- unstable or
negative modes, giving a total of 15 modes.  Using the JNR
parametrization the minimum energy hedgehog has
\begin{equation}
\rho ={1\over r^2 +\tau^2} + {\Lambda^\prime\over r^2 +(\tau -T^\prime )^2}
+ {\Lambda^\prime\over r^2 +(\tau +T^\prime )^2}
\end{equation}
where $\Lambda^\prime ={\Lambda\over T^2}$ and $T^\prime =(T^2
+2\Lambda )^{1\over 2}$ and with $T^2 =84.6, \Lambda = 2.6427$.

This configuration corresponds to a line of cubics with collinear
roots.  The perturbations which break the collinearity gives the
vector instability while the perturbations that rotate the line gives
the axial one. These perturbations of the instanton configuration
give rise to exactly the same perturbations of the corresponding
Skyrmions which reduce the energy according to the analysis of Bang
and Wirzba\cite{Bang-Wirzba}. 
We will return to these unstable modes in section 4.3.

\begin{figure}
\centering
\mbox{\epsfig{figure= 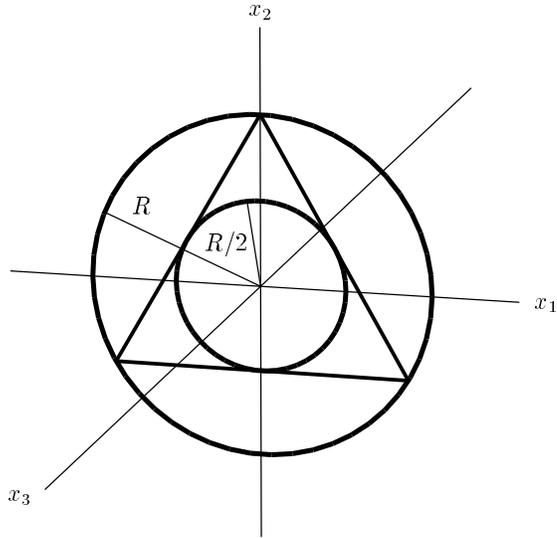,width=3.0truein}}
\caption{Reflection symmetry: $x_1\rightarrow - x_1$ and $x_2\rightarrow -x_2$,
and ${\mathrm{O(2)}}$ symmetry about the $x_3$ axis.}
\label{fig:fig9}
\end{figure}

The most attractive channel instantons and hence Skyrmions, in the 
Hartshorne description, are distinguished by
concentric circles and ellipses. For high eccentricities we have two well
separated Skyrmions with relative iso-rotation of $180^\circ$. The minimum 
energy configuration appears when the ellipse
degenerates to a circle of radius  $R\over 2$
(see figure $\ref{fig:fig9}$). 
Evidently
the porism of triangles is given by a set of equilateral triangles, 
obtained from one another by a simple rotation. This Hartshorne configuration
exudes ${\mathrm{O(2)}}$ symmetry, which indeed 
the instanton configuration, and the 
subsequent Skyrmion field also exhibit.

\begin{figure}
\centering
\mbox{\epsfig{figure= 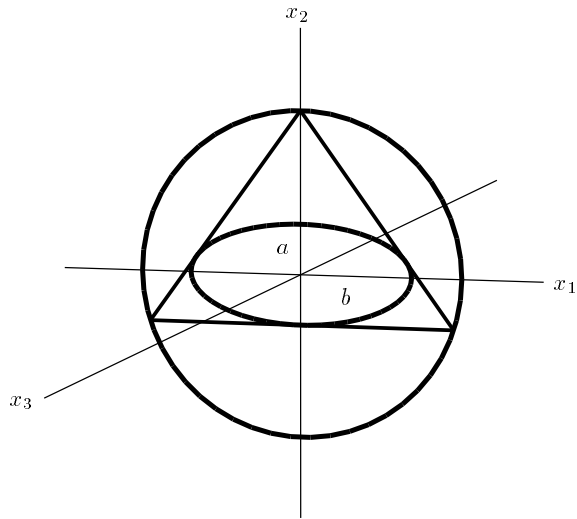,width=3.0truein}}
\caption{Reflection symmetry about the three orthgonal axes: 
$x_{\mathrm{i}}\rightarrow - x_{\mathrm{i}}$.}
\label{fig:fig10}
\end{figure}

\begin{figure}
\centering
\mbox{\epsfig{figure= 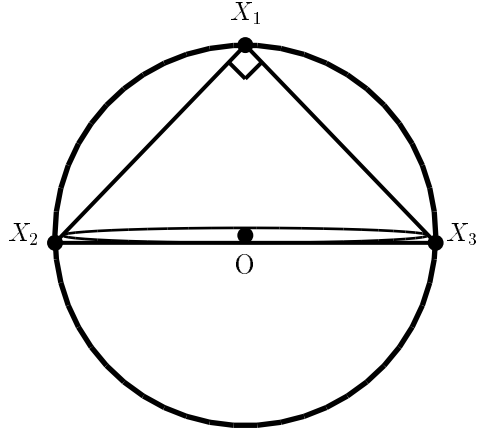,width=2.5truein}}
\caption{Hartshorne diagram 
corresponding to two well separated Skyrmions in the most 
attractive channel.}
\label{fig:fig11}
\end{figure}

If the Hartshorne ellipse is only
concentric with the circle then the  ${\mathrm{O(2)}}$ 
symmetry reduces to reflection symmetry with respect to the three axes. The 
Poncelet condition requires that $a+b=R$, where $a$ and $b$ are the semi-major
and semi-minor axes respectively while $R$ is the radius of the circle
(see figure $\ref{fig:fig10}$).
When the ellipse becomes extremely eccentric (but always 
remains concentric with the circle), it is 
easy to verify that the configuration corresponds to well separated Skyrmions
with a relative iso-rotation of $180^\circ$. Indeed the triangle degenerates 
to a right isoceles triangle with hypotenuse approximately a diameter
(see figure $\ref{fig:fig11}$).
The configuration corresponds to Skyrmions localized at positions $X_2$ and 
$X_3$ with iso-rotation factors $\widehat{(X_2-X_1)}\cdot\vec\tau$ and 
$\widehat{(X_3-X_1)}\cdot\vec\tau$ respectively ($\widehat{(X_2-X_1)}$ is a 
unit vector in the $X_2-X_1$ direction).
The configuration is
\begin{eqnarray}
U &=& \biggl({\mathrm{i}} \widehat{(X_2-X_1)}\cdot\vec\tau\biggr)^\dagger
U(x-X_2)
{\mathrm{i}} \widehat{(X_2-X_1)}\cdot\vec\tau\nonumber
\\
&&\qquad\qquad\times
\biggl({\mathrm{i}} \widehat{(X_3-X_1)}\cdot\vec\tau\biggr)^\dagger
U(x-X_3)
{\mathrm{i}} \widehat{(X_3-X_1)}\cdot\vec\tau
\end{eqnarray}
where $U(x-X_2)$ is a Skyrmion at $X_2$,
hence the relative iso-orientation is
\begin{eqnarray}
R &=& {\mathrm{i}} \widehat{(X_2-X_1)}\cdot\vec\tau
(-{\mathrm{i}}) \widehat{(X_3-X_1)}\cdot\vec\tau\nonumber
\\&=& {\mathrm{i}} \biggl( \widehat{(X_2-X_1)}\times
\widehat{(X_3-X_1)}\biggr) \cdot\vec\tau
\end{eqnarray}
for the Skyrmion located at $X_3$. However the vector 
\begin{equation}
\widehat{(X_2-X_1)}\times\widehat{(X_3-X_1)}
\end{equation}
is a unit vector (because the angle at $X_1$ is $90^\circ$), orthogonal 
to the plane defined by $X_1$, $X_2$ and $X_3$, 
and hence also orthogonal to the separation axis $X_2-X_3$. Furthermore since
$R$ is of the form ${\mathrm{i}}\,\hat n\cdot\vec\tau$ it effects an 
iso-rotation by exactly $180^\circ$ as desired, about the axis 
$\widehat{(X_2-X_1)}\times\widehat{(X_3-X_1)}$.
The discrete symmetries of the concentric ellipse expand to the continuous
${\mathrm{O}}(2)$ symmetry of the concentric circles and implies 
that this configuration should correspond to the minimum energy 
configuration which is known to have toroidal symmetry
\cite{Verbaarschot,Mantontore,Kopeliovich}. After
minimization with respect to the (1) free parameter $R$ one actually finds the 
minimal torus to within a few percents. 

The manifold of attractive channel Skyrme fields forms an 11 dimensional 
sub-manifold of the full 15 dimensional manifold of instanton generated 
Skyrme fields. The ellipse simply has 2 degrees of freedom, its eccentricity 
and its orientation, instead of 4, and the orientation of the 2-plane must 
be orthogonal to the time axis implying only the action of ${\mathrm{O}}(4)/
({\mathrm{O}}(2)\times{\mathrm{O}}(2))$ yielding $6-(1+1)=4$ parameters instead
of 6, reducing by another 2. Hence 15 goes to 11 parameters. As 9 parameters
correspond to the action of the global symmetry group of translations, 
rotations and iso-rotations, we must find the gradient flow curves in a two 
dimensional subspace, parametrized by $a$, $b$ and $R$ subject to the
constraint $a+b=R$. These gradient flow curves would start approximately at 
the asymptotic critical point of the two infinitely separated, minimal energy, 
isolated Skyrmions, and arriving at the minimal energy toroidal configuration. 
Since the product ansatz tells us that asymptotically this is a 10
dimensional manifold, the gradient flow must also yield a 10 dimensional 
manifold. This is intuitively reasonable, minimizing in a two dimensional 
manifold will typically yield a one dimensional ``valley'' or ``path'' of 
steepest descent linking together the critical points. Hence we do indeed 
obtain a $9+1=10$ dimensional manifold of most attractive channel instanton 
generated Skyrme fields. 

The gradient flow has not to date been calculated. Hosaka \etal
\cite{Hosaka} have exhibited 
a qualitatively similar manifold of 
constrained minima. It is obtained simply by letting 
the set of triangles vary from that of the right isoceles triangle 
corresponding to widely separated Skyrmions, to the 
equilateral triangle of the toroid through intermediate symmetric isoceles 
triangles. The Hartshorne ellipse starts at 
very high eccentricity and varies until it degenerates 
to the circle. The energy is minimized for each intermediate triangle,
fixing the value of $R$. The minimal energy for fixed eccentricity decreases
monotonically until the circle is reached at $a/b=1$. The constrained energies 
are always within $1\%$ or $2\%$ of the similar but 
fully numerical computations 
of Verbaarschot \etal\cite{Verbaarschot-Walhout-Wambach-Wyld} and of 
Walhout\cite{Walhoutmulti}. 

Geometrically the 10 dimensional most attractive channel manifold consists of 
the direct product of a
6 dimensional (global) manifold, which is generated by 3 independent spatial 
translations and 3 independent isospin rotations, with
a 4 dimensional (relative) manifold. The 4 dimensional relative manifold 
consists of ``centered''  
Skyrme fields, which can be acted upon by the group of spatial rotations. 
The manifold is parametrized by a coordinate depicting separation and 3 
angular coordinates taken without loss of generality to be the Euler angles 
specifying a frame of 
unoriented Cartesian axes. The generic ${\mathrm{SO}}(3)$ orbit is actually
only ${\mathrm{SO}}(3)/V$, where $V$ is the group of $180^\circ$ rotations
about the 3 axes (and the identity), which of course leave an unoriented 
Cartesian frame invariant. At minimal separation the orbit 
degenerates to $\Rset P_2$ which is the same as 
the sphere $S^2$ with antipodal 
points identified, physically it is the orbit of the symmetry axis of the 
toroid, taking into account that this axis is unoriented.

The Atiyah-Hitchin manifold\cite{Atiyah-Hitchin}, 
corresponding to the moduli space of centered 2
BPS-monopole configurations has exactly the same orbit structure. Hence the 
Atiyah-Hitchin manifold is a good candidate describing centered, attractive 
channel Skyrmions. The Atiyah-Hitchin manifold has an implicit metric, the 
Atiyah-Hitchin metric. This metric, however, is not appropriate 
for Skyrmions. The Atiyah-Hitchin encodes in it velocity-dependent Coulomb 
interactions between monopoles, which are absent for Skyrmions. The true
metric must be calculated using the Skyrme energy functional. This has been 
done by Leese \etal\cite{Leese-Manton-Shroers}.

%% file: section33.tex
The study of high baryon number solutions of the Skyrme model has always been 
a very difficult problem to tackle. What is lacking is a good ansatz which 
captures the symmetries and simplifies the equations of motion. 

\begin{figure}
\centering
\mbox{\epsfig{figure=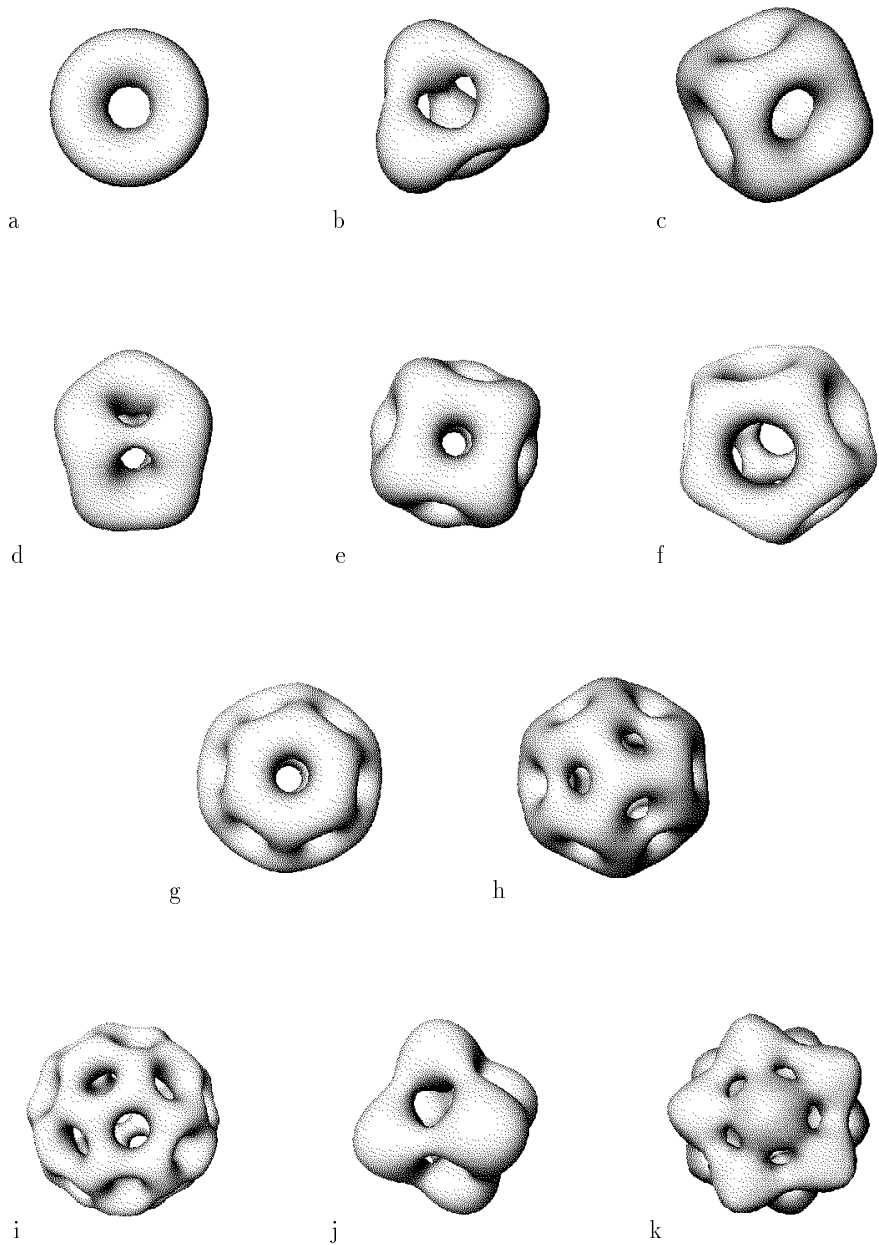,width=5.0truein}}
\caption{Various baryonic density for Skyrme states of baryon number 2 to 
12.}
\label{fig:fig12}
\end{figure}

A first glance at higher baryon number 
was given by Braaten \etal\cite{Braaten-Townsend-Carson} almost 10 years 
ago. Using relaxation methods on a highly (in those days) powerful Cray 
super-computer, they isolated states for the sectors $B=2$ to $B=6$ and 
computed their energies. Quite surprisingly, the configurations 
for $B=3$, 4, 5 and 6 
took very geometrical shapes, tetrahedral ($B=3$), octahedral ($B=4$), and 
less symmetrical ones for $B=5$ and 6. Obviously, these look nothing like 
nuclei as described by shell or droplet models from traditional nuclear 
physics. Of course, the doughnut shaped $B=2$ state, the Skyrme model's 
deuteron state, looks very little like a pair of weakly interacting nucleons. 
However, as the classical binding energy of the doughnut (100 MeV) is 
much greater than the deuteron's real binding energy (around 2 MeV), one 
could hope that the $B=2$ case is not typical for the Skyrme model and that 
things would settle down for higher baryon number. We will see that 
it is not the case, and one gets the impression from the literature that 
the results for ${\mathrm{B}}=2$ to $6$ were both unexpected and not 
understood: consequently they were left pretty 
much alone for the interim. In 1996 Battye and Sutcliffe
\cite{Battye-Sutcliffe-PRL} 
confirmed these results (except the $B=6$ state which seems 
to have been misidentified) using state-of-the-art software and hardware, as 
well as found the structure of probably minimum energy states with $B=7$, 8 
and 9 (see figure $\ref{fig:fig12}$ taken from reference 
\cite{Battye-Sutcliffe-PRL}). 
This indicates that indeed the Skyrme model favours regular geometric 
configurations instead of shell model-type lumps. These shapes in fact, 
are very similar to those found in chemistry. Only time and further 
investigations will tell if indeed and why such configurations are physically 
relevant. 

One must note here that all these states 
have been found using numerical relaxation methods, with the algorithm 
feeling its way down the slopes of configuration space in search of the 
lowest energy possible in each 
given baryon number sector. Of course, it is 
impossible to be absolutely sure that the configurations 
obtained so far are absolute 
minima of the energy in each baryon number sector, and not merely local 
minima, without using other methods. Thus there is still a 
possibility that the spectrum of real states might be different, even though 
the fact that the energy per baryon number is very close to the Bogomolny 
bound (about 10\% or so over the bound) 
makes this possibility quite remote. 
Oscillations and perturbations around those solutions have been considerd to 
test (at least locally) the stability of the solutions, and to compare the 
resulting spectrum with known excited states of nuclei. So far the results 
look encouraging but there is still a lot of work to be done before 
comparison with nature can be made in a serious manner
\cite{Baskerville,Baskerville2}.  

As impressive as these numerical results might look, they are still a long way 
ahead of actually doing simulations of Skyrmion scattering 
processes for several initial Skyrmions or for large energy. Even though 
it actually is possible to do so, it still takes strong numerical 
skills to do the simulation and extract and interpret the results. To be able 
to push forward the current study of the model, and also to avoid that 
the study of soliton 
scattering is absorbed into purely numerical or computational 
physics, one needs some kind of analytical handle on the problem, even an 
approximate one (for a flavour of the difficulties encountered in the study 
of soliton-soliton scattering see the article by Crutchfield and 
Bell\cite{Crutchfield-Bell}). 
The instanton method which we have 
just described does exactly 
this but it is of little help for the study of large baryon number solutions, 
except under very special conditions\cite{Leese-Manton}.
The gradient flow curve method 
which we will present in section 4 does not help much either since to 
apply it we need to know the manifold connecting the critical points of the 
system for low energy in a given baryon number sector. However we do not know 
this manifold, and we actually deduce its structure from numerical 
simulations. 

Help comes our way from the study of the BPS monopole system. It is not our 
intention to discuss this system here (see the review by
Sutcliffe\cite{SutcliffeIJMP}). Pioneers of this model 
are Bogomolgny\cite{Bogomolny}, Prasad and 
Sommerfield\cite{Prasad-Sommerfield}, 
't Hooft\cite{tHooftmono}, Polyakov\cite{Polyakov}, 
Atiyah and Hitchin\cite{Atiyahmono,Hitchin}, 
Manton\cite{Mantonmono}, Gibbons\cite{Gibbons}, Ward\cite{Ward}
and Nahm\cite{Nahm}, to mention a few. The BPS monopole is a topological 
soliton of a massless Higgs type model with an SU(2) gauge symmetry, which 
saturates the Bogomolny bound of the theory\cite{Goddard-Olive,Colemanmono}. 
Over the years, the spectrum of 
the model has been studied, and states with each value of the magnetic charge 
(which represents the winding number associated to the soliton in this 
model) have been isolated. Quite surprisingly, there exists a state 
with magnetic charge $N$ with symmetries identical to those of the 
Skryme model for 
baryon number $N$\cite{Houghton}
(this fact has been verified for $N$ ranging from 1 to 
9): toroidal for $N=2$, tetrahedral for $N=3$, etc. This of course 
does not indicate that BPS fields should be used to study baryonic systems, 
but it does indicate 
that the mathematical arsenal used in the study of the BPS 
system could perhaps be generalized to Skyrmion systems\footnote{In fact, the 
similarity between BPS and Skyrme systems could run deeper than it appears. 
During his talk at the CRM-Fields-CAP 1997 workshop ``Solitons'' in 
Kingston, Canada, N.S. 
Manton conjectured that the moduli space of vortices is a submanifold 
of the moduli space of BPS monopoles, which is itself a submanifold of 
that of Skyrmions, itself being included in the moduli space of SU(2) 
instantons. 
This fascinating conjecture remains to be established.}. 
The tools of interest here are 
called rational maps and were introduced in the study of BPS systems by 
Donaldson\cite{Donaldson}, 
and recently elaborated upon by Jarvis\cite{Jarvis}. 
We will not explain in 
detail the method nor show how it works. Instead we will give a taste of how 
it can be 
applied to Skyrmion systems following the article of Houghton 
\etal\cite{Houghton}.  

BPS monopoles are solutions of a model possessing a 
symmetry breaking term which breaks an SU(2) symmetry via a triplet of Higgs
fields to a U(1) symmetry, interpreted as electromagnetism. 
Inside the monopole, the 
SU(2) symmetry is intact, but it has to be broken to U(1) on the outside so 
as to give the soliton finite energy (because of a potential term in the 
Lagrangian density). Restricting the Higgs field to its broken symmetry value 
outside the soliton fixes its length, but not its 
direction: it can take any direction in 
$\Rset^3$ and in fact describes an $S^2$. Infinity in flat 3-dimensional 
space also consists of a 2-sphere, so the Higgs field of the monopole is a 
map from $S^2$ to $S^2$. Such maps are divided in disjoint homotopy 
classes numbered by the number of times the first $S^2$ is wrapped around the 
second $S^2$. This topological winding number 
is proportional to the magnetic charge of 
the field. The vacuum has the Higgs field pointing the same way everywhere 
(thereby having winding number and magnetic charge 0), while the unit 
magnetic charge monopole looks like a hedgehog from afar. By stereographic 
projection one can transform the map from $S^2$ to $S^2$ to another which 
maps the complex plane in another complex plane (if we identify a given point 
of the spheres with infinity in the usual way). Donaldson\cite{Donaldson} 
showed that 
there is a one to one relation between the field of an $N$ monopole and 
rational maps $R(Z)$ of degree N. A rational map of degree $N$ from
$\Cset\rightarrow \Cset$ is defined as:
\begin{equation}
Z\mapsto R(Z) = {p(Z)\over q(Z)}\label{eq:rat}
\end{equation}
where $p(Z)$ and $q(Z)$ are polynomials of at most degree $N$, with at least 
one being of degree $N$ and with no common roots. The parameters of the 
polynomials generate (much like in the case of the parameters of the 
instantons of the previous subsection) a finite dimensional manifold of 
configurations and can be chosen so as to give the soliton some desired 
symmetries, and fine tuned to lower the energy as much as possible. We refer 
the reader to the literature for further details\cite{Houghton}. 
Experience shows that once 
a particular set of symmetries has been chosen and implemented in the 
rational map, the method gives a good approximation 
to the exact (numerical) result. 

\begin{figure}
\centering
\mbox{\epsfig{figure= 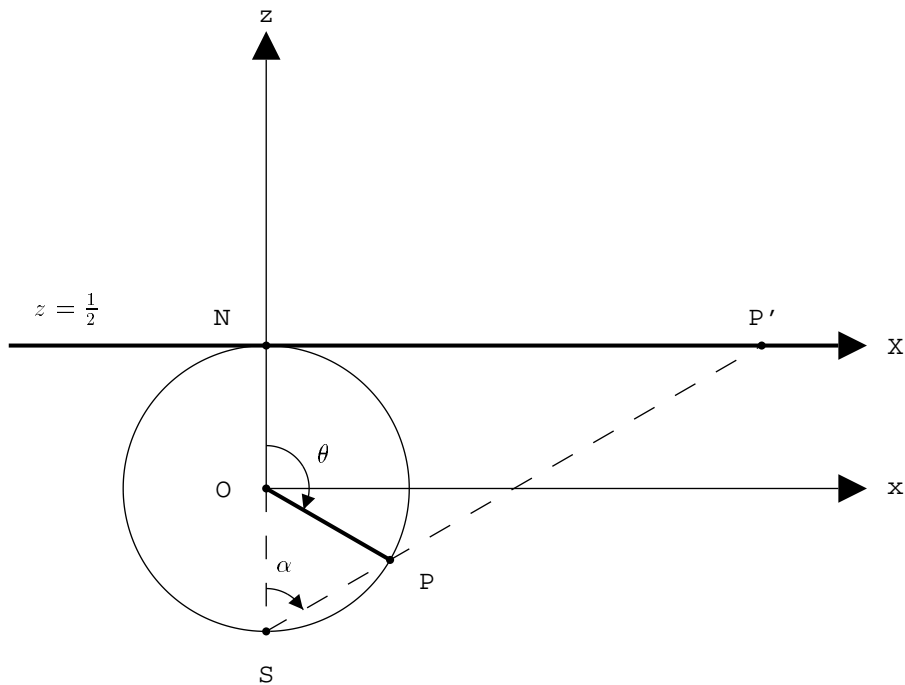,width=5.0truein}}
\caption{Stereographic projection of an $S^2$ of radius $1/2$ on plane $(XY)$.
It gives $Z=X + {\mathrm{i}}Y = \tan{(\theta/2)} 
{\mathrm{e}}^{{\mathrm{i}}\phi}$, and the projection angle $\alpha = 
\theta/2$.}
\label{fig:fig13}
\end{figure}

Houghton \etal\cite{Houghton} generalized this method to the Skyrme model by 
choosing the following ansatz for Skyrme fields:
\begin{equation}
U(\vec x) = e^{ \mathrm{i} f(r) \hat n_{\mathrm{R(Z)}}\cdot\vec\tau}
\label{eq:urat}
\end{equation}
where $\hat n_{\mathrm{R(z)}}$ is a function of $Z$, or 
in other words of the usual spherical 
angles $\theta$ and $\phi$ only. $f(r)$ depends solely on the distance $r$ 
from 
the origin. This ansatz is interesting in that it singles out the distance 
to the origin from the angular coordinates. As Houghton
\etal noted, one can understand this ansatz as mapping 
the two-spheres centered on the origin of space
onto the 2-spheres which correspond to
latitudes in the $S^3$ (of $\mathrm{SU(2)}$). 
To do this the two-spheres
of the $\Rset^3$ of space are mapped via stereographic projection onto the 
complex plane (with complex infinity identified to a single point of the 
sphere) as shown in figure \ref{fig:fig13}. 
Elements $Z$ 
of this plane are then functions of the spherical angles $\theta$ and 
$\phi$: 
\begin{equation}
Z = \tan{(\theta/2)} {\mathrm {e}}^{ {\mathrm{i}} \phi}.\label{eq:defzproj}
\end{equation} 
We are now on familiar 
ground: elements of the initial complex plane $Z$ are then mapped by the 
rational map $R(Z)$ into another complex plane, itself obtained by 
stereographic projection of the latitudes of the ${\mathrm{SU(2)}}$ 
three-sphere:
\begin{equation}
\matrix{
\Rset^3 & \longrightarrow & \Rset & \times & S^2 & & S^2 & ``\times'' & S^1 & 
\longrightarrow & S^3
\cr
\{r,\theta,\phi\} & & & & \{\theta,\phi\} & & \{\hat n_{\mathrm{R(Z)}}\} 
& & & & \{U(\vec x)\}
\cr
 & & & & \big\downarrow & & \big\uparrow & & & 
\cr 
 & & & & R: \Cset & \longrightarrow & \Cset & & & 
\cr
 & & & & \{Z\} & & \{R(Z)\} & & & 
}
\end{equation}
$\hat n_{R(Z)}$ is expressed as follows:
\begin{equation}
\hat n_{\mathrm{R(Z)}} = {1\over |1+|R|^2|} ( 2\Re{(R)},2\Im{(R)},
1-|R|^2 )
\label{eq:nrat}
\end{equation}
where $\Re(R(Z))$ and $\Im(R(Z))$ represent the real and imaginary parts of 
$R(Z)$ respectively. $\hat n_{\mathrm{R(Z)}}$ depends on the rational map 
$R(Z)$ chosen, and gives, for 
example, the ordinary radial vector $\hat n$ if $R(Z)=Z$. This particular map
reproduces of course the ordinary Skyrmion.
 
The parametrization $(\ref{eq:urat})$
chosen for $U$ can seem a bit restrictive. It is 
especially suited to reproduce configurations localized around the origin. 
It probably does not reproduce accurately, well separated Skyrmions (although 
it works surprisingly well and gives an idea of some processes). The main 
advantage of this parametrization is that it 
decouples contributions from the radial (\ie ones related to $f(r)$) and 
from the angular part (\ie related to $R(Z)$). This way, it is possible 
to first impose a given symmetry to the configuration and then 
minimize the angular contribution to the energy. Subsequently by
minimizing with respect to $f(r)$ we obtain the minimum energy possible
within the ansatz. 
We note that in certain 
cases, varying the value of some parameters of $R(Z)$ gradually and 
minimizing with respect to $f(r)$ gives ``snapshots'' of certain scattering 
processes. The case of $B=3$ is a good example. 

Choosing the 
parameters in the polynomials $p(Z)$ and $q(Z)$ so as to get solutions 
with the 
right symmetries is a technical but important point which we will illustrate 
by the following example.
Let us compute the expression for the rational map $R(Z)$ which gives the 
Skyrmion field $U(\vec x) = u_0 + {\mathrm{i}}\;\vec u\cdot\vec\tau$ with 
lowest energy in the $B=2$ sector. 
As we will see in section 4.3 the following three symmetries are 
characteristic of a pair of Skyrmions 
converging
together at the origin to form a tightly bound minimum energy 
toroidal solution of the equations of motion:
\begin{eqnarray}
I_1: x\rightarrow - x&:&\quad u^1 \rightarrow u^1\label{eq:symbpa1}
\\
y \rightarrow y & &\quad u^2 \rightarrow -u^2\nonumber
\\
z \rightarrow z& &\quad u^3 \rightarrow u^3\nonumber
\end{eqnarray}
\begin{eqnarray}
I_2: x\rightarrow x&:&\quad u^1 \rightarrow u^1\label{eq:sympa2}
\\
y \rightarrow -y & &\quad u^2 \rightarrow -u^2\nonumber
\\
z \rightarrow z& &\quad u^3 \rightarrow u^3\nonumber
\end{eqnarray}
\begin{eqnarray}
I_3: x\rightarrow x&:&\quad u^1 \rightarrow u^1\label{eq:sympa3}
\\
y \rightarrow y & &\quad u^2 \rightarrow u^2\nonumber
\\
z \rightarrow -z& &\quad u^3 \rightarrow -u^3.\nonumber
\end{eqnarray}
The most general expression for the rational map
$R$ of degree 2 is given by:
\begin{equation}
R(Z) = { \alpha Z^2 + \beta Z + \gamma\over
\lambda Z^2 + \mu Z + \nu}\label{eq:rgen}
\end{equation}
where $\alpha$, $\beta$, $\gamma$, $\lambda$, $\mu$ and $\nu$ are constants
(real or imaginary) to be determined by imposing the symmetries $I_1$, $I_2$
and $I_3$. To do this, we need to parametrize them using $Z$ and $R(Z)$.
Using $(\ref{eq:defzproj})$ and
\begin{equation}
\vec u = \sin{f(r)} \hat n_{\mathrm{R(Z)}}
\end{equation}
we see that the transformation $x\rightarrow -x$ of $I_1$ is equivalent to 
$Z \rightarrow - \bar Z$ after projection on the complex plane. 
Similarly, $y\rightarrow -y$ and $z\rightarrow -z$ just translate
to $Z \rightarrow \bar Z$ and $Z \rightarrow 1/\bar Z$, respectively. For 
our needs we will only consider the new transformations 
${\cal I}_1\equiv I_1 I_2$ and ${\cal I}_2 \equiv I_2 I_3$ (the product 
$I_3 I_1$ being discarded since it brings no new constraint on $R$). 
The symmetry ${\cal I}_1$ is given then by:
\begin{eqnarray}
{\cal I}_1: Z\rightarrow - Z&:& \quad \hat n^1 \rightarrow \hat n^1
\\
&:& \quad \hat n^2 \rightarrow \hat n^2\nonumber
\\
&:& \quad \hat n^3 \rightarrow \hat n^3.\nonumber
\end{eqnarray}
The field $\hat n_{\mathrm{R(Z)}}$ being then invariant under the 
transformation, this imposes the constraint 
\begin{equation}
R(-Z) = R(Z)\label{eq:rpaire}
\end{equation} 
on $R$. Similarly, ${\cal I}_2$ is 
written as:
\begin{eqnarray}
{\cal I}_2: Z\rightarrow 1/Z&:& \quad \hat n^1 \rightarrow \hat n^1
\\
&:& \quad \hat n^2 \rightarrow -\hat n^2\nonumber
\\
&:& \quad \hat n^3 \rightarrow -\hat n^3.\nonumber
\end{eqnarray}
A short calculation shows that $\hat n$ transforms under this transformation
like $\hat n_{\mathrm{R(Z)}} \rightarrow\hat n_{\mathrm{1/R(Z)}}$. So $R$ also 
has to satisfy  
\begin{equation}
R(1/Z) = 1/R(Z).\label{eq:rfrac}
\end{equation}
We now apply the conditions $(\ref{eq:rpaire})$ and $(\ref{eq:rfrac})$
to the general form of $R$, $(\ref{eq:rgen})$ to fix the constants 
$\alpha$, $\beta$, $\gamma$, $\lambda$, $\mu$ and $\nu$. Equation 
$(\ref{eq:rpaire})$ sets $\beta$ and $\mu$ to 0, while $(\ref{eq:rfrac})$ gives
the following constraints on the remaining parameters:
\begin{equation}
\alpha \gamma = \nu \lambda
\qquad {\mathrm{and}} \qquad
\alpha^2+\gamma^2 = \lambda^2 + \nu^2.
\end{equation}
We can set $\nu=1$ by scaling each variable then we find:
\begin{equation}
R(Z) = {Z^2 - \gamma\over -\gamma Z^2+1}
\quad
{\mathrm{or}}
\quad{-Z^2 - \gamma\over \gamma Z^2+1}.
\end{equation}
The second possibility corresponds to a rotation by $90^\circ$, 
$Z\rightarrow {\mathrm{i}} Z$.
One then replaces this expression in the energy of the
ansatz $(\ref{eq:urat})$ given by
\begin{equation}
E = 4\pi \int \Biggl(
r^2 f'^2 + 2 N (f'^2+1) \sin^2{f} + {\cal I} {\sin^4{f}\over r^2}
\Biggr) {\d}r
\end{equation}
where
\begin{equation}
{\cal I} = {1\over 4\pi} \int \Biggl(
{1 + |Z|^2\over 1+|R|^2} \biggl|{{\d}R\over{\d}Z}\biggr|
\Biggr)^4
{ 2{\mathrm{i}} {\d}Z {\d}\bar Z\over
(1+|Z|^2)^2}.
\end{equation}
As seen earlier, the contribution to the energy coming from the angular part
of the ansatz decouples from the radial part, and ${\cal I}$ can be minimized 
as a function of the parameter $\gamma$. One finds the extremal value ${\cal I}
= \pi + 8/3$ for $\gamma=0$. This simply corresponds to the map 
$R(Z) = Z^2$ which
possesses the cylindrical symmetry
\begin{equation}
R({\mathrm{e}}^{ {\mathrm{i}}\alpha }) =
{\mathrm{e}}^{ 2{\mathrm{i}}\alpha } R(Z),
\end{equation}
which contains the discrete symmetries $I_1$ and $I_2$. This is 
the toroidal
configuration which we will discuss more when we describe the low 
energy manifold
for the sector $B=2$. Replacing the minimum value of ${\cal I}$ in the
expression of the energy and minimizing further relative to the radial
function $f$, one finally obtains a value only $3\%$ greater than 
the mass of the
torus obtained by a fully numerical computations. Rational maps with more 
complicated symmetries generalise the previous discussion and are presented in
reference \cite{Houghton}.

We will close this section with the following comments about the number 
of ``holes'' (regions with zero baryon number density) which 
are present in a given configuration. $\d R/\d z$ 
is zero if the Wronskian is zero
\begin{equation}
W(z) = p'(Z) q(Z) - p(Z) q'(Z)=0\label{eq:defw}
\end{equation}
which is 
the numerator of ${\d} R/\d Z$. The baryon number for the ansatz 
$(\ref{eq:urat})$ is given by
\begin{equation}
N = {1\over 4\pi} \int \Biggl[
\Biggl({1 + |Z|^2\over 1+|R|^2}\Biggr)\, \biggl|{{\d}R\over{\d}Z}\biggr|
\Biggr]^2
{ 2{\mathrm{i}} {\d}Z {\d}\bar Z\over
(1+|Z|^2)^2}.
\end{equation}
The baryon number density is then 
proportional to ${\d} R/\d Z$ and vanishes where the Wronskian is zero. 
This means  
that it will be zero along rays pointing to infinity from the origin, and 
whose directions are given by the roots of $W(Z)$. $R(Z)$ being of degree $N$,
$W(Z)$ is generically of degree $2N - 2$ (the naively leading power of $Z$
cancels in $(\ref{eq:defw})$), and so a configuration of this baryon 
number should have $2N - 2$ holes in it. For the case $N=2$, $q(Z)=1$ so the
Wronskian is in fact linear in $Z$ and only has one zero, which is consistent
with the trivial fact that a torus only has one hole.

%% file: section34.tex
Morse theory\cite{Morse,Milnor,Taubes,Baal}, 
otherwise known as global variational analysis, relates the 
topology of a manifold to the number of and types of critical points of a 
function defined on the manifold. For application in field theory, the 
manifold in question is the (infinite dimensional) space of all field 
configurations and the ``function'' defined on this manifold is generally a 
functional, typically the energy functional or perhaps the action functional. 
The generalization of Morse theory to the infinite dimensional arena goes 
under the name of Ljusternik-Snirelman theory\cite{Ljusternik}.

\begin{figure}
\centering
\mbox{\epsfig{figure= 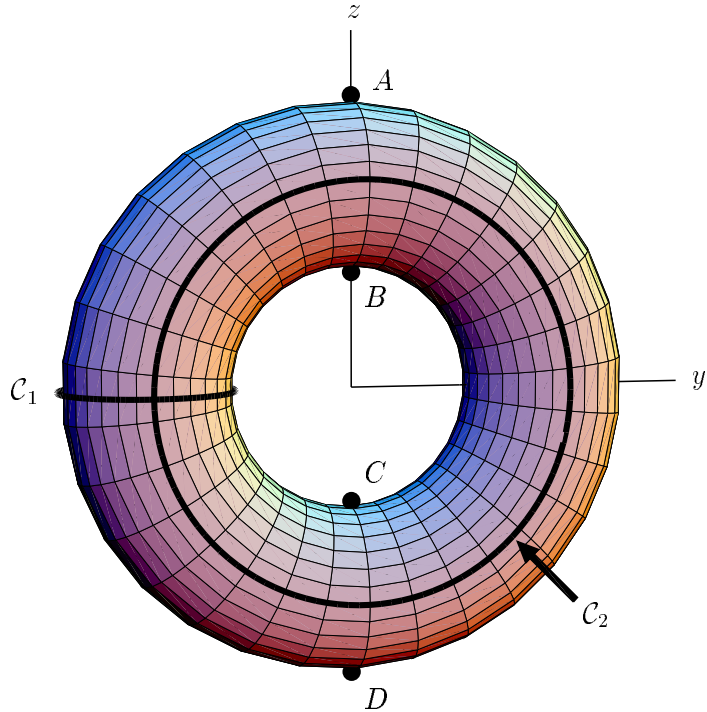,width=4.0truein}}
\caption{The altitude fonction $f(x,y,z)=z$ defined on an ordinary torus 
possesses 4 critical points: the global maximum $A$, the global minimum $D$,
and two minimax's $B$ and $C$. Morse theory identifies the presence of the 
minimaxima with the existence on the torus of two types of non-contractible 
loops ${\cal C}_1$ and ${\cal C}_2$. }
\label{fig:fig14}
\end{figure}

\begin{figure}
\centering
\mbox{\epsfig{figure= 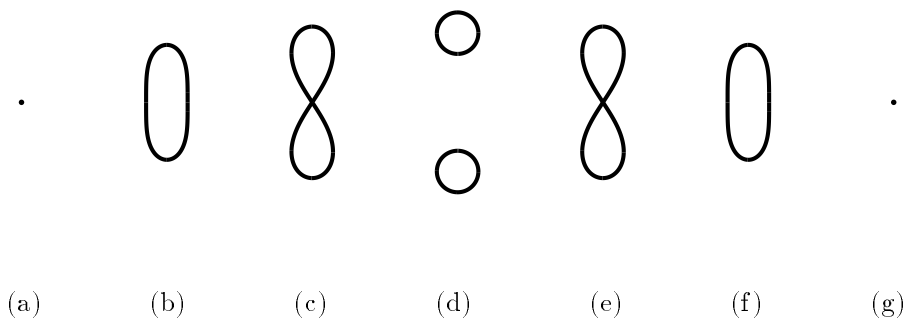,width=5.0truein}}
\caption{Intersection of the torus with planes (a) $z=R+r$, (b) $z=R$, 
(c) $z=R-r$, (d) $z=0$, (e) $z=-R+r$, (f) $z=-R$ and (g) $z=-R-r$.}
\label{fig:fig15}
\end{figure}

The classic, illustrative example of the application of Morse theory is 
furnished by a function defined on a torus. We take the 
outer radius to be $R$ and the inner radius to be $r<R$, with the symmetry 
axis pointing along the $x$-axis. The function in question should be a 
``Morse function'', the definition of which we will address below. We 
will consider the function defined by the value of the $z$ coordinate of the 
Cartesian coordinates of each point of the torus, which happens to be a Morse 
function. Physically, the function in question is the altitude from the $z=0$ 
plane for each point on the torus. Now let us consider the critical points of 
this function on the torus. It is well known that every function defined on a 
compact manifold achieves its global maximum and its global minimum somewhere 
on the manifold. Clearly the function will be critical at these two points, 
hence, in fact the compactness of the manifold, which is is a topological 
characterization, has implied the existence of two critical points. There are 
however, even more. These are predicted by Morse theory, due to the 
non-simple connectedness of the torus. The torus admits two different 
non-contractible closed loops  (see figure $\ref{fig:fig14}$). 
Morse theory implies that 
there are at least two other critical points, which are minimax's. Physically 
we can just see them: the points $B$ and $C$ in figure $\ref{fig:fig14}$. 
If we consider the intersection of the torus 
with successive planes of constant $z$, the two planes for which the 
topological nature of this intersection changes correspond to the positions 
of the minimax's. Successively, as we sweep the plane through the 
torus the intersection (see figure $\ref{fig:fig15}$)
will commence as a point (a very degenerate ``circle'' 
or loop), then a normal loop, but then at one point the loop will pinch in and 
touch itself and then break up into two disjoint loops. 
This is the first minimax or saddle point. Then the two loops will separate a 
little but again come back and touch. When they touch is another saddle 
point, and further they will separate to form a single loop, which will 
finally terminate by degenerating to a point. 

To prove the existence of an additional critical point due to non-trivial 
topology, consider the somewhat more general
example of a smooth function $f(x)$ defined on a compact 
manifold with a non-contractible loop. We wish to prove that there exist at 
least three critical points for the function. It is immediate that the global 
minimum $f(x_{\mathrm{min}})$ and the global maximum $f(x_{\mathrm{max}})$ 
exist. Without loss of 
generality $f(x_{\mathrm{max}}) > f(x_{\mathrm{min}})$, 
since if $f(x_{\mathrm{max}}) = f(x_{\mathrm{min}})$, the function is a 
constant. Also without loss of generality we may assume that the maximum and 
minimum are achieved as unique, individual, isolated points. This means that 
there are already two critical points hence if the minimum or maximum is 
achieved elsewhere there would be at least three critical points and we would 
be done. 

\begin{figure}
\centering
\mbox{\epsfig{figure= 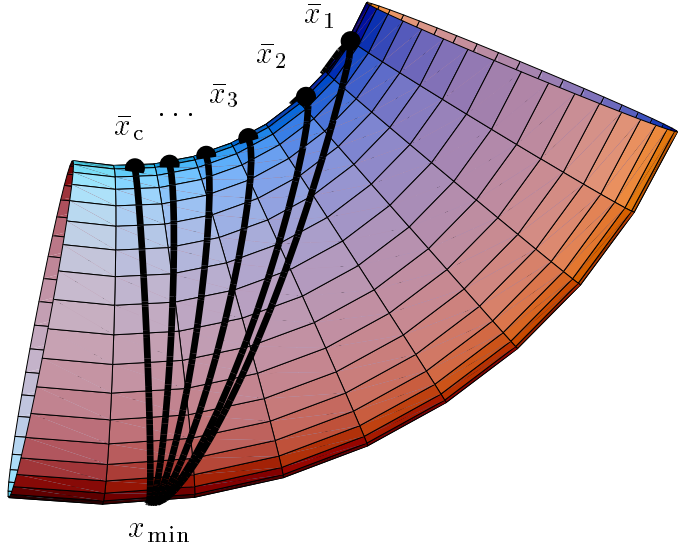,width=4.0truein}}
\caption{Drawing of the successive paths passing through $x_{\mathrm{min}}$
and the local maxima $\bar x_{\mathrm{i}}$ for $i=1,2,3,\ldots$ of the curves
which converge to the real minimax $\bar x_{\mathrm{c}}$.}
\label{fig:fig16}
\end{figure}

Now consider the set of non-contractible loops in the manifold which go 
through the minimum. For each loop find the point at which the function is 
maximum, the set $\{\bar x\}$. If this occurs at several points for any
one loop, choose the 
point $\bar x$ where $|\vec\nabla f|^2$ is minimal (since we are looking for a 
critical point where $|\vec\nabla f|=0$). The maximum $f(\bar x)$ 
is necessarily greater than $f(x_{\mathrm{min}})$ since if it were not
then $f(x)$ would have to be critical at $\bar x$, and then would 
already admit a third critical point at $\bar x$. 
Now we find the minimum of the set $\{f(\bar x)\}$ by varying the loop. 
The point where this occurs must exist and 
will correspond to a critical point of the function, a saddle point. It must 
exist because a bounded, monotone, 
sequence in a compact manifold always admits a limit 
point. We consider the sequence of points obtained by finding 
$\bar x$ for a given curve and then varying successively the 
curve to yield $\bar x_1$, $\bar x_2$, etc. such that $f(\bar x_{\mathrm{i}}) 
\ge \bar f(x_{\mathrm{i+1}})$ (see figure $\ref{fig:fig16}$).
Let $\bar x_{\mathrm{c}}$ be defined by $\bar x_{\mathrm{c}}\in \{\bar x\}$
such that $f(\bar x_{\mathrm{c}}) ={\mathrm{min}} \bigl(f(\bar x)\bigr)$
and we call the curve passing through $\bar x_{\mathrm{c}}$ the critical 
curve, 
$\vec x_{\mathrm{c}}(t)$.
$\vec\nabla f(\bar x_{\mathrm{c}})$ must vanish. If 
$|\vec\nabla f(\bar x_{\mathrm{c}})| >\delta$ ($\delta>0$)
then consider the curve 
obtained by deforming the points of the critical curve 
in a neighborhood of $\bar x_{\mathrm{c}}$ along the direction 
opposite to the gradient:
\begin{equation}
\vec x_{\mathrm{c}}(t) \rightarrow \vec x_{\mathrm{d}}(t) =
\vec x_{\mathrm{c}}(t) - 
\epsilon(t) \;\vec\nabla f(\vec x_{\mathrm{c}}(t)),
\end{equation}
where $\epsilon(t)$ is non-zero only around $\bar x_{\mathrm{c}}$.
For $\epsilon(t)$ small enough, the deformed curve is 
still a non-contractible curve, however the maximum of the function along 
$\vec x_{\mathrm{d}}(t)$ 
is clearly less than the maximum at $\bar x_{\mathrm{c}}$
because the gradient $-\vec\nabla f$ points along the direction of decreasing 
$f$. This 
is a contradiction since $\bar x_{\mathrm{c}}$ is the minimum of 
$\{ \bar x \}$. Hence $\vec \nabla f(\bar x_{\mathrm{c}})=0$. 

Morse theory goes on to give a set of inequalities relating the number of 
critical points to the changes in topology of the manifold. We refer the 
reader to the literature
\cite{Nash,Mantonsph,Milnor,Taubes,Ljusternik,Klinkhamer} 
for more details. We note that nothing depended 
critically on the fact that we were considering a non-contractible loop, it 
could well have been any non-contractible compact manifold, for example a 
sphere $S^2$. Hence non-trivial homotopy groups $\Pi_{\mathrm{m}}$ can imply 
the existence of non-minimal critical points. The crucial ingredient for the 
success of the minimax procedure was that the minimal critical point was 
non-degenerate. In fact, a function which only admits non-degenerate critical 
points is called a Morse function. To apply Morse 
theory however we do not truly 
require Morse functions. The function should simply be non-degenerate along 
every non-contractible loop. 

The application of Morse theory to infinite dimensional manifolds was 
analyzed by Ljusternik-Snirelman\cite{Ljusternik}. 
A very readable account of the use of this 
theory was done by Taubes for the case of magnetic 
monopole\cite{TaubesCargese}. The idea is to 
first establish the existence of non-contractible loops in the configuration 
space $M_0$, the space with net monopole number zero, \ie $\Pi_1(M_0)\ne 0$. 
Secondly consider the configurations of an infinitely separated 
monopole-anti-monopole pair. This configuration is in $M_0$, and the 
non-contractible loop corresponds to rotating the monopole relative to the 
anti-monopole in iso-space by one complete revolution. The minimal critical 
point corresponds to the situation where the monopole and anti-monopole have 
annihilated and all radiation has dissipated off to infinity leaving the 
quiescent (symmetry broken) vacuum behind. This critical point of zero energy 
is non-degenerate along every non-contractible loop. Taubes searched for the 
existence of a non-minimal critical point in the configurations bounded above 
in energy by the asymptotic critical point of the infinitely separated 
monopole-anti-monopole pair and bounded below by the energy (zero) of the 
vacuum. First it was shown that this infinite dimensional subset of the 
configuration space is in fact compact. With this information it is 
sufficient to exhibit a single non-contractible loop, with energy everywhere 
less than the energy of the infinitely separated pair, to be able to conclude 
that the minimax procedure will converge. It will of course converge to 
a different critical point than the trivial vacuum, since the vacuum is 
non-degenerate. The existence of this loop was shown\cite{Taubes} 
proving the existence of a non-minimal critical point.

Bagger \etal\cite{Bagger} attempted to mimic this procedure for the 
Skyrme model. The configuration space for the Skyrme model corresponds to 
maps from $\Rset^3$ into the group $\mathrm{SU(2)}$:
\begin{equation}
{\cal C}:\{ U(\vec x): \Rset^3 + \infty \rightarrow {\mathrm{SU(2)}}\}
\end{equation}
$\Rset^3 + \infty \equiv S^3$, hence $\cal C$ separates into disjoint sectors 
labeled by the homotopy classes $\Pi_3(\mathrm{SU(2)})=\Zset$, the integer 
being of course the baryon number. Hence
\begin{equation}
{\cal C} = {\cal C}_0 + {\cal C}_1 + {\cal C}_2 + \cdots
+ {\cal C}_{-1} + {\cal C}_{-2} + \cdots.
\end{equation}
The existence of non-contractible loops in $\cal C$
\begin{equation}
\Pi_1({\cal C}) \ne 0
\end{equation}
implies that
\begin{equation}
\Pi_1({\cal C}_{\mathrm{n}}) \ne 0,
\end{equation}
and that they are actually all equal. Each sector actually contains each 
other sector, simply by constructing the requisite number of 
baryon-anti-baryon pairs, and moving, say, the anti-baryons as far away as is 
required. 

$\Pi_1({\cal C}_0)$ can be seen to be exactly the same as 
$\Pi_4({\mathrm{SU(2)}})$. We can see this through the following construction.
$\Rset^3+\infty$ can be thought of as the end cap 
of a 4-cube. The parameter along the fourth dimension corresponds to the
\begin{figure}
\centering
\mbox{\epsfig{figure= 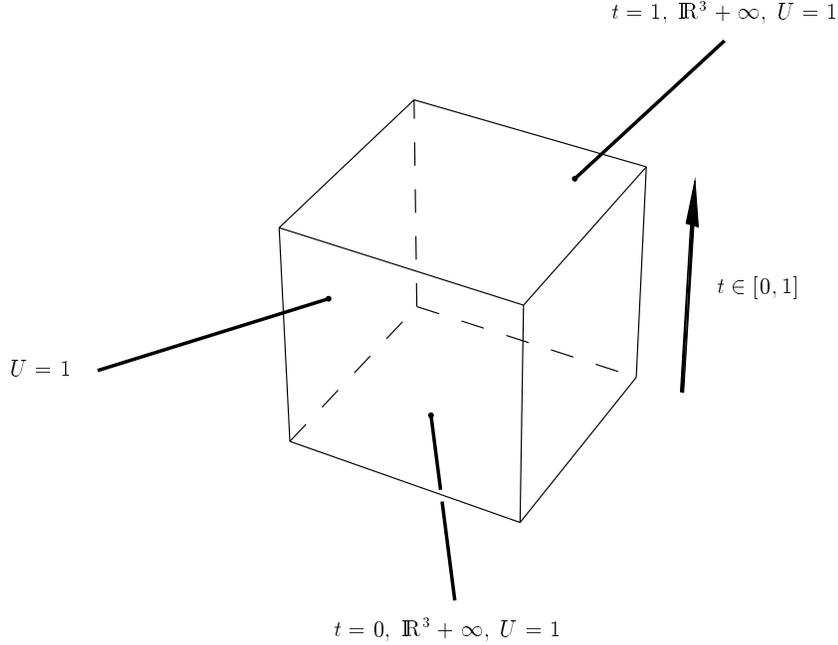,width=4.5in}}
\caption{Values of $U$ defined on the hypercube $\Rset^4$ of space.}
\label{fig:fig17}
\end{figure} 
parameter along the loop (see figure \ref{fig:fig17}).
If we start at $t=0$ with $U=1$, impose that $U=1$ along the vertical faces
(because these faces also correspond to the point at infinity of $\Rset^3$), 
and $U=1$ at $t=1$, we define a loop in ${\cal C}_0$. However a 4-cube with 
the surfaces identified is the same as $S^4$, $\Rset^4 + \infty \equiv S^4$, 
hence we equally well define a mapping 
\begin{equation}
U(\vec x,t): S^4 \rightarrow \mathrm{SU(2)},
\end{equation}
\ie an element of $\Pi_4(\mathrm{SU(2)})$. It is known that\cite{Steenrod}
\begin{equation}
\Pi_4(\mathrm{SU(2)}) \equiv \Zset_2.
\end{equation}
Hence there exist non-contractible loops in the space of configurations of 
the Skyrme model. These loops can most physically be realized as corresponding 
to a Skyrmion-anti-Skyrmion pair, as in the monopole situation, the two 
solitons are rotated relative to one another through one complete revolution. 
The difference between the Skyrme model and the monopole situation is that 
while loops involving further complete relative revolutions are distinct
non-contractible loops for monopoles, for Skyrmions
all loops with an odd number of 
complete revolutions are equivalent to each other and non-contractible while 
an even number of complete revolutions yields a contractible loop. 
Bagger \etal\cite{Bagger} 
found, at least to first order in perturbation theory 
for large separation of the two solitons, that there were no 
non-contractible 
loops where the energy was everywhere less than the energy of the asymptotic 
critical point of the infinitely separated Skyrmion-anti-Skyrmion pair. 
Augmenting the Skyrme model with an electromagnetic interaction, however, 
gave a sufficiently attractive Skyrmion-anti-Skyrmion potential for them to 
conclude the existence of a non-minimal critical point in this somewhat 
modified theory. 

The problem was analyzed in greater detail by Isler \etal\cite{Isler}, for the 
$B=0$ and $B=2$ situation together. Let 
$U_{\mathrm{S}}(\vec x)$ be the field of a Skyrmion. Then for well 
separated Skyrmions or Skyrmion-anti-Skyrmion pair, the product ansatz 
suffices. Let
\begin{eqnarray}
U_{\mathrm{B=2}}(\vec x) &=& R(t) U_{\mathrm{S}}(\vec x - \vec x_1) 
R^\dagger(t)\; U_{\mathrm{S}}(\vec x - \vec x_2),
\label{eq:ub2}
\\
U_{\mathrm{B=0}}(\vec x) &=& R(t) U_{\mathrm{S}}(\vec x - \vec x_1) 
R^\dagger(t)\; U^\dagger_{\mathrm{S}}(\vec x - \vec x_2)
\label{eq:ub0}
\end{eqnarray}
where $R(t)$ is an ${\mathrm{SU(2)}}$ matrix that introduces a relative 
iso-rotation. 
If $R(t)$ varies from any point in $\mathrm{SU(2)}$ to its antipodal point, 
\ie
\begin{equation}
R(t)\biggl|_{t=1} = -R(t)\biggl|_{t=0}\qquad\qquad\qquad t \in [0,1]
\end{equation}
then $t$ parametrizes a non-contractible loop in ${\cal C}_0$ or ${\cal C}_2$.
This is the usual topology which demonstrates $\mathrm{SU(2)}$ as the simply 
connected (double) cover of $\mathrm{SO(3)}$. It is a reasonably 
straightforward computation to find the energy to lowest order in the 
separation $d = ||\vec x_1 - \vec x_2||$,
\begin{eqnarray}
E_{\mathrm{B=0}} &=& 2\; E_{\mathrm{S}}  - 
4 \pi f_\pi^2 \kappa^2 {(1-\cos\theta) (3 (\hat n\cdot\hat d)^2 - 1)\over d^3}
\\
E_{\mathrm{B=2}} &=& 2\; E_{\mathrm{S}}  + 
4 \pi f_\pi^2 \kappa^2 {(1-\cos\theta) (3 (\hat n\cdot\hat d)^2 - 1)\over d^3}
\end{eqnarray}
for
\begin{equation}
R = e^{\mathrm{i} \theta \hat n\cdot \vec\tau/2},
\end{equation} 
$E_{\mathrm{S}}$ is the energy of a Skyrmion
and $\kappa$ 
is the coefficient of the $1/r^2$ fall off in the Skyrmion profile 
function $f(r)$, $f(r)\sim \kappa/r^2 + O(1/r^6)$ (see equation 
$(\ref{eq:fsk})$). The potential $V$ serves to 
separate the (reduced) configuration space consisting of relative 
iso-rotations 
into two disjoint parts. Indeed this reduced space of configurations is 
isomorphic to a 3-ball of radius $2\pi$ modulo one identification. 
$\theta$ plays the role of the 
radius, $\hat n$ the unit vector giving the direction. Furthermore antipodal 
points are identified, $(\theta,\hat n)\equiv (2\pi -\theta,-\hat n)$. This 
identification is particularly reductive for $\theta=2\pi$, the whole sphere 
at $\theta=2\pi$ corresponds to $R=-1$ which is identified with the origin 
where $R=1$. The factor $1-\cos\theta$ is positive semi-definite, and equal to 
zero only at $\theta=0$ or $2\pi$. The function $3(\hat n\cdot\hat d)^2 - 1$ 
varies from 2 when $\hat n\cdot\hat d = 1$ to $-1$ when $\hat n\cdot \hat 
d = 0$. Hence when the direction of the relative rotation is chosen such that 
$\hat n\cdot\hat d=\pm 1/\sqrt{3}$, this defines a double cone that passes 
through the 
origin, demarcating the boundary between 
\begin{figure}
\centering
\mbox{\epsfig{figure= 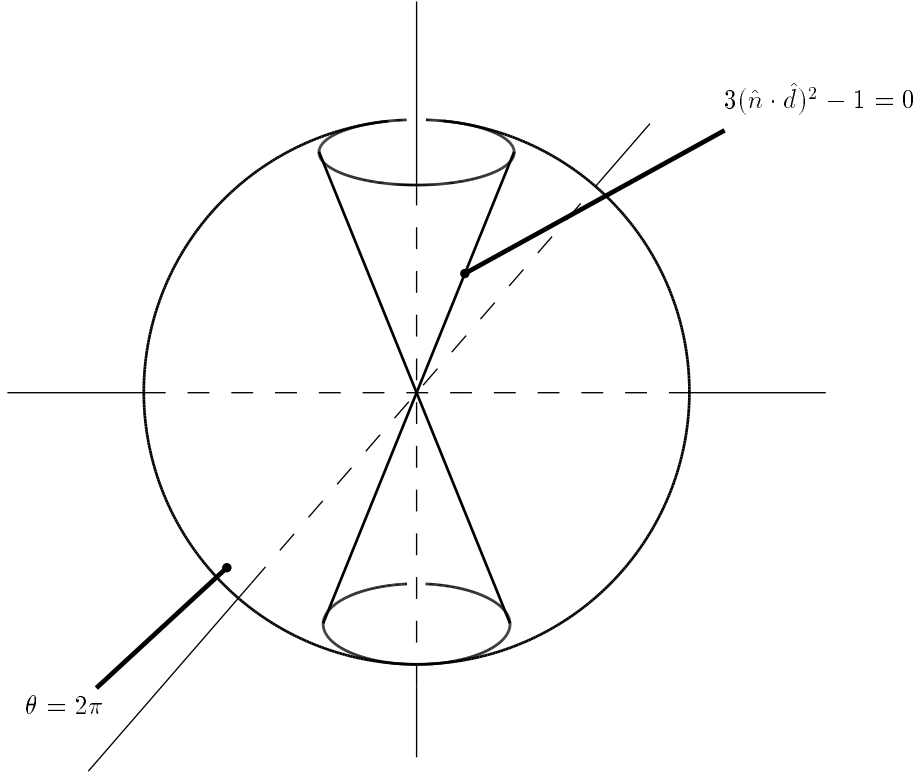,width=5.0in}}
\caption{Diagram of the attractive region for the Skyrmion-Skyrmion 
$B=2$ potential.}
\label{fig:fig18}
\end{figure}
regions of relative attraction and relative repulsion (see 
figure \ref{fig:fig18}). The 
regions of attraction have less energy than $2\;E_{\mathrm{S}}$, 
while the regions of repulsion have more energy than $2\; 
E_{\mathrm{S}}$. 

If the attractive region is in the ``time-like'' region of these cones,
which is the case for the case $B=0$, there 
is no curve which can pass from a point within this region to its antipode 
without either crossing over to the region of repulsion or touching the cone 
at its vertex. This is evident since a point and its antipode find themselves 
in opposite sides of the ``forward'' or ``backward'' light cone. All paths 
linking them must pass through the vertex or enter into the repulsive region 
which is not desired since the energy here is greater than that of two 
infinitely separated Skyrmions (or Skyrmion-anti-Skyrmion pair). 
Hence we cannot conclude the existence of a non-minimal critical point, we 
need to find a non-contractible curve where the energy is everywhere less than 
$2\;E_{\mathrm{S}}$. Even though our curves are never greater 
than $2\;E_{\mathrm{S}}$ (when they pass through the vertex)
we cannot be assured that the minimax 
procedure will just converge to the asymptotic critical point
(of energy $2\, \epsilon_{\mathrm{Skyrmion}}$). The situation 
just described is that which applies to the case $B=0$, \ie a 
Skyrmion-anti-Skyrmion  pair. Here, there is no question that the minimum 
energy configuration is non-degenerate along every non-contractible curve. 
Hence it remains an open question whether indeed there are non-minimal 
critical points in the sector $B=0$. 

Considering the case $B=2$ (see figure $\ref{fig:fig18}$), 
the attractive and repulsive regions exchange 
with respect to $B=0$. Hence the attractive region is the ``space-like'' 
region relative to the cone. It is evident that now there exist 
non-contractible loops, which remain everywhere in the ``space-like'' region. 
They simply skirt around the cone to the other side of the origin where lies 
the antipode. Hence we show the existence of non-contractible loops which are 
everywhere lower in energy than $2\;E_{\mathrm{S}}$. 
However this is still not enough to conclude the existence of non-minimal 
critical points. Indeed in the sector $B=2$ 
the energy functional fails to be a Morse function to a sufficient extent. 
The minimal 
critical point has been demonstrated to be a toroidal configuration with 
axial symmetry, in all but a rigorous, analytical mathematical proof. The 
axis of symmetry has no direction, the toroidal configuration rotated by 
180$^\circ$, about an axis orthogonal to the symmetry axis, 
is identical to the 
starting configuration. This implies that the minimal energy configuration is 
degenerate along a non-contractible loop. Hence it is the strongest 
possibility that the minimax procedure will converge simply to the minimum 
energy toroid.

Even though our exercise with Morse theory has led to no new solutions it is a 
worthwhile analysis allowing us to understand the model in a more profound 
way. There are several open questions raised by the analysis, two evident 
ones are do there exist sphalerons in the $B=0$ sector and do the existence 
of non-trivial higher homotopy groups imply existence of sphalerons. 
Indeed Morse theory has been used in relation with rational maps
in a more recent article\cite{Houghton}.

%% file: section41.tex
The dynamics of solitons is an extremely interesting and complicated problem. 
There are many different modes of excitation for a single soliton itself. 
There are certain modes, the zero modes of the classical small oscillation 
problem about the soliton, which properly belong to the soliton itself. They 
are usually treated semi-classically via Bohr-Sommerfeld type quantization 
rules. There are also regular vibrational and resonant modes which correspond 
to excited states of the soliton. Additionally there are modes which 
correspond to the scattering of (non-solitonic) waves off the soliton itself,
for example, pions scattering from nucleons. 

Next one can consider the interaction of two solitons with each
other. Here one can support many forms of reactions: scattering,
deformation, bound states, annihilation among others. Soliton-anti-soliton 
annihilation is particularly problematic because of the coexistence of both 
perturbative and non-perturbative regimes. A physical example is evidently
nucleon-anti-nucleon annihilation. 
The potential between a nucleon and an anti-nucleon has been obtained 
by Lu and Amado\cite{VNNbar-AP} using the product ansatz for 
large distances, and by Lu \etal\cite{VNNbar-Num} for distances
larger than 0.8 fm using numerical methods. Physically there emerges a 
critical distance $d_0$ which is of the order of 0.8--1.2 fm, 
outside of which the interaction between the particles is essentially 
repulsive. On the other hand if they attain this critical distance, 
they will quickly 
combine together into a lump of mesonic matter of baryon number zero. 
Numerical
simulations of the classical system are essential to understand 
this complicated process\cite{VWWW-NNbar}.
It was shown that the reaction for transforming the pair of
particles into a single lump of zero baryon number happens at the limit of 
causality and that the energy
left by the disintegration remains localized for a relatively long time (until
pion radiation waves disperses it into the 
vacuum)\cite{NNbar-Sim}.
The remaining part of the process, the emission of pion waves from the
lump of mesonic matter, has been studied in the most detail, 
using path integral
methods\cite{NNbar-IC}, or coherent state 
methods\cite{Amado-et-cie}. 
The results reproduce well the experimental phase shifts. 
We will not treat soliton-anti-soliton annihilation any further.

Even more complicated situations arise as we  
increase the soliton number. We can go to the point where one has an
infinite number of solitons and all their various phases, fluid or
solid, with possible crystalline structures. In this review we shall only 
discuss the
low energy interaction of solitons, specifically Skyrmions, corresponding 
to the low energy scattering of nucleons.

Scattering of nucleons within the true microscopic theory, the
standard model, is impossible to treat satisfactorily, essentially
because of our inability to compute anything in the low energy
domain. Even for very high energy scattering the final processes
leading to hadronization are not computable from the microscopic
theory. For low energy processes we have effective field theories,
such as the Skyrme model, which afford more tractable descriptions of
the physics involved. But even here, the baryons are represented as
solitons and an exact quantum description of soliton states is still
lacking. The only perturbative expansion feasible seems to be the
semi-classical approximation.

The semi-classical approximation serves well to describe constituent
properties of individual solitons. Essentially, the procedure is to
identify the low energy, collective modes of the soliton and to
quantize them. The interactions between solitons can only be treated
perturbatively at large separations, by computing the effective
interaction potential between them.

In the past ten years there have been substantial advances in treating the
part of the interaction at short distances and how to describe the scattering 
within the semi-classical approximation. Even the semi-classical 
approximation is not exactly solvable: soliton-soliton scattering involves 
an infinite number of degrees of freedom except for some very special cases 
of integrable models. The problem could be tractable if there were some way 
of truncating to a finite number of degrees of freedom. Exactly such a 
truncation was suggested by Manton\cite{Mantongfc}.

In very general terms, one is interested in finding the low energy 
degrees of freedom of the two-soliton system. Typically one finds only a 
finite number of relevant degrees of freedom. The low-energy motion can then
restrict itself self-consistently to these degrees of freedom.

The canonical example of such a situation was provided by the case of magnetic
monopoles in the so-called BPS limit\cite{Gibbons,Atiyah-Hitchin}. 
Here the inter-monopole force vanishes
exactly, the magnetic Coulomb repulsion being exactly cancelled by an 
attractive force due to the existence of a massless scalar exchange. Hence 
there exist static solutions with two monopoles situated at arbitrary 
relative orientations and positions. This set of configurations corresponds 
to a sub-manifold of the set of all configurations and is called the moduli 
space. Indeed the characterization in terms of positions and orientations 
makes sense for monopoles when they are well separated, but as they come 
close together they lose their identity. What is preserved is the 
dimension of the space of moduli. For large separation, the moduli describe 
the position and orientation of each monopole: there are three degrees of 
freedom for the position of each monopole and one internal phase (related to
the residual $\mathrm{U}(1)$ gauge symmetry) giving  
a total four degrees of freedom per monopole. 
As the two monopoles approach one another, only 
the dimension of the relative moduli space remains, the monopoles deform 
completely and fuse into a single entity. Clearly, since there are no forces, 
the moduli describe an equipotential surface. This surface is also the set of 
minimal energy configurations in the two monopole sector.

\begin{figure}
\centering
\mbox{\epsfig{figure=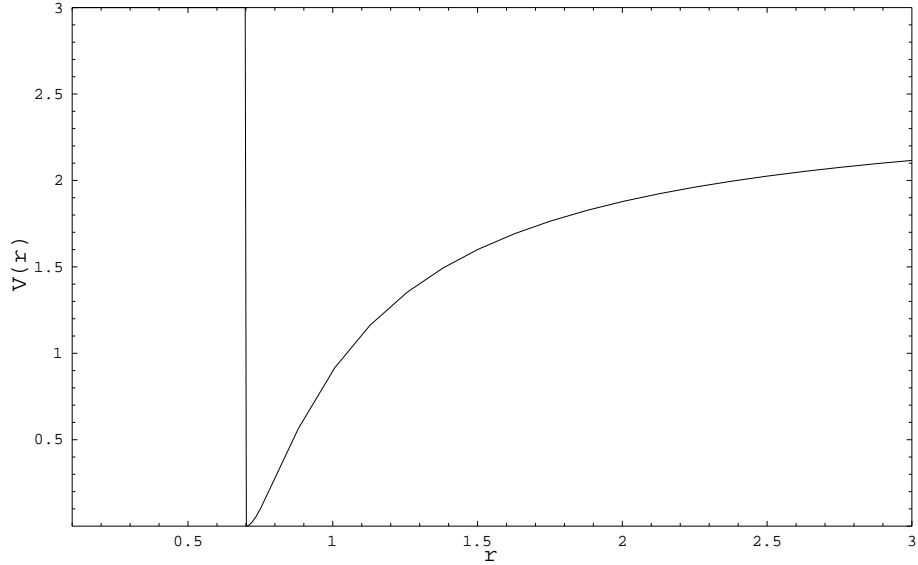,height=3.0in}}
\caption{Potential $V(r)$ on a hard, frictionless sphere as a function of the 
radius $r$. $R_{\mathrm{e}}\simeq 0.7$ is the radius of the sphere.}
\label{fig:fig19}
\end{figure}

We can make an intuitive analogy with the surface of the earth. If this 
surface were perfectly spherical and frictionless, the potential as a 
function of radius would be (effectively) 
infinite at the radius of the earth and the 
equal to the usual gravitational potential for larger radii (figure 
$\ref{fig:fig19}$). 
The radial motion, although not simple harmonic motion, 
corresponds to highly energetic modes. If we start at some point on this 
idealized earth, with an initial velocity that is tangent to the surface
and arbitrarily small in amplitude, it is intuitively clear that the motion 
will remain very close to the surface of the earth. It is easy to prove in 
this case that the motion will follow geodesic curves on the surface of the 
earth (we will clarify somewhat the notion of geodesic later on). 

The problem of the monopoles is analogous. If the initial conditions 
correspond to being on a point of the moduli space of two monopole solutions, 
and the initial velocity is tangent to this surface and arbitrarily small, 
it was proposed\cite{Mantongfc,Gibbons,Atiyah-Hitchin} 
that the subsequent movement follows 
the appropriate 
geodesic in the manifold of the moduli space. This produces a concrete 
example of the truncation of the infinite number of original degrees
of freedom to a finite number of relevant, effective degrees of freedom.

The general situation with solitons cannot be adequately described in terms of
a moduli 
space as in the case of BPS monopoles. For example, Skyrmions or non-BPS 
limit (non-zero Higgs mass) monopoles experience inter-solitonic forces; 
hence no static solutions exist corresponding to arbitrary relative position 
of the solitons. Typically there do exist several low-lying 
critical points of the energy functional which should be involved and are 
important in the low-energy dynamics. In models where the solitons are not 
confined, one such critical point corresponds to infinitely-separated 
solitons. In addition, if configurations of energy lower than twice the energy
of one soliton exist, then the minimum energy configuration represents
another critical point corresponding to a bound state of two solitons.
Furthermore there could be  
other metastable solutions such as the dibaryon of the Skyrme 
model\cite{Jackson}, or the sphaleron solutions of the Weinberg-Salam 
model\cite{Mantonsph,Klinkhamer}, among others. The low-energy
dynamics will restrict itself to these critical points and certain
paths linking them together, as we will see in the next subsection.

\subsection{General formalism}
Manton\cite{Mantongfc} suggested a possible truncation of the degrees of 
freedom to describe 
the low energy dynamics of the soliton in these more general situations. He 
suggested that the dynamics would truncate self-consistently to the union of 
all the low energy critical points and a set of curves which pass between 
these various critical points. These curves are alternatively taken to be the 
paths of steepest descent or the gradient flow curves linking the critical 
points together. The gradient flow
method or steepest descent method gives rise to only slightly different sets of
configurations if the critical points are truly low lying and the gradients 
are small, as we will now discuss. 
There is also another formalism called the valley method 
which also serves to give the paths which connect the critical points, 
however, we shall not discuss this method here\cite{Valley}.

{\em Gradient flow curves\/} 
are mathematically described as the integral curves of the
vector field corresponding to the gradient. This gives a first order 
differential equation:
\begin{equation}
g_{\mathrm{ij}}\;{{\d} x^{\mathrm{j}}(\lambda)\over {\d} \lambda} 
= - {\partial V(x^{\mathrm{i}}(\lambda))\over \partial x^{\mathrm{i}}}
\label{eq:gfc}
\end{equation}
where $x^{\mathrm{i}}(\lambda)$ are the coordinates of a point along the 
curve. $g_{\mathrm{ij}}$ is
the metric on the space of all configurations and $V(x^{\mathrm{i}})$ is 
the potential
defined on it. The initial directions taken corresponding to the unstable
directions (negative modes) are
extracted from the matrix of second derivatives of
the potential at the position of the critical points. 

The {\em steepest descent curves\/}, 
however, are defined in Manton\cite{Mantongfc} 
by the equation
\begin{equation}
g_{\mathrm{ij}}  {{\d}^2 x^{\mathrm{j}}\over {\d} \lambda^2} = 
- {\partial V\over \partial x^{\mathrm{i}}}
\label{eq:psd}
\end{equation}
augmented with the boundary condition that 
\begin{equation}
\lim_{\lambda\to-\infty} x^{\mathrm{i}}(\lambda)  = x^{\mathrm{i}}_0
\end{equation}
where $x^{\mathrm{i}}_0$ are the coordinates of the unstable critical point. 
These 
correspond, actually, to the approximate dynamical trajectories followed by 
the system starting at $t=-\infty$ at the critical point and moving in the 
unstable directions while neglecting terms that are quadratic in the 
velocities. $t$ corresponds exactly to $\lambda$. 
The {\em exact dynamics\/} would follow from the equation
\begin{equation}
g_{\mathrm{mj}}\; \ddot x^{\mathrm{j}} + 
\biggl( {\partial g_{\mathrm{ml}}\over\partial x^{\mathrm{k}}}
-{1\over 2} {\partial g_{\mathrm{kl}}\over\partial x^{\mathrm{m}}} \biggr) 
\dot x^{\mathrm{k}} \dot x^{\mathrm{l}} = - 
{\partial V\over \partial x^{\mathrm{m}}}\label{eq:eqmvt} 
\end{equation}
which is written in differential geometry as the geodesic equation
\begin{equation}
g_{\mathrm{mj}} \biggl 
( \ddot x^{\mathrm{j}} + \Gamma^{\mathrm{j}}_{\mathrm{kl}}\; \dot 
x^{\mathrm{k}} \dot x^{\mathrm{l}} \biggr) = -
{\partial V\over\partial x^{\mathrm{m}}}
\end{equation}
where the Levi-Civita connection is
\begin{equation}
\Gamma_{\mathrm{kl}}^{\mathrm{j}} = {1\over 2}\, g^{jm} \biggl(
{\partial g_{\mathrm{ml}}\over\partial x^{\mathrm{k}}} +
{\partial g_{\mathrm{mk}}\over\partial x^{\mathrm{l}}} -
{\partial g_{\mathrm{kl}}\over\partial x^{\mathrm{m}}}\biggr).
\end{equation}
It is evident that the approximation is valid when 
$\dot x^{\mathrm{i}}$ are very small.

If we consider the same initial conditions with the dynamics augmented by a
damping term representing friction, the steepest descent curves will 
naturally go over to the gradient flow curves
\begin{equation}
g_{\mathrm{ij}} {{\d}^2 x^{\mathrm{i}}\over {\d}\lambda^2} + 
g_{\mathrm{ij}} \Gamma^{\mathrm{i}}_{\mathrm{kl}} 
{{\d} x^{\mathrm{k}}\over {\d} \lambda}
{{\d} x^{\mathrm{l}}\over {\d}\lambda} = - {\partial V\over\partial 
x^{\mathrm{i}}} - 
b\, g_{\mathrm{ij}} {{\d} x^{\mathrm{i}}\over {\d}\lambda}\label{eq:damping}
\end{equation}
($b\gg 0$ measures the amount of damping in the system).
Reparametrizing the curves with $\lambda\rightarrow b\lambda$ and taking
$b\rightarrow +\infty$ we get
\begin{equation}
0 = {\partial V\over \partial x^{\mathrm{i}}} + 
g_{\mathrm{ij}} {{\d} x^{\mathrm{j}}\over {\d}\lambda}\label{eq:damped}
\end{equation}
which is just the gradient flow equation.

We believe that all three equations yield essentially the same manifold to 
which the dynamics should truncate for a large variety of dynamics. What is 
required is that variations between the trajectories which determine the 
sub-manifolds go to zero. 
This will occur if the potential in the ``transverse" directions is very 
steep. All three trajectories will be pushed together because of 
energy considerations: the trajectories cannot vary too far from the gradient 
flow trajectories because it costs too much in energy. In 4.2.3 we will
discuss a specific example wherein the three methods will be compared.

We will show with several examples how the truncation of the dynamics comes
about and also some examples of how it can fail.

\subsection{Examples of truncation for a few systems}

\subsubsection{Particle subject to a potential with $S^2$ symmetry} 
First we treat the simplest, non-trivial example conceivable, which we
have already mentioned: a particle on a spherical surface. 
Consider a Lagrangian in $\Rset^3$ with Cartesian coordinates 
$x^{\mathrm{i}}$ and Pythagorian metric $\delta_{\mathrm{ij}}$:
\begin{eqnarray}
L & = & {1\over 2} \delta_{\mathrm{ij}} \dot x^{\mathrm{i}} \dot 
x^{\mathrm{j}} - \lambda (r^2-a^2)^2
\nonumber
\\
& = & {1\over 2} ( \dot r^2 + r^2 \dot\theta^2 + r^2 \sin^2\theta \dot\phi^2) -
\lambda (r^2-a^2)^2\label{eq:lsphrad}
\end{eqnarray}
where $r^2 = \delta_{\mathrm{ij}} x^{\mathrm{i}} x^{\mathrm{j}}$, 
$\theta$ and $\phi$ are 
the usual spherical angular coordinates, $\lambda$ 
and $a^2$ are real and positive constants. For $\lambda =0$ the space of 
static solutions is just $\Rset^3$ and the general motion is along straight 
lines. For any non-zero $\lambda$ the set of static solutions is the set of 
sub-manifolds of $\Rset^3$ corresponding to critical points of the potential. 
For this potential the minimum corresponds to the manifold of the sphere of 
radius $a$. In addition there is an unstable critical point at the origin. 
The typical motion for any value of $\lambda$ depends on the total energy. 
The motion is of course bounded since the potential energy rises without 
bound for large $r$. The effective potential for the radial motion is
\begin{eqnarray}
U(r) & = & V(r) + {l^2 \over 2 r^2}
\nonumber
\\
& = & \lambda (r^2-a^2)^2 + {l^2 \over 2 r^2}
\end{eqnarray}
where $l$ is the (conserved) angular momentum. This
potential has only one minimum, pushed further out from $r=a$ (its position for
$l=0$) because of the angular momentum barrier. If the total energy of the
system is fixed while $\lambda\rightarrow+\infty$ then 
the motion is energetically 
bounded to stay within a region that is arbitrarily close to the minimum of 
$V(r)$. This minimum corresponds to the surface $r=a$. 

\begin{figure}
\centering
\mbox{\epsfig{figure= 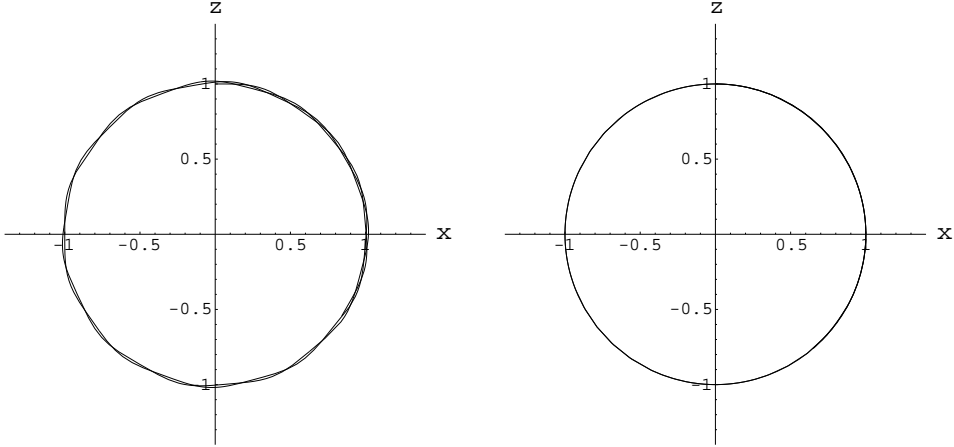,height=2.5in}}
\caption{Comparison of the ``exact'' (left) and approximate solution 
given by the path of steepest descent (right) for a particle subject to 
potential $(\ref{eq:lsphrad})$. They are both very close to the 2-sphere 
$S^2$ ($a=1$, $\lambda=3$, $r_0=a$, $\dot r_0=0$ and $\dot\theta_0=0.5$ here).}
\label{fig:fig20}
\end{figure}

Indeed, the Lagrangian $(\ref{eq:lsphrad})$ after the substitution 
$r = a + \epsilon/\sqrt{\lambda}$ where $\epsilon = \epsilon(t)$ and
$a\gg\epsilon/\sqrt{\lambda}$, becomes
\begin{eqnarray}
L &=& {1\over 2} {\dot\epsilon^2\over \lambda} -
\lambda ( {2 a \epsilon\over \sqrt{\lambda}} + 
{\epsilon^2\over \lambda} )^2 +
{1\over 2} (a + {\epsilon\over\sqrt{\lambda}})^2 (\dot\theta^2
+\sin^2\theta\dot\phi^2)\nonumber
\\
&=& {1\over 2} 
(a^2 + {\epsilon^2\over\lambda} +{2 a \epsilon\over \sqrt{\lambda}})
(\dot\theta^2 + \sin^2\theta\dot\phi^2)
+ {1\over 2} {\dot\epsilon^2\over \lambda} 
- \lambda ( {4 a^2 \epsilon^2\over\lambda} + 
{\epsilon^4\over\lambda^2}
+ {4 a \epsilon^3\over \lambda \sqrt{\lambda}})\nonumber
\\
&\simeq& {a^2\over 2} (\dot\theta^2 + \sin^2\theta\dot\phi^2) +
{1\over 2} {\dot\epsilon^2\over \lambda} - 4\, a^2\,\epsilon^2
\end{eqnarray}
after applying the condition $a\gg\epsilon/\sqrt{\lambda}$.
We see that the radial motion and the angular motion completely decouple as
$\lambda\rightarrow+\infty$. To leading order the energy in the radial 
oscillations is
\begin{equation}
E_{\mathrm{r}} = {1\over 2}\, 8 a^2\, \epsilon_0^2
\end{equation}
where $\epsilon_0$ is the amplitude of the solution $\epsilon(t) = 
\epsilon_0\,\sin{ (2\sqrt{2} a \lambda\,t)}$ to the equation of motion of
$\epsilon$, and which should be taken
to be independent of $\lambda$. $r$ then oscillates between 
$a-\epsilon_0/\sqrt{\lambda}$ and $a+\epsilon_0/\sqrt{\lambda}$, while the
energy in the angular motion is
\begin{equation}
E_{\theta,\phi} = {1\over 2} a^2 ( \dot\theta^2 + \sin^2\theta\dot\phi^2).
\end{equation}
The movement of the system with arbitrarily small initial velocity tangent to 
the sphere will be governed exclusively by the kinetic term of the 
Lagrangian. The dynamical constraint of being forced to live on the sphere of 
radius $a$ will be expressed by the appearance of the metric induced on the 
spherical sub-manifold of $\Rset^3$ by the Pythagorian metric already 
existing in the ambient space. This is obtained by choosing appropriate 
coordinates on the sub-manifold, $u^1=\theta$, $u^2=\phi$, the standard 
spherical coordinates, and to replace for $x^{\mathrm{i}} = 
x^{\mathrm{i}}(u^{\mathrm{j}})$ in the metric of the ambient space
\begin{eqnarray}
{\d} s^2 & = & \delta_{\mathrm{ij}}\; {\d} x^{\mathrm{i}} {\d} x^{\mathrm{j}}
\nonumber
\\
& = & \delta_{\mathrm{ij}}\; {\partial x^{\mathrm{i}}\over\partial 
u^{\mathrm{k}}} 
{\partial x^{\mathrm{j}}\over\partial u^{\mathrm{l}}} 
{\d} u^{\mathrm{k}} {\d} u^{\mathrm{l}}
\nonumber
\\
& = & a^2 {\d}\theta^2 + a^2 \sin^2\theta\,{\d}\phi^2.\label{eq:metrsph}
\end{eqnarray}
So the effective Lagrangian will then be
\begin{eqnarray}
L & = & {1\over 2}\, g_{\mathrm{ij}}\; {{\d} u^{\mathrm{i}}\over {\d} t} 
{{\d} u^{\mathrm{j}} \over {\d} t}
\nonumber
\\
& = & {1\over 2} \bigl( a^2 \dot\theta^2 + a^2 \sin^2\theta\,\dot\phi^2\bigr). 
\end{eqnarray}
The orbits of such a Lagrangian are great circles around the sphere, which are 
in fact exactly the geodesics of the metric $(\ref{eq:metrsph})$, 
and are traced 
out by the motion at essentially constant velocity. We compare in figure 
$\ref{fig:fig20}$ 
the ``exact'' numerical solution with the approximate solution $r=a$ and find 
a good agreement even for several revolutions around the sphere.  

\subsubsection{Particle on logarithmically deformed two-dimensional space}
A second instructive example is analogous to a scattering problem. 
This example shows how simple changes in the metric
can induce radical changes in the motion.

\begin{figure}
\centering
\mbox{\epsfig{figure= 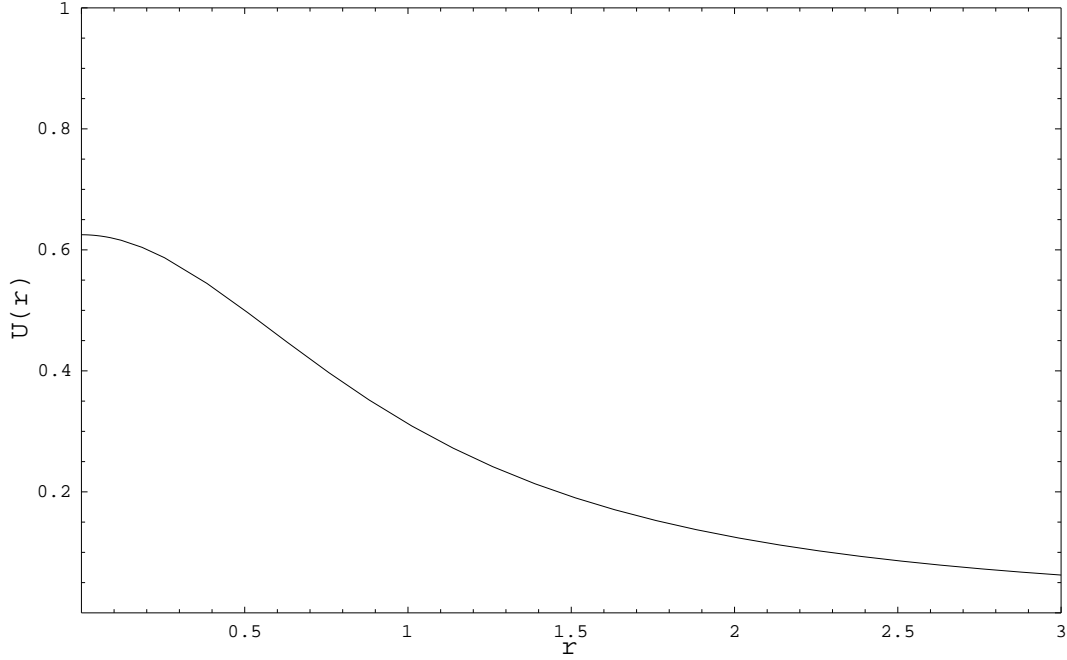,height=3.5in}}
\caption{Plot of the radial effective potential $U(r)$ for a particle 
restricted to move on the logarithmic surface $z=\ln r$
($v_0=1$ and $b=0.5$).}
\label{fig:fig21}
\end{figure}

Consider again the motion of a particle in $\Rset^3$ with
Pythagorian metric and a potential corresponding to a Lagrangian
\begin{eqnarray}
L & = & {1\over 2} \delta_{\mathrm{ij}}\, \dot x^{\mathrm{i}} 
\dot x^{\mathrm{j}} - \lambda g^2(x^{\mathrm{i}})\nonumber
\\
& = & {1\over 2} [ \dot x^2+\dot y^2+\dot z^2] - \lambda g^2(x,y,z).
\label{eq:lavecg}
\end{eqnarray}
We consider for the present purposes a function $g(x^{\mathrm{i}})$ which 
admits solutions to the equation
\begin{equation}
g(x^{\mathrm{i}}) = 0.\label{eq:condg0}
\end{equation}
Then as $\lambda\rightarrow +\infty$ the low energy motion will be restricted 
to the sub-manifold described by $(\ref{eq:condg0})$. Take the example
\begin{equation}
g(x^{\mathrm{i}}) = z - \ln r = 0, \qquad\qquad {\rm where }\; r = 
\sqrt{x^2 + y^2}.
\end{equation}
The sub-manifold is the surface of revolution obtained by turning the function 
$z = \ln(x)$ about the $z$ axis (essentially it is a horn with a
singular vertex at 
$x=y=0$ and $z=-\infty$). A good set of coordinates to use are the 
usual polar coordinates in the $(xy)$ plane. As $\lambda\rightarrow+\infty$  
we can neglect the motion transverse to the surface, in which case 
$z=\ln r$ and 
the potential becomes zero while the kinetic term gives
\begin{eqnarray}
L & = & {1\over 2} (\dot z^2 + \dot r^2 + r^2 \dot\theta^2)
\nonumber
\\
& = & {1\over 2} \biggl[ \biggl(1 + {1\over r^2} \biggr) \dot r^2 + 
r^2 \dot\theta^2 \biggr]
\nonumber
\\
& \equiv & {1\over 2}\, \Bigl( g_{\mathrm{rr}} \dot r^2 + g_{\theta\theta} 
\dot \theta^2 \Bigr).
\end{eqnarray}
Hence the dynamics corresponds to geodetic motion on the 2 dimensional
manifold with metric
\begin{equation}
{\d} s^2 = \biggl(1 + {1\over r^2}\biggr) {\d} r^2 + r^2 {\d}\theta^2.
\end{equation}
It is clear that as $r\rightarrow+\infty$ the manifold becomes flat. The 
equations for the motion are
\begin{eqnarray}
& {{\d}\over {\d} t} & \biggl[ \biggl(1+{1\over r^2}\biggr)\dot r^2\biggr] 
+ {\dot r^2\over r^3} - r \dot\theta^2 = 0
\\
& {{\d}\over {\d} t} & \bigl( r^2 \dot\theta\bigr) = 0.
\end{eqnarray}
This is just a ``central force problem" with an $r$-dependent mass. The second
equation corresponds to the conservation of angular momentum, $l$,
\begin{equation}
\dot\theta = {l\over r^2}.
\end{equation}
Conservation of energy yields a second integral of motion
\begin{equation}
\epsilon = {1\over 2} \biggl[ \biggl(1+{1\over r^2}\biggr) \dot r^2 + 
{l\over r^2}\biggr].
\end{equation} 
The orbit equation is
\begin{eqnarray}
{{\d} r\over {\d}\theta} & = & \pm {r^2\over l\sqrt{r^2+1}} 
\sqrt{ 2 \epsilon r^2 - l^2}
\nonumber
\\
& = & \pm {r^2\over \sqrt{r^2 + 1}} \sqrt{ {r^2\over b^2} - 1}
\end{eqnarray}
where $\epsilon = v_0^2/2$, $l = b\,v_0$, $v_0$ is the initial velocity and 
$b$ is actually the distance of closest approach. 
We see that $r\ge b$. We can define the effective
potential\cite{Goldstein} $U(r) = V(r) + l^2/2 r^2$ by
\begin{equation}
{r^2\over l}\sqrt{2 (\epsilon - U(r))} = {r^2\over l \sqrt{r^2+1}} \sqrt{2 
\epsilon r^2 - l^2}.
\end{equation}
This yields
\begin{equation}
U(r) = {1\over r^2+1} \biggl(\epsilon + {l^2\over 2}\biggr) = 
{v_0^2\over 2} {1+b^2\over 1+r^2}
\end{equation}
which gives the graph in figure $\ref{fig:fig21}$.

\begin{figure}
\centering
\mbox{\epsfig{figure= 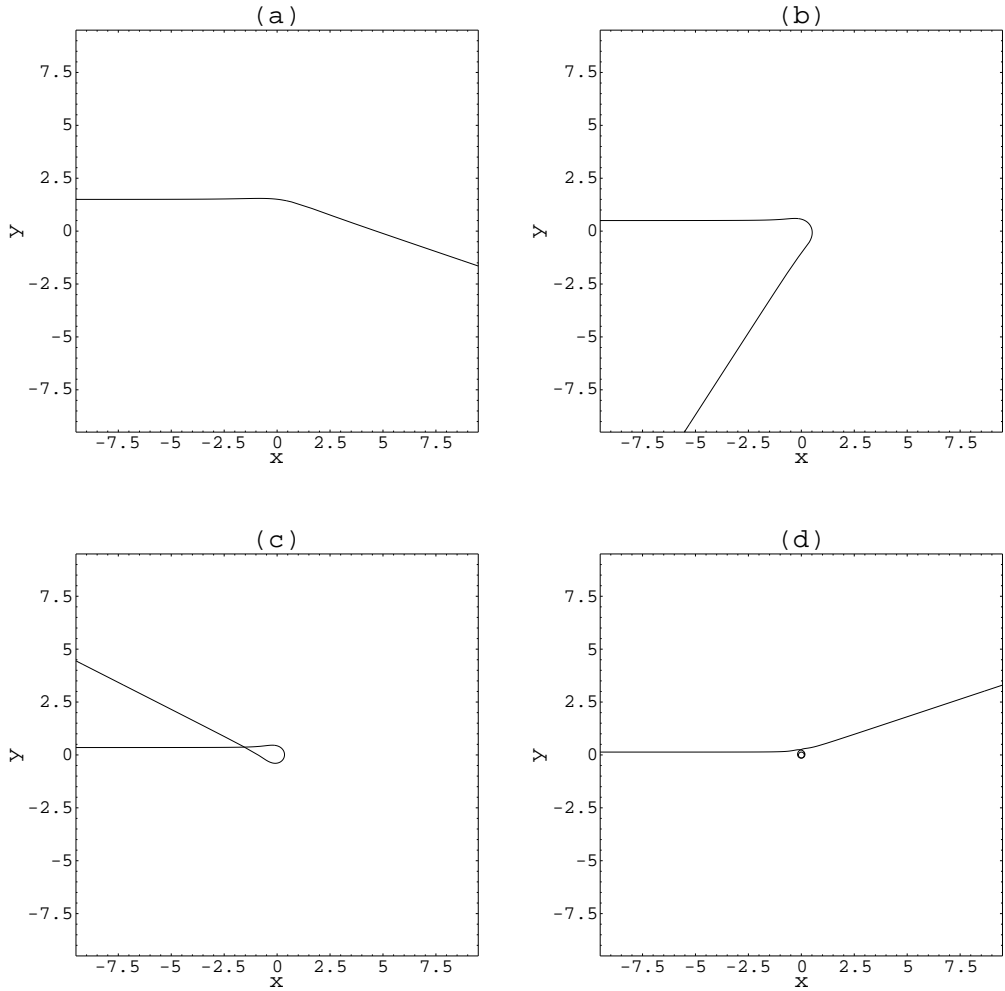,width=5.5in}}
\caption{Trajectories for the particle trapped on the logarithmic surface 
for different values of the impact parameter $b$: (a) $b=1.5$, (b) $b=0.5$, 
(c) $b=0.35$ and (d) $b=0.135$ (The particle is coming from the left and its 
velocity is irrelevant).}\label{fig:fig22}
\end{figure}

There is no bound orbit; the turning point of the scattering occurs for $r=b$ 
when $U(b) = v_0^2/2$, \ie there is no more kinetic energy in the radial 
motion. The orbit equation is integral in terms of elliptic integrals and 
yields
\begin{equation}
\theta(r) = {b\over E({\mathrm i} b, 1/r)}
\end{equation}
where $E$ is the elliptic integral of the second kind.
This gives a mild attractive potential. Typical geodesics bend slightly towards
the singularity at $r=0$ for large impact parameter $b$, however, for $b$ near
zero the particle makes several revolutions about the singularity before 
returning to infinity (see figure $\ref{fig:fig22}$). 
The $b=0$ geodesic is of course singular and the 
particle falls into the hole at $r=0$. 

\subsubsection{Systems without a continuous set of static solutions}
Returning to the first example of the sphere, we augment it with a potential
in the $z$ direction. Physically this could correspond to a uniform 
gravitational field.

The Lagrangian that we consider is
\begin{equation}
L = {1\over 2} \bigl(\dot r^2 + r^2( \dot\theta^2 + 
\sin^2\theta\dot\phi^2)\bigr) - \lambda (r^2-a^2)^2 - \Delta r \cos\theta.
\end{equation}
with $\Delta>0$ and $\Delta\ll\lambda$. 
Now the critical points of the potential correspond only to the absolute 
minimum located near the south pole of the sphere ($r=a$ and $\theta=\pi$), 
and a minimax near the north pole at $r=a$ and $\theta=0$. 
The conditions for the critical points are
\begin{eqnarray}
& & 4 \lambda (r^2 - a^2) r - \Delta \cos\theta = 0
\\
& & \Delta r \sin\theta = 0
\end{eqnarray}
which give $\theta=0$, $\pi$ and 
\begin{equation}
r = a \pm {\Delta\over 8\lambda a^2} + 
O\Biggl({\Delta^2\over\lambda^2}\Biggr).
\end{equation}
This case falls directly into the category for which Manton proposes that the
motion truncates to the unstable manifold formed from the union of the curves 
of steepest descent or the gradient flow curves connecting critical points. 
The proposal corresponds to the following equations:

the gradient flow curve $(\ref{eq:gfc})$:
\begin{eqnarray}
& &\dot r  =  - 4 \lambda r (r^2-a^2) - \Delta \cos\theta
\\
& & r^2 \dot\theta  =  \Delta r \sin\theta 
\\
& &r^2 \sin^2\theta \dot\phi = 0
\end{eqnarray}
or the paths of steepest descent $(\ref{eq:psd})$
\begin{eqnarray}
& & \ddot r  =  - 4 \lambda r(r^2-a^2) - \Delta\cos\theta
\\
& &r^2\ddot\theta  =  \Delta r \sin\theta
\\
& &r^2 \sin^2\theta \ddot\phi = 0
\end{eqnarray}
or the complete equations of motion $(\ref{eq:eqmvt})$
\begin{eqnarray}
& &\ddot r  =  r(\dot\theta^2 + \sin^2\theta\,\dot\phi^2) - 4 
\lambda r(r^2-a^2) - \Delta\cos\theta
\\
& &2 r \dot r + r^2 \ddot\theta  =  \Delta r \sin\theta +
r^2\sin\theta\cos\theta \dot\phi^2
\\
& &{{\d} \over {\d} t} \bigl[ r^2 \sin\theta \dot \phi\bigr]  = 0.
\end{eqnarray}
As $\lambda/\Delta\rightarrow+\infty$, all three methods simply reduce to the
sphere $r=a$, and the motion is constrained to this sphere with an effective
potential
\begin{equation}
V_{\mathrm{eff}}(\theta) = a\,\Delta\,\cos\theta.
\end{equation}
The difference in the three method gives rise to surfaces which vary by a
thickness of the order of $\Delta/\lambda$. This can be important if this 
ratio is not small. 

\begin{figure}
\centering
\mbox{\epsfig{figure= 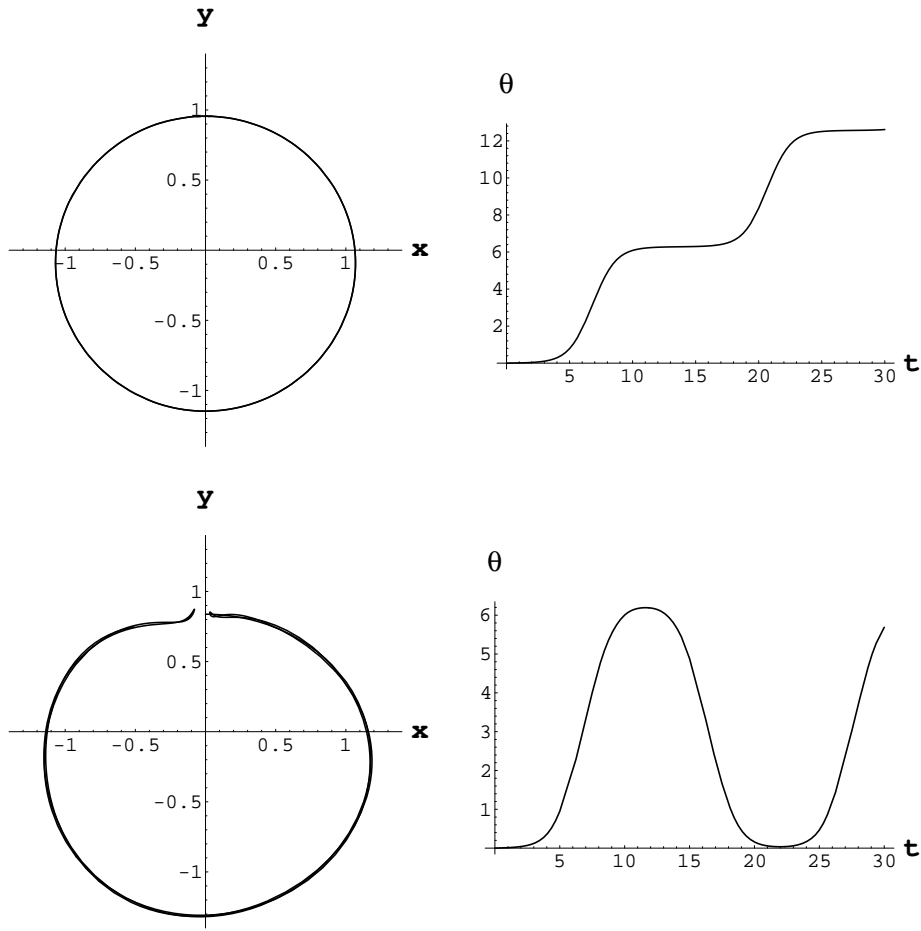,height=7.0in}}
\caption{Trajectories of the particle energetically constrained near a 
circle, and also subject to a linear, vertical gravitational 
potential (left) and a plot of 
the polar angle $\theta$ as a function of time $t$ during the process (right). 
For certain values of the parameters, the 
particle excites the radial mode sufficiently
and does not quite make it over the 
top (even though it has more than enough energy to do so), in complete 
contradiction with the truncation method. Figures shown are obtained
for $\lambda = 3$ and $\delta = 1$ (top), and for $\lambda = 1$ and 
$\delta = 1$ (bottom).}
\label{fig:fig23}
\end{figure}

Numerical studies show that the motion is well approximated by motion on the
truncated sub-manifold, with an important caveat: if
the particle approaches the saddle point too
closely the approximation can be misleading. To see this 
consider the case where the particle starts at the unstable
critical point with a finite initial velocity. The 
movement when restricted to the unstable manifold will evidently correspond to 
revolutions about vertical great circles. Conservation of energy implies that 
the particle will always have enough energy to rise up to the saddle (point) 
and pass over the top. The actual motion however will necessarily excite the 
radial degree of freedom. If the dynamics conspire such that as the particle 
approaches the saddle point enough energy has been transferred to the radial 
degree of freedom to energetically prohibit the particle to pass over the 
saddle, the subsequent motion will fall back down the sphere on the same side
in complete 
disaccord with the prediction of the truncated dynamics. We have actually 
numerically discovered this kind of deviation from the expected
behaviour from the truncated dynamics (see figure $\ref{fig:fig23}$).
It is clear that the motion is very sensitive to the initial conditions and 
to the coupling between the low 
energy degrees of freedom and the high energy modes which govern the transfer 
of energy between these modes.

Another situation where the motion on the truncated manifold is very 
susceptible to small transverse oscillations occurs when the truncated 
set of configurations 
ceases to be a manifold. For example if the potential in 
$(\ref{eq:lavecg})$ 
is of the form
\begin{equation}
V(x^{\mathrm{i}}) = \lambda \bigl( g(x^{\mathrm{i}}) f(x^{\mathrm{i}}) 
\bigr)^2,
\end{equation}
as $\lambda\rightarrow+\infty$ the solution of
\begin{equation}
g(x^{\mathrm{i}})=0
\end{equation}
and
\begin{equation}
f(x^{\mathrm{i}})=0
\end{equation}
will give in general two different sub-manifolds for the truncated dynamics. 
This is fine if the two manifolds are disjoint; then they will be separated by 
large energy barriers and the dynamics will behave independently in each 
sub-manifold. However if they are tangent or even cross, the dynamics becomes
very sensitive to transverse oscillations.

\begin{figure}
\centering
\mbox{\epsfig{figure= 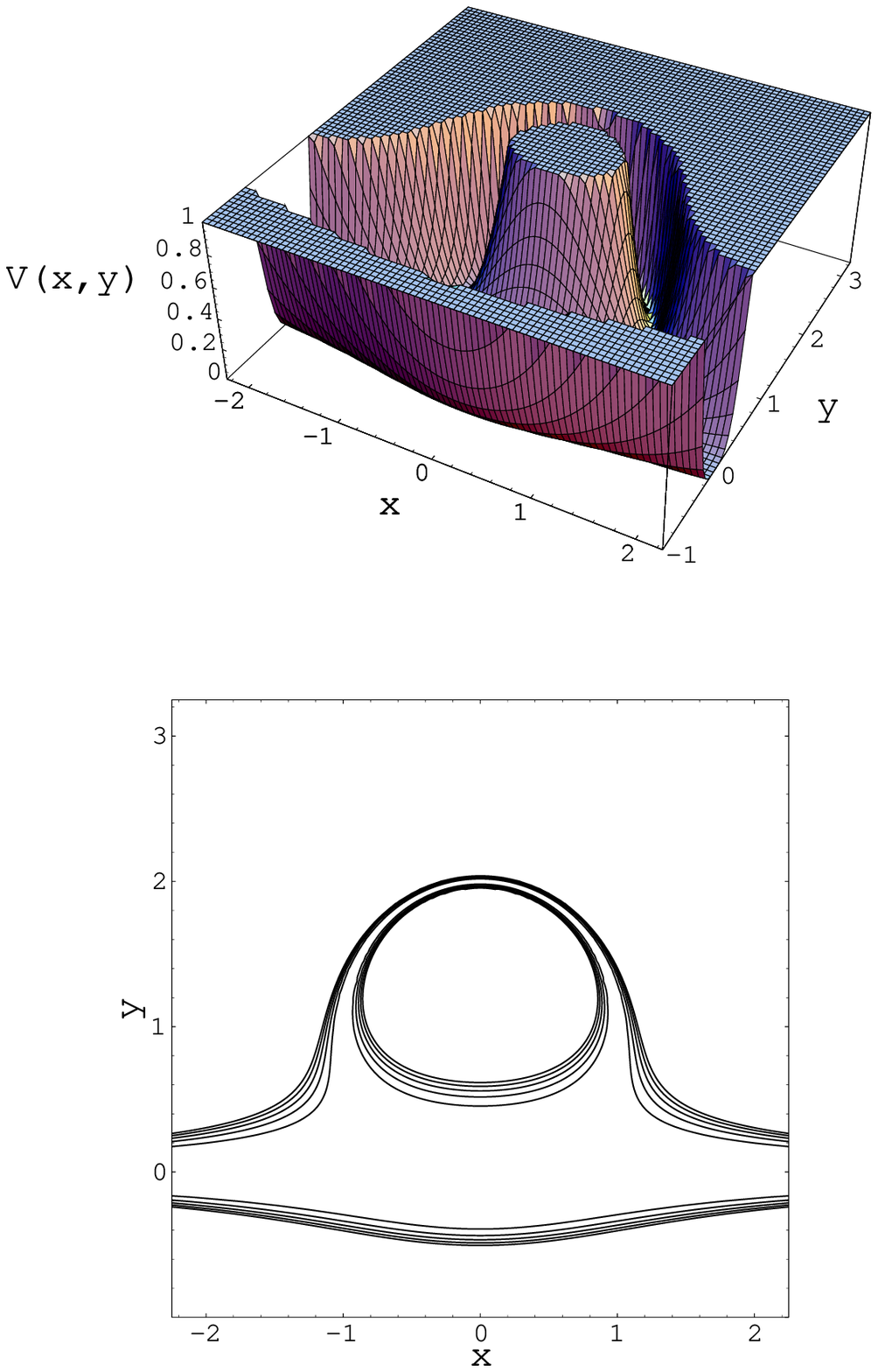,height=7.0in}}
\caption{Three-dimensional (top) and contour plot (bottom) of the potential 
of the Lagrangian $(\ref{eq:lagrbiff})$ for $\lambda=20$ and $a=1$.}
\label{fig:fig24}
\end{figure}

\begin{figure}
\centering
\mbox{\epsfig{figure= 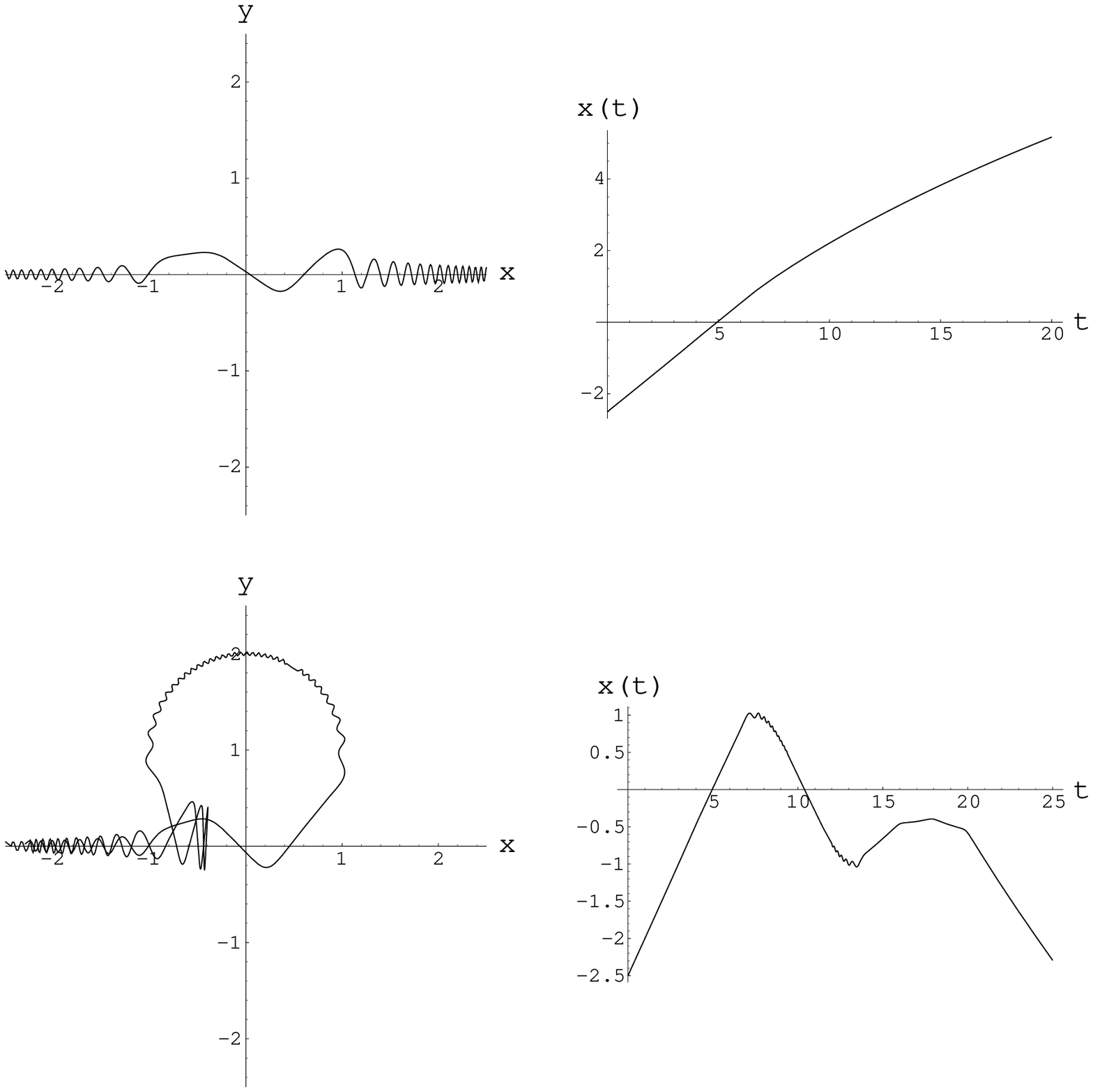,height=5.in}}
\caption{Trajectories of a particle moving on a space made of a straightline 
with a circle tangent to (left) and a plot of its 
position along the $x$ axis as a function of time (right). The final 
state is extremely sensitive to initial conditions, with the particle 
either going through the intersection area at the first try (top) or getting 
caught in the loop and retracing its steps on the $x$ axis (bottom).}
\label{fig:fig25}
\end{figure}

Take for example the two dimensional Lagrangian
\begin{equation}
L = {1\over 2} (\dot x^2+\dot y^2) 
- \lambda \biggl[ y \biggl( x^2 +(y-{1\over 2})^2 -1\biggr)\biggr]^2.
\label{eq:lagrbiff}
\end{equation}
As $\lambda\rightarrow+\infty$ the low energy dynamics will truncate to the
curves
\begin{equation}
y = 0
\end{equation}
or
\begin{equation}
x^2 + \biggl(y-{1\over 2}\biggr)^2 = 1.
\end{equation}
These curves describe the $x$ axis and a circle of radius 1 which is just 
tangent to the $x$ axis at the origin (see figure $\ref{fig:fig24}$). 
A low energy particle, in general, 
will effect tiny oscillations transverse to these curves and translate 
along these curves at roughly constant velocity. Suppose we start at a point 
on the circle with velocity tangent to the circle. Depending on the phase 
and amplitude of the radial oscillation that will necessarily be excited, at 
the moment that the particle passes the point of contact, the particle can 
easily be trapped by the valley along the $x$ axis and move on to 
$x\rightarrow+\infty$. Conversely, a particle moving along the $x$ axis, 
having a slight transverse 
oscillation, can be trapped in the circle. It is clear 
that the large scale low energy dynamics is not independent of the
excitation of the high energy, modes however small they may be (see figure 
$\ref{fig:fig25}$).

%% file: section42.tex
We now apply Manton's formalism of gradient flow curves to the Skyrme model,
for which the method was first put forward\cite{Mantongfc}. The first step is 
to single out the relevant critical points of the energy functional (in fact, 
the critical points of the potential energy of the model). Unlike the simple 
examples of the previous subsection, this task is quite hard, 
since a field theory is a system with an infinite number of degrees of 
freedom living (in this case) in a 3 dimensional space. Furthermore, 
non-linearity complicates the problem, and it 
is not surprising that coordinated numerical and analytical methods have 
to be used to 
find solutions to the equations of motion and to study their stability.
We then describe how the global and relative degrees of freedom factorize.
Finally we present the construction of the full unstable manifold linking 
together the critical points of this sector.

\subsubsection{Critical points of the $B=2$ sector} 

The first solution of the Skyrme Lagrangian in the $B=2$ sector was 
already found by Skyrme\cite{Skyrme} in the 60's. It is the product ansatz 
defined as follows:
\begin{equation}
U_{\mathrm{PA}}(\vec x) = A_1^\dagger U_{\mathrm{S}}(\vec x-\vec R_1)A_1 \; 
A_2^\dagger U_{\mathrm{S}}(\vec x-\vec R_2)A_2
\end{equation}
where $U_{\mathrm{S}}$ is the Skyrmion field, $A_1$ and $A_2$ are 
$\mathrm{SU(2)}$ matrices defining the orientation of Skyrmion 1 and 2 
respectively, and $\vec R_1$ and $\vec R_2$ represent their positions. 
The product ansatz is a solution of the equations of motion only in the limit 
where the distance between the particles $R\equiv ||\vec R_1 - \vec R_2||$ is 
infinite, since then, in the neighborhood of each Skyrmion, the other field 
is the identity 
and hence the product satisfies the equations of motion (and vice versa 
in the neighbourhood of 
the other Skyrmion). When $R$ is finite, the product ansatz is not a solution 
of the theory, but it still represents a good approximation to the 
configurations along gradient flow curves  
for large separations. This is true even though ``Skyrme matter'', like 
typical soliton fields, is quite soft and the mutual interactions of 
the fields deform the Skyrmions at a distance: they are no longer exact 
hedgehogs. However the ansatz is good 
enough to obtain an approximation to the potential between 
the solitons. This 
potential depends strongly on the relative orientation $A_1^\dagger A_2$ of 
the Skyrmions, and this allows it to be repulsive or attractive. Hence
the solution at $R\rightarrow+\infty$ is a saddle point 
of the energy functional and necessarily 
a state of lower energy of most likely 
different structure exists. We will return to the potential and the 
product ansatz later since they are useful to understand the structure 
of the manifold of gradient flow curves of the system.

One expects to find in the $B=2$ sector a state which represents the deuteron, 
and indeed it would be another great triumph of the Skyrme model if such a 
state was found, with the right quantum numbers, binding energy, etc. The
first localized state with baryon number two, which did not give the deuteron, 
is an immediate generalisation 
of the $B=1$ Skyrmion and was probably already contemplated by Skyrme himself 
in the 60's, although the result was first published by Jackson\cite{Jackson} 
in the 80's. The $B=2$ hedgehog is defined (as we have seen in section 3.2.3)
as 
\begin{equation}
U(\vec x) = {\mathrm{e}}^{\mathrm{i} f_2(r) \hat\cdot\vec\tau}
\label{eq:dib}
\end{equation}
where the profile function $f_2(r)$ has to be computed numerically like in the 
$B=1$ sector, but obeys the following modified boundary conditions: 
$f_2(0)=2\pi$ and $f_2(+\infty)=0$. This solution is a straightfoward 
generalization of the $B=1$ hedgehog, and we will call it the 
dibaryon in what follows to avoid any confusion with the ordinary hedgehog. 
The mass of the dibaryon happens to be about $1.855\times 24 \pi^2$ in natural 
units, which translates roughly to 2.5 to 4.4 GeV 
(the value depends on the ones 
chosen for the parameters of the model $f_\pi$ and $e$) or about 3 times the 
mass of a single Skyrmion. This state can be shown to be 
unstable. It has much more energy than the product ansatz itself, and one 
expects the dibaryon to disintegrate at least 
into a pair of widely separated Skyrmions. 
We will see that exactly this has been observed in numerical 
simulations\cite{Waindzoch1}. 
Hence the dibaryon is also a saddle point. Thus we have, so far, two 
types of saddle points from which to construct gradient flow 
curves. We are still missing, however, the minimum energy $B=2$ state to 
where these curves lead. 

\begin{figure}
\centering
\mbox{\epsfig{figure= 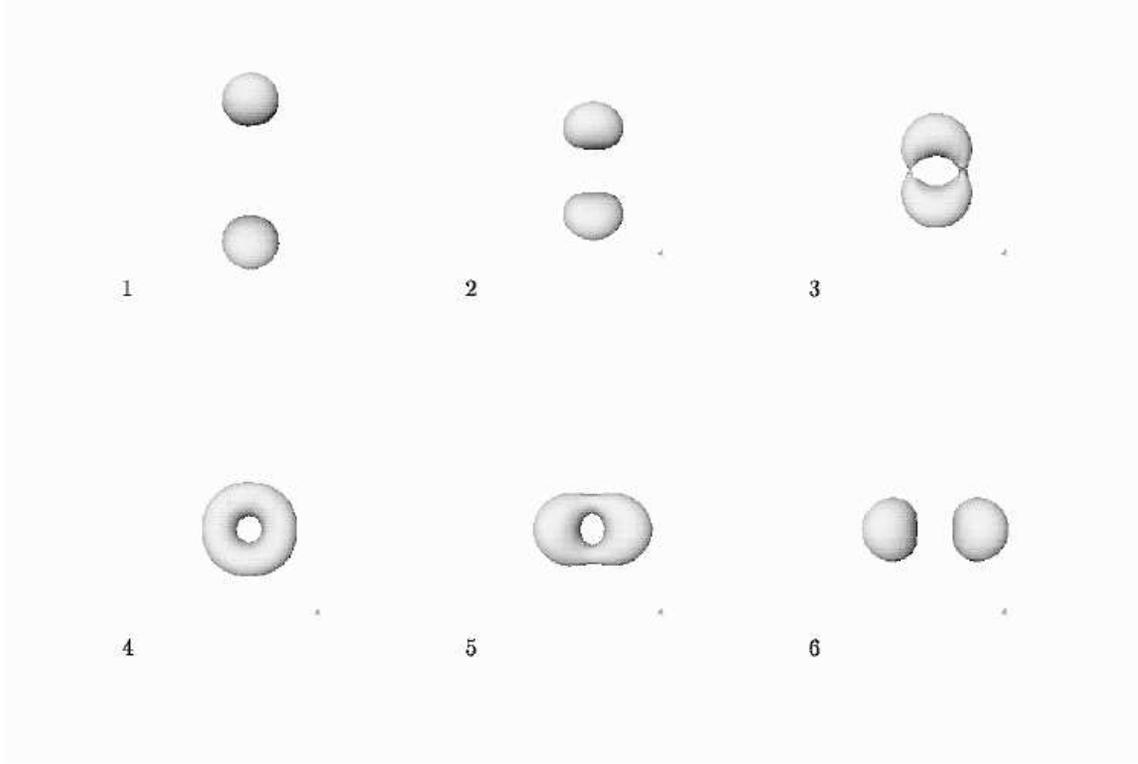,height=4.0in}}
\caption{Numerical simulation of the scattering of two Skyrmions via the $B=2$ 
torus.}
\label{fig:fig26}
\end{figure}

The state which is generally accepted as having the lowest energy in the 
$B=2$ sector was discovered independently by Verbaarschot\cite{Verbaarschot} 
and Kopeliovich\cite{Kopeliovich} using numerical methods, and actually 
proposed by 
Manton\cite{Mantontore} by indirect methods related to dipole-dipole 
dynamics and 
symmetries. Its energy is about $1.18\times 24\pi^2$ in 
natural units which is about $4\%$ less than a pair of free 
Skyrmions. 
Quite intriguingly, the minimum energy state is toroidal. The deuteron, being 
the least bound of all nuclei, comprises of a proton and neutron that are 
quite spatially
separated and distinct most of the time. The toroidal form well 
represents this spatial separation, however, it fails to reproduce the 
distinct character of the nucleons since the Skyrmions are completely 
deformed and have lost their separate identities. On the other hand, this 
state is expected to take an active part in the 
scattering processes of two Skyrmions, especially those with zero 
impact parameter and with the relative orientation which generates maximum 
attraction. In such a situation, 
the Skyrmions move toward each other, deform and 
come together in the toroidal state before dividing into two Skyrmions 
which depart along a direction perpendicular to the initial one (the 
scattering plane is fixed by the initial (iso)orientations of the Skyrmions)
\cite{Mantontore,Battye-Sutcliffe} (see figure $\ref{fig:fig26}$
taken from reference \cite{Battye-Sutcliffe}). 
This $90^\circ$ scattering is typical of head on collisions of identical 
solitons\cite{Rbmack}. The interpretation of this configuration as the 
deuteron is 
still controversial given its large binding energy (of the order of 100 MeV 
instead of the 2 MeV or so observed in nature), small radius and 
``strange shape''.\footnote{One way to 
resolve this issue might be to compare the cross section 
for nucleon-nucleon scattering computed from the Skyrme model with actual 
experimental measurements, and look for a signature predicted by the Skyrme 
model for Skyrmions going through the toroidal configuration. This is still 
an open problem since no general cross section for nucleon-nucleon scattering, 
neither classical nor quantum mechanical, has been computed so far (We present 
in the last subsections of this review our own semi-classical computations 
of nucleon-nucleon scattering angles with the Skyrme model under certain
conditions.)}
Although it is not certain that this state is the absolute  lowest energy 
state of the $B=2$ sector, it does seem reasonable that it is so. No other 
states with lower energy have been found so far, numerical simulations of 
Skyrmion-Skyrmion scattering have not uncovered another intermediate 
state\cite{Battye-Sutcliffe}, 
and numerous modes of oscillation around 
the toroid have been considered without finding any negative 
modes\cite{Baskerville}. We will assume here that it is in fact the 
state with lowest energy and that the manifold of low energy dynamics of the 
$B=2$ sector consists of path joining the minimum and the two different 
families of saddle points. 

To visualize the unstable manifold of the $B=2$ we need
to know how the different configurations presented above are linked together 
by the dynamics. The number of degrees of freedom in the unstable and zero 
modes of each state will be most helpful. 
A solitonic system has an 
infinite number of degrees of freedom or modes that it can excite. Modes can 
be categorized according to their role in the stability of a particular 
configuration: negative modes generate paths followed by the system when it  
moves to another configuration of lower energy, positive modes are 
modes associated with oscillations around soliton 
configurations and zero modes 
correspond to rigid motions of the solitons arising from 
symmetries of the energy functional which are not respected by 
a given solution. If the system has low 
enough energy, only a finite number of these modes will be 
excited. In fact the system will first explore 
only the zero and negative modes 
(if there are any); higher energy excites the positive modes starting with 
those with the lowest frequencies. The parameters corresponding to zero 
modes are particularly important, and are called collective coordinates 
of the system. They form what is called the moduli space for the case 
where there are no negative modes. For example, in any BPS system, the 
collective coordinates corresponding to the zero modes are sufficient to 
describe the 
low energy motion of the system, which then corresponds 
to the geodesics on the moduli space.
In cases where there is a weak 
potential acting between the solitons (like the Skyrme model), one has to 
also  
include the negative modes and the union of the corresponding 
gradient flow curves, to have 
enough freedom to let the system evolve in a satisfactory fashion. 
The number of degrees of freedom rises with the baryon 
number, since new collective coordinates enter to describe 
the relative motion of the particles, however the number can
decrease if new symmetries 
arise via the dynamical evolution of the system.

\subsubsection{Factorization of global and relative coordinates} 

Let $U(\vec x)$ be a completely general configuration of the system obeying 
the conditions that $U(\vec x)$ goes to the identity at infinity
fast enough so that 
the 
total energy associated to the configuration is finite, localized in space, 
and the baryon number associated to the configuration is an integer. Such a 
configuration possesses generic collective coordinates corresponding to the 
invariance of the energy functional under the following 3 transformations:
\begin{eqnarray}
\vec x &\rightarrow& \vec x + \vec R_{\mathrm{T}}
\\
\vec x &\rightarrow& D(B)\cdot \vec x
\\
U(\vec x) &\rightarrow& A U(\vec x) A^\dagger.
\end{eqnarray}
In general a configuration is not invariant under these transformations, 
giving rise to nine zero modes in the fluctuation spectrum.
$\vec R_{\mathrm{T}}$ is the position of the center of mass of the system, 
and is 
related to invariance of the energy under translation, the rotation matrix 
$D(B)$ associated to the $\mathrm{SU(2)}$ matrix $B$ parametrizes the 
rotational invariance of the system under global rotation, and finally the 
$\mathrm{SU(2)}$ matrix $A$ parametrizes the invariance of the system under 
global isospin rotation. Because of the absence of special symmetries of 
$U(\vec x)$, the 9 parameters are completely independent, and if one gives 
them time dependence and quantizes the system treating them as dynamical
variables, 
they give, respectively, the conserved total momentum $\vec P$, 
the conserved total 
angular momentum $\vec L$ and the conserved total isospin $\vec T$.  
The 9 global collective coordinates do not participate in the interesting 
dynamics of the system.  

We will now apply these ideas to the $B=2$ sector of the Skyrme model
in turn considering the product ansatz, the dibaryon and the torus.
As we saw earlier, the product ansatz 
\begin{equation}
U_{\mathrm{PA}}(\vec x) = A_1 U_{\mathrm{S}}(\vec x-\vec R_1)A_1^\dagger \; 
A_2 U_{\mathrm{S}}(\vec x-\vec R_2)A_2^\dagger\nonumber
\end{equation}
is a solution of the theory only in the limit where $R=||\vec R_1-\vec R_2||$
is infinite; we will consider it to be arbitrarily large. 
The product ansatz is a solution described 
by 12 collective coordinates, or twice what is needed to describe a single 
Skyrmion: 3 parameters to define the position of the Skyrmion, and another 3
for its orientation or iso-orientation. This last fact comes from the 
spherical symmetry of the Skyrmion:
\begin{eqnarray}
A U_{\mathrm{S}}(\vec x) A^\dagger &=& \cos{f(r)} + {\mathrm{i}} \sin{f(r)} 
A \vec\tau A^\dagger\cdot\hat x\nonumber
\\
&=& \cos{f(r)} + {\mathrm{i}} \sin{f(r)} \tau^{\mathrm{a}} D_{\mathrm{ab}}(A) 
\hat x^b\label{eq:ada}
\end{eqnarray}
hence 
\begin{equation}
A U_{\mathrm{S}}(\vec x) A^\dagger = U_{\mathrm{S}}(D(A)\cdot\vec x)
\label{eq:symsph}
\end{equation}
where $D(A)$ is the rotation matrix defined by $(\ref{eq:du})$ associated 
to the 
$\mathrm{SU(2)}$ matrix $A$. So iso-rotating the Skyrmion by a certain amount 
is exactly equivalent to rotating it in space by the same amount. Thus one 
requires only three parameters instead of six to describe the orientation 
(and iso-orientation). Another way to see this is to consider the following 
expression:
\begin{equation}
U(\vec x;A,B) = A U_{\mathrm{S}}(D(B)^{-1}\cdot\vec x) A^\dagger
\end{equation}
where the iso-orientation $A$ and the orientation $B$ are considered 
independent. This state is invariant under the following transformation of $A$ 
and $B$:
\begin{eqnarray}
A &\rightarrow& AC
\\
B &\rightarrow& BC
\end{eqnarray}
since
\begin{eqnarray}
U(\vec x;AC, BC) &=& AC U_{\mathrm{S}}(D(BC)^{-1}\cdot\vec x)\nonumber
C^\dagger A^\dagger
\\
&=& A U_{\mathrm{S}}(D(B)^{-1}\cdot\vec x) A^\dagger\nonumber
\\
&=& U(\vec x;A,B)\label{eq:redsph}
\end{eqnarray}
by equation $(\ref{eq:symsph})$. So $U(\vec x;A,B)$ has a continuous 
redundancy parametrized by 3 angles, which arises from the spherical 
symmetry of the Skyrmion. The remaining 3 angles and 3 coordinates of the 
center of mass parametrize a manifold which can be written 
$S^3\times\Rset^3$. The product ansatz therefore has 6 independent 
parameters to describe each of the two particles, for a total of 12. 

We will now show how to re-express these 12 parameters $A_1$, $A_2$, 
$\vec R_1$ and $\vec R_2$ in terms of global and relative ones. 
To do this we will reproduce the discussion of Walhout 
and Wambach\cite{Walhout}, who 
use the analogy with the treatment of a rigid body: the global coordinates 
will represent the transformation from the laboratory frame to the body fixed 
frame, and the relative coordinates will describe the system in the body 
fixed frame. Walhout and Wambach\cite{Walhout} chose the Skyrmions to be 
separated along the $z$ 
axis in the body fixed frame by a distance $R$, and the Skyrmion at the 
position $\vec x= -(R/2)\; \hat e_3$ to be isorotated by $C$ relative to 
the other:
\begin{eqnarray}
A_1^{\mathrm{bf}} &=& 1
\\
A_2^{\mathrm{bf}} &=& C
\\
\vec R_1 &=& {R\over 2} {\hat e}_3
\\
\vec R_2 &=& - {R\over 2} {\hat e}_3.
\end{eqnarray}
In the lab frame, we will use $\vec R_{\mathrm{T}}$ to denote the 
position of 
the center of mass of the system, and we will position the Skyrmions in space 
using the inverse of the rotation matrix corresponding to the $\mathrm{SU}(2)$
matrix $B$, and a matrix $A$ will rotate 
the whole system in isospace.
This gives us the product ansatz
\begin{eqnarray}
U^{\mathrm{lab}}_{\mathrm{PA}}(\vec x) &=& 
A_1 U_{\mathrm{S}}(\vec x - \vec R_1) A_1^\dagger
A_2 U_{\mathrm{S}}(\vec x - \vec R_2) A_2^\dagger\nonumber
\\
&=& A U_{\mathrm{S}}[D(B)\cdot(\vec x - \vec R_{\mathrm{T}} - 
D(B)^{-1}\cdot {R\over 2} \hat e_3) ]\times\nonumber
\\
&&\qquad C U_{\mathrm{S}}[D(B)\cdot(\vec x - \vec R_{\mathrm{T}} +
D(B)^{-1}\cdot {R\over 2} \hat e_3) ] C^\dagger A^\dagger
\label{eq:apglrel}
\end{eqnarray}
with
\begin{eqnarray}
A_1 &=& AB
\\
A_2 &=& ACB
\\
\vec R_1 &=& \vec R_{\mathrm{T}} + D(B)^{-1}\cdot {R\over 2} \hat e_3
\\
\vec R_2 &=& \vec R_{\mathrm{T}} - D(B)^{-1}\cdot {R\over 2} \hat e_3
\end{eqnarray}
which is clearly of the form:
\begin{equation}
U(\vec x) = A U\bigl(D(B)\cdot(\vec x-\vec R_{\mathrm{T}})\bigr) A^\dagger
\end{equation}
where the global degrees of freedom are singled out. 
If we count the parameters we 
find that there are 13 of them instead of just the 12 which
we started with. Walhout and Wambach\cite{Walhout} 
(inspired by some work by Verbaarschot\cite{Verb-anom}) show that one 
angle, which we 
shall define shortly, parametrizes a symmetry of the ansatz 
$(\ref{eq:apglrel})$ and is therefore 
redundant. In the body fixed system, the product ansatz 
$(\ref{eq:apglrel})$ obeys the following symmetry:
\begin{equation}
U^{\mathrm{bf}}_{\mathrm{PA}}(\vec x;C) = C (\mathrm{i}\vec\tau\cdot\hat n)
U^{\mathrm{bf}}_{\mathrm{PA}}(D(\mathrm{i}\vec\tau\cdot\hat n)\cdot\vec x;C)
(-\mathrm{i}\vec\tau\cdot\hat n) C^\dagger\label{eq:pasymc}
\end{equation}
where $\hat n$ is any vector perpendicular to the $z$ axis, and $C$ is
defined as follows:
\begin{equation}
C(\vec\tau\cdot\hat n) = (\vec\tau\cdot\hat n)C^\dagger.\label{eq:defc}
\end{equation}
If one writes $C$ as $e^{\mathrm{i} \vec\gamma\cdot\vec\tau}$, then this 
implies that $\hat n = \hat\gamma\times\hat e_3$. Let us check that 
$(\ref{eq:pasymc})$ is 
indeed a symmetry of the product ansatz. Applying the transformation to 
the product ansatz in the body fixed frame one gets:
\begin{eqnarray}
U^{\mathrm{bf}}_{\mathrm{PA}}(\vec x;C) &=&
U_{\mathrm{S}}(\vec x - {R\over 2}\hat e_3) 
C U_{\mathrm{S}}(\vec x + {R\over 2}\hat e_3) C^\dagger\nonumber
\\
&\rightarrow& C U_{\mathrm{S}}(\vec x + {R\over 2}\hat e_3) C^\dagger
U_{\mathrm{S}}(\vec x - {R\over 2}\hat e_3)
\end{eqnarray}
since $D(\mathrm{i}\vec\tau\cdot\hat n)$ is a rotation by $\pi$ around an axis
perpendicular to the $z$ axis, and using $(\ref{eq:defc})$. We then see that 
the Skyrmions
change places, but in the limit where $R\rightarrow+\infty$, which is a 
necessary condition for the product ansatz to be a solution of the equations 
of motion, the order of the Skyrmion matrices is irrelevant since the two 
commute. Then $(\ref{eq:pasymc})$ is really a 
symmetry of the solution within the product
ansatz. It is also a symmetry of the Lagrangian (since it only consists of 
global rotation and isorotation) and of the subsequent evolution.
The symmetry $(\ref{eq:pasymc})$ is valid for any vector $\hat n$
as long as it is perpendicular to the $z$ axis: the angle which defines the 
orientation of $\hat n$ in the $(xy)$ plane actually spans a whole family of 
discrete symmetries of the ansatz. Since $\hat n = \hat\gamma\times\hat e_3$,
rotating $\hat n$ in the $(xy)$ plane by the angle $\theta$ also rotates 
$\hat\gamma$ by the same amount in the same plane, which in turn is equivalent 
to acting on the matrix $C$ as follows:
\begin{equation}
C\rightarrow e^{\mathrm{i} {\theta\over 2}\tau^3} C 
 e^{-\mathrm{i} {\theta\over 2}\tau^3}.
\end{equation}
It is easily checked that for the product ansatz this symmetry takes the
following form:
\begin{equation}
U^{\mathrm{bf}}_{\mathrm{AP}}(\vec x;C) =
e^{-\mathrm{i} {\theta\over 2}\tau^3} 
U^{\mathrm{bf}}_{\mathrm{AP}} 
(D(e^{\mathrm{i} {\theta\over 2}\tau^3})\cdot\vec x;
e^{\mathrm{i}{\theta\over 2}\tau^3} C e^{-\mathrm{i}{\theta\over 2}\tau^3}
)
e^{\mathrm{i} {\theta\over 2}\tau^3}\label{eq:redondance}
\end{equation}
or that, in the laboratory reference frame, the product ansatz is invariant 
under the combined transformations:
\begin{eqnarray}
A &\rightarrow& A e^{-\mathrm{i} {\theta\over 2} \tau^3}
\\
B &\rightarrow& e^{\mathrm{i} {\theta\over 2} \tau^3} B
\\
C &\rightarrow& e^{\mathrm{i} {\theta\over 2} \tau^3} C
e^{-\mathrm{i} {\theta\over 2} \tau^3}.
\end{eqnarray}
This shows that the angle of rotation $\theta$ around the $z$ axis implicitly 
contained in $C$ is redundant and does not play a role in the describing the
configuration. $C$ is then only parametrized by 2 angles: one for the 
orientation of $\hat \gamma$ in the $(xz)$ plane, and the magnitude 
$|\vec\gamma|=\gamma$, by which the second Skyrmion is rotated 
relative to the other. We then have, as we should, 9 global collective 
coordinates ($A$, $B$ and $\vec R_{\mathrm{T}}$) and 3 relative ones
($C$ and $R$) giving a total of 12 instead of 13. One should carefully 
note that equation $(\ref{eq:redondance})$ by no means 
implies that the product ansatz possesses a continuous symmetry: it only does 
if one writes it using too many parameters as collective coordinates (exactly
like the case for the ordinary Skyrmion described above by equation
$(\ref{eq:redsph})$). Once 
the correct 
number of collective coordinates has been used, the only symmetries 
that the product ansatz possesses are 3 reflection symmetries relative to the 
$(xy)$, $(yz)$ and $(xz)$ planes (all those discrete symmetries are exact only
in the limit where $R\rightarrow +\infty$). 

\subsubsection{The construction of the unstable manifold}

Let us suppose that the Skyrmions always keep the same hedgehog shape no matter
how strongly they interact with each other. Then one could obtain the 
Lagrangian of the system by giving time dependence to the collective 
coordinates, and replacing the product ansatz in the Skyrme Lagrangian and
computing the requisite integration over all space. These 12 degrees of freedom
then describe a manifold, which we shall note $M_{12}$, with a metric and a 
potential induced by the kinetic and 
potential parts of the Lagrangian, respectively\cite{Irwin}. 
The Skyrmion-Skyrmion or
nucleon-nucleon dynamics is then obtained by doing classical or quantum 
mechanics in the curved space of the manifold $M_{12}$. Unfortunately,
this program is not correct since the Skyrmions deform and do not stay 
in the form of a hedgehog
when they interact (obviously since they deform enough to merge into a torus),
and $M_{12}$ has no hope of describing the full dynamics in the $B=2$
sector. But one expects that the unstable manifold (\ie that obtained
by gradient flow curves) of the $B=2$ sector, which we 
shall call ${\cal M}_{12}$, and $M_{12}$
will become approximately equal for $R$ sufficiently large. 

Numerical studies have shown that the product ansatz give a surprisingly good 
approximation to ${\cal M}_{12}$ up to a separation of just a few 
$\mathrm{fm}$ (depending on the initial relative orientations of the 
Skyrmions). This enables one to extract useful information on
the Skyrmion-Skyrmion dynamics relatively easily. One can for 
instance compute an approximation
to the Skyrmion-Skyrmion potential even for moderate values of the separation
$R$. The separation is not 
a zero mode of the energy any longer, but it is a relatively ``soft'' 
mode such
that the potential energy does not vary by large values as a function of $R$,
as long as $R$ is large. 
In low energy scattering, the $R$ mode can be considered a ``slow'' mode
compared to the angular variables representing the rotation of particles 
which are ``fast'' variables. We will describe exactly such an approach in 
the next subsection. 

The first calculation of the Skyrmion-Skyrmion potential was obtained
by Skyrme himself and is defined by:
\begin{equation}
V_{\mathrm{SS}} = E_{\mathrm{B}=2} - 2 M_{\mathrm{S}}
\end{equation}
where $E_{\mathrm{B}=2}$ is the static energy of the system. This involves 
integrating Skyrmion profiles over all space, which is best done using an 
expansion in inverse powers of the separation $R$. To leading order one finds 
\begin{equation}
V_{\mathrm{SS}}(R,C) = - 4 \pi f_\pi^2 \kappa^2 
{
D(C)_{\mathrm{ii}} - 3 \hat R^{\mathrm{i}} D(C)_{\mathrm{ij}} 
\hat R^{\mathrm{j}}
\over R^3}
\end{equation}
which only depends on the relative coordinates: 2 angles and the 
distance $R$. 

Subsequently, there has been a great deal of work on 
the extraction of the nucleon-nucleon potential from the Skyrme model and
its comparison with traditional nuclear 
potentials\cite{Jackson-Jackson-Pasquier,Vinh,Walhout,Hosaka}. Early 
computations, mainly using the product ansatz, quickly showed that the 
general tensorial form of the potential so obtained was 
in good agreement with what is known of the nuclear force (1-2 pion exchange, 
repulsive core, etc.). However, a particularly 
disturbing defect haunted the problem 
for many years: the absence of a central attraction at medium range. This was
all the more disturbing since it is precisely this part of the nuclear force 
which binds nucleons together into nuclei. Many ans\"atze were tried 
generalizing the product ansatz: the modified product 
ansatz\cite{Jackson-Jackson-Pasquier} (the Skyrmions were
given the freedom to change their radii as they interacted, but with little
gain in the central attraction), the symmetrized product 
ansatz\cite{RiskaMPA} (a step in
the right direction since it was symmetric under the exchange of the particles,
and provided some central attraction but at the expense of the long range
behaviour), and more recently the instanton 
method\cite{Hosaka} (which combines the 
advantages of all the previous ans\"atze), but the crucial central attraction
still eluded all efforts. 

It was in 
two\cite{Verbaarschot-Walhout-Wambach-Wyld,Walhoutpot} 
exact, numerical analyses 
that a central attraction was found. The two approaches bifurcated in the way
that
they included physical, quantum nucleonic states. In the approach of 
Verbaarschot \etal\cite{Verbaarschot-Walhout-Wambach-Wyld} 
the potential is computed using numerical relaxation 
methods on a $20\times 20\times 40$ regular lattice for fixed separation and 
imposing only the discrete symmetries of the most attractive channel (
see below). The separation was defined to be the distance between the 
respective centers of baryon number density. They found a reasonable attraction
(roughly 70 MeV) in the central channel. In their approach, the physical 
nucleonic states where incorporated by treating the angular and iso-angular
coordinates as rigid quantum rotors, and the radial motion was treated
semi-classically
using the WKB method. They found that the kinetic energy of the angular and 
iso-angular motion was sufficient to destabilize any possible bound state. 
However the numerical analysis was recently repeated using a more refined 
method\cite{Wong},
the finite element approach, which permits an irregular lattice. In this way,
for the same amount of computing time, the algorithm can sample the critical 
regions with a finer grid, and consequently improve the accuracy. This work 
shows that indeed the central channel is sufficiently attractive to support a
bound state, the deuteron, within the WKB method for the radial degree of 
freedom.
On the other hand in the work of Walhout and Wambach\cite{Walhoutpot}, although
the potential was
again computed for static Skyrmionic configurations, the 
nucleon-nucleon potential was extracted by projection onto asymptotic quantum
nucleon states {\it \`a la} Jackson, Jackson and 
Pasquier\cite{Jackson-Jackson-Pasquier}. Here also
a central attraction was found, however the further analysis necessary to 
establish a bound state was not presented.
Walet and Amado\cite{Walet} further 
improved the results by including the $\Delta$ resonance as intermediate 
state, as well as some gluonic corrections. With all these corrections,
the thus obtained Nucleon-Nucleon potential
is indeed quite close to traditional (phenomenological) 
nuclear potentials, even though the 
arbitrariness in the definition of the distance between a pair of particles
when they are very deformed (at small $R$) makes it hard to judge.
Here we will only be interested in the long range part of the potential which
takes the form of the one-pion exchange potential. 

The tensorial form of the potential reveals how Skyrmions will react to 
their relative orientations. The different situations can be 
catalogued in three cases or channels. In the hedgehog-hedgehog channel (HH), 
$C=1$ and the potential is zero (no one-pion exchange potential). As
$R$ gets smaller then 
other interactions take over and build a repulsive core. In this channel, the
Skyrmions stay roughly hedgehog-like until $R$ is of the order of 1 
$\mathrm{fm}$. If the second Skyrmion is rotated by $\pi$ around the axis of
separation, $C=\mathrm i \tau^3$, then a strong repulsive interaction 
keeps the 
Skyrmions apart and the potential energy rises quickly as $R$ decreases. In 
this channel (REP), Skyrmions only stay roughly spherically symmetric for $R$ 
greater than about 1.5 $\mathrm{fm}$. The most interesting case is 
the so-called most attractive channel (MAC), where one Skyrmion is rotated 
relative to the other by $\pi$ around a direction perpendicular to the axis of
separation. In this channel, the Skyrmions always attract and come closer and 
closer together until they fuse into the toroidal configuration. This is the 
most
studied and phenomenologically interesting case since it is believed to contain
the nucleon-nucleon binding dynamics. Of course, in 
a realistic general collision
of nucleons, one does not stay in any one of these channels 
(spins are generally 
not correlated during real scattering) and situations should be hybrids of
the ones described so far. 

By freezing the relative iso-orientations, the two collective
coordinates represented by $C$ disappear, and the configurations then span a 
10-dimensional manifold: ${\cal M}^{\mathrm{HH}}_{10}$,
${\cal M}^{\mathrm{REP}}_{10}$ and ${\cal M}^{\mathrm{MAC}}_{10}$. 
${\cal M}^{\mathrm{HH}}_{10}$ and
${\cal M}^{\mathrm{REP}}_{10}$ are actually not very interesting since they 
would not represent very accurately the dynamics of the system. This is
because the HH and REP channels lead to relatively energetic states: 
the gradient on the manifold is large and the probability of exciting modes 
perpendicular to ${\cal M}^{\mathrm{HH}}_{10}$ and 
${\cal M}^{\mathrm{REP}}_{10}$ is not negligeable. 
It seems clear to us that more than
10 collective coordinates are needed to describe the system in these channels.
On the other hand, it seems reasonable that if the system has low enough 
energy,
it will all by itself steer clear of these regions and will not ``feel''
the existence of these higher energy states. It also seems 
physically sound that the system will spend most of its time near 
${\cal M}^{\mathrm{MAC}}_{10}$. There the energy gradients are much smaller 
(there is about a 4\% difference in energy between the torus state and a pair
of infinitely separated hedgehogs). The 
manifold ${\cal M}^{\mathrm{MAC}}_{10}$
has been extensively studied by Leese \etal\cite{Leese-Manton-Shroers} 
in connection with the study of the
deuteron system. 

The product ansatz in the most attractive
channel possesses the following discrete symmetries at infinite separation
(for two well separated Skyrmions on the $z$ axis, equidistant from the origin
 with one rotated relative to the other by 180$^\circ$ about the $y$ axis):
\begin{equation}
\vec \pi(-x,y,z) = \left(
\matrix{-1&0&0\cr
        0&1&0\cr
        0&0&1\cr}\right) 
\vec \pi(x,y,z)\label{eq:symdpa1}
\end{equation}
\begin{equation}
\vec \pi(x,-y,z) = \left(
\matrix{1&0&0\cr
        0&-1&0\cr
        0&0&1\cr}\right) 
\vec \pi(x,y,z)\label{eq:symdpa2}
\end{equation}
\begin{equation}
\vec \pi(x,y,-z) = \left(
\matrix{-1&0&0\cr
        0&1&0\cr
        0&0&1\cr}\right) 
\vec \pi(x,y,z).\label{eq:symdpa3}
\end{equation}
The last symmetry is actually valid for any separation while the first two 
are approximate and are exact only
in the $R\rightarrow+\infty$ limit. These discrete symmetries have been shown
to be exact for all configurations of ${\cal M}^{\mathrm{MAC}}_{10}$ including 
the torus, 
because they are conserved by the dynamics starting from $R=+\infty$ in the 
most attractive channel.

As the Skyrmions merge into the torus, the separation coordinate $R$ 
approaches a minimum value which depends
on the definition for the separation chosen (which is always arbitrary to a 
certain point, since when the Skyrmions deform and come close together, the
definition of separation becomes blurred). Another collective coordinate is
effectively eliminated by the appearance of the axial symmetry of the 
torus\cite{Braaten-Carson}:
\begin{equation}
U_{\mathrm{torus}}(\vec x) = e^{\mathrm{i} \theta\tau^3}
U_{\mathrm{torus}}(D( e^{-\mathrm{i} {\theta\over 2}\tau^3})\cdot\vec x)
e^{-\mathrm{i} \theta\tau^3}.
\end{equation}
This axial symmetry is created by the dynamics of the system and generalises 
the
discrete symmetries of equations $(\ref{eq:symdpa1})$, $(\ref{eq:symdpa2})$
and $(\ref{eq:symdpa3})$. This implies that the torus is
described by only 8 collective coordinates which parametrize an 8-dimensional 
manifold named ${\cal M}_8$. Manton shows that one can describe the 
topology of ${\cal M}_8$ somewhat more precisely. Let us factor out 
global translations
and global isorotations from ${\cal M}_8$, which are represented by a
factor of $\Rset^3\times S^3$. We are then left with 2 Euler angles, which
parametrize a 2-sphere. The appearance of such an $S^2$ on a manifold is called
a Bolt, in the theory of gravity\cite{Eguchi,Atiyah-Manton-2,Mantonskbps}. 
A Bolt is a kind of ``soft'' singularity
on a manifold. Atiyah and Manton further argue that because of the reflection
symmetry of the torus relative to the 
plane perpendicular to the axis of the torus
(the plane which ``slices'' the torus 
in two identical halves modulo a reflexion),  
the 2-sphere is really an $\Rset P_2$: a sphere with its antipodal points 
identified. This means that the two Euler angles over-define the orientation of
the axis of the torus. 
This is topologically similar to the BPS case where a torus 
state also exists in the winding number (magnetic charge) 2 sector.

The dibaryon is at the center of the most exotic dynamics of the $B=2$ sector.
Because of its hedgehog structure, see equation $(\ref{eq:dib})$, it has 
the same 6 
zero modes as the $B=1$ Skyrmion. Like all solutions of the theory, it also has
positive modes corresponding to small oscillations about the critical point. 
However, it also possesses negative modes since it is a saddle point of the
theory. It has long been suspected that the dibaryon is able to disintegrate
into a pair of Skyrmions or into the torus itself; this was numerically
checked only recently by Waindzoch and Wambach\cite{Waindzoch1}. 
In their 
articles they use a discretized version of the dibaryon as the initial 
configuration and then perturbed it
in order to study how the dibaryon is connected to the rest of the $B=2$ 
manifold. The negative modes of the dibaryon had already been investigated
analytically by Bang and Wirzba\cite{Bang-Wirzba} on a 3-sphere of 
radius $L$. The limit 
$L\rightarrow +\infty$ of their findings is in good agreement with those of
the numerical simulations. They solve a Schr\"odinger-type
equation for the perturbation of the dibaryon field in order to find the 
perturbations $\delta U$ which give the maximum negative energy gradients. This
gave them three magnetic modes and three electric modes, whose names 
refer to the properties of transformation of the perturbation under 
rotations. We have already discussed these negative modes in section 3.2.3.
The magnetic mode along the direction $z$ is 
parametrized as follows\cite{Waindzoch1}:
\begin{eqnarray}
\delta U^{\mathrm{M}}_{\mathrm{z}}&:& \delta\pi^0 = 0
\\
& & \delta\pi^1 = -g(r) {y\over r}
\\
& & \delta\pi^2 = g(r) {x\over r}
\\
& & \delta\pi^3 = 0
\end{eqnarray}
where $g(r)$ is a function computed numerically, the modes along the $x$ and
$y$ axes being obtained by cyclic permutations. This mode possesses a simple 
interpretation when one recognizes that the dibaryon configuration is very 
close to a pair of Skyrmions in the product ansatz, placed one on top of the
other. There are two ways to lower the energy of the resulting configuration: 
by translating or by rotating one Skyrmion relative to the other. 
The former represents the magnetic mode; there are three independent 
magnetic modes corresponding to the three orthogonal translation
directions present in $\Rset^3$. 
The latter represents the electric mode which is parametrized by a 
somewhat more complicated form\cite{Waindzoch1}:
\begin{eqnarray}
\delta U^{\mathrm{E}}_{\mathrm{z}}&:& \delta\pi^0 = - 3 a(r) {z \over r}
\\
& & \delta\pi^1 = \Biggl( c(r) - {b(r)\over 2}\Biggr) {z x\over r^2}
\\
& & \delta\pi^2 = \Biggl( c(r) - {b(r)\over 2}\Biggr) {z y\over r^2}
\\
& & \delta\pi^3 = 3 \Biggl( c(r) - {b(r)\over 2}\Biggr) {z^2\over r} +
{3 \over 2} b(r) 
\end{eqnarray}
for a perturbation along the $z$ axis (the profile functions $a(r)$, $b(r)$ 
and $c(r)$ are 
obtained numerically). The other two can be obtained by cyclic
permutations, $\delta U^{\mathrm{E}}_{\mathrm{x}}$ and 
$\delta U^{\mathrm{E}}_{\mathrm{y}}$ corresponding to rotations along the
$x$ and $y$ axis respectively. 

Ordinary time evolution of the dibaryon subject to these modes exhibits a 
fission process taking place. It can be numerically seen that for the 
perturbation 
$\delta U^{\mathrm{M}}_{\mathrm{z}}$, the dibaryon disintegrates in two 
Skyrmions moving along the $z$ axis. This negative mode is associated with a 
large energy gradient: the Skyrmions separate relatively quickly and oscillate
strongly. The electric mode $\delta U^{\mathrm{E}}$ is 
slightly less energetic and the Skyrmions
although moving away also along the $z$ axis, do so in a ``twisting'' fashion.
Large oscillations are also present here. 
Waindzoch and Wambach also studied these 
same disintegrations using the gradient flow   
method. The oscillations, which are transverse to the low energy unstable
manifold, are damped (see equations $(\ref{eq:damping})$ and
$(\ref{eq:damped})$) and it appears then that the
magnetic mode is tangent to the path leading to the product ansatz
in the HH channel, while the electric mode leads to the REP channel. This is
still a delicate issue since the numerical computations are very hard to do 
and the precision
required to confirm these results is quite high\cite{Waindzoch1}. 
Confirmation by analytic
methods would be most welcome. Waindzoch and Wambach
also show that the right superposition of perpendicular magnetic and 
electric
perturbations can direct the system along the path leading straight to the 
toroidal
configuration.

\begin{figure}
\centering
\hspace*{-1.5cm}
\mbox{\epsfig{figure= 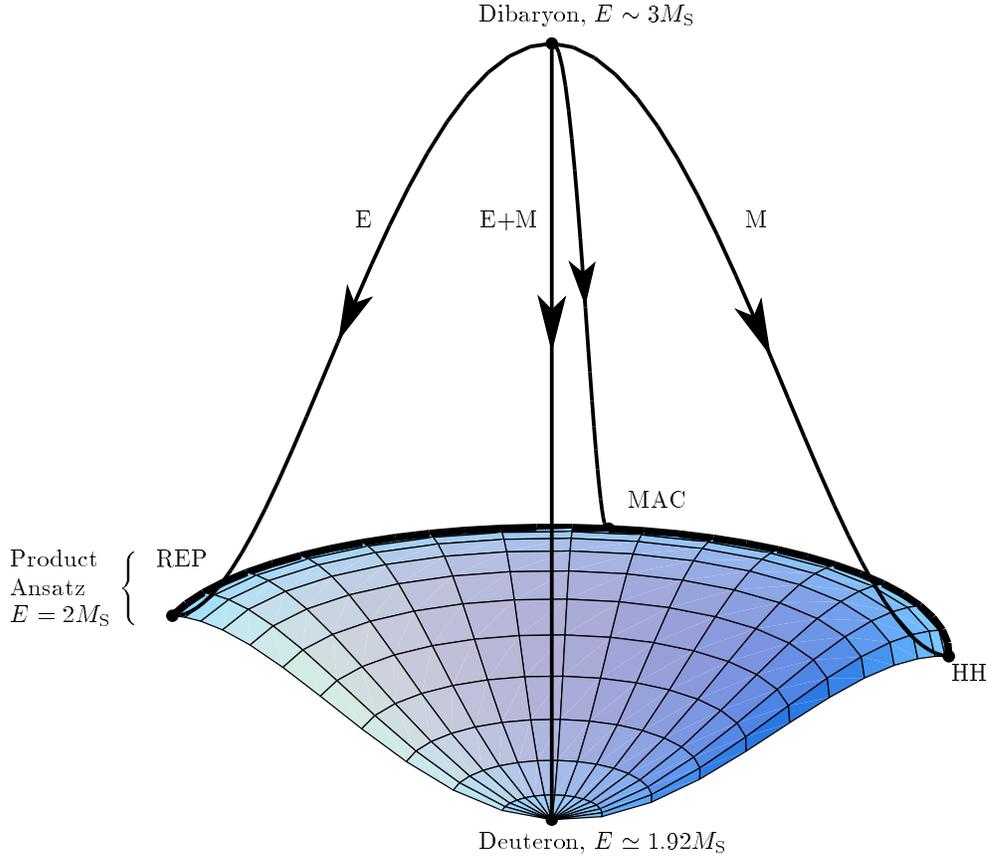,height=5.0in}}
\caption{Diagram of the $B=2$ manifold showing paths in configuration space
connecting the dibaryon to the product ansatz in the HH and REP channels
with paths tangent to the magnetic (M) and electric (E) modes respectively. 
The deuteron and the product ansatz MAC are connected to the dibaryon by 
paths comprising of an orthogonal superposition of E and M modes.} 
\label{fig:fig27}
\end{figure}

We are now ready to put the various parts together. Figure  
$\ref{fig:fig27}$ shows\break
schematically what the configuration space for the $B=2$ sector looks like. 
The 12-dimensional manifold
${\cal M}_{12}$ is the minimum required to describe the dynamics of a pair
of Skyrmions or nucleons. As we saw earlier, for Skyrmions infinitely far from
each other, these 12 collective coordinates can be identified as the individual
positions and iso-orientations of the separate solitons. These states can be
well represented by the product ansatz saddle point. The product ansatz comes
in three different iso-orientations, if one chooses to freeze such a degree of
freedom: hedgehog-hedgehog channel, repulsive channel  and the most attractive
channel whose collective coordinates parametrize the manifolds
${\cal M}^{\mathrm{HH}}_{10}$, ${\cal M}^{\mathrm{REP}}_{10}$ and
${\cal M}^{\mathrm{MAC}}_{10}$ respectively. The product ansatz in the most
attractive channel is linked to the toroidal configuration and its manifold
${\cal M}_{8}$ by gradient flow curves, while the HH and REP channels are
connected to the dibaryon by paths whose directions near the dibaryon are
tangent to its magnetic and electric modes respectively. The number of
collective coordinates along those paths changes from 10 to 6 because of the
appearance of the spherical symmetry of the dibaryon as well the freezing 
of the separation between the particles, removing 4 collective coordinates. 
The dibaryon is also directly connected to the torus as mentioned
earlier by paths with initial directions spanned by the superposition of
orthogonal magnetic and electric modes. Intermediate paths 
are generated by different combinations of these
6 modes. At the current rate of progress in this area, the relationship between
the initial conditions and the final states attained should soon be clear.
In the adiabatic approximation for Skyrmion dynamics, the paths can be 
followed in any direction
(there is no pion wave emission here) but this is not physically plausible
when energies differences of more than 1 GeV are concerned.

%% file: section43.tex
We now turn to the problem of nucleon-nucleon scattering as reported in 
Gisiger and Paranjape\cite{Gisiger1,Gisiger2} following the methods proposed 
by Manton\cite{Mantongfc}.

Scattering of Skyrmions at low energies should truncate to a dynamics
taking place on an unstable 
manifold linking together all of the low energy critical points. 
We consider the scattering in a particularly simplifying approximation. One 
observes that as the Skyrmions are separated to infinity, their low-energy 
behaviour corresponds to two independent, free Skyrmions, each with 3 
translational degrees of freedom and 3 rotational or isorotational degrees of 
freedom. The moduli space is simply two copies of $\Rset^3 \times S^3/Z_2$ 
and the induced metric is also the natural metric on these manifolds. As the 
separation between the Skyrmions decreases, interactions develop 
between the Skyrmions, both in the potential and in the kinetic energy 
(,\ie, the metric). These interactions can be written 
in a systematic perturbative expansion in 
inverse powers of the separation. 
In what follows, we will only consider scattering at large impact parameter,
for which the separation between the Skyrmions is always large. The 
perturbative expansion can therefore be truncated, and indeed 
our main approximation is to take only the 
leading term in this expansion. Surprisingly, this term comes from the 
expansion of the 
kinetic part of the Lagrangian and not the potential, and it has 
actually been overlooked by previous investigations.  
A secondary assumption that has been made and which we want to 
underline is that the pion mass is taken to be zero. This could perhaps 
explain why the interaction that we find has been missed: it comes from the 
two-pion exchange part of the interaction which is generally neglected
when considering massive pions. We complete our analysis 
by using the Bohr-Sommerfeld 
quantization method and a modification of the method of the variation of 
constants to calculate the scattering trajectories of the nucleons.

%% file: section431.tex
Written using the left-invariant one form already introduced previously 
(equation $(\ref{eq:1fru})$), the Skyrme Lagrangian takes the form:
\begin{eqnarray}
{\cal L}_{sk} = {f_\pi^2\over 2} {\cal R}(U)_\mu\cdot{\cal R}(U)^\mu
-{1\over{4 e^2}}\biggl [& {\cal R}(U)_\mu & \cdot{\cal R}(U)^\mu\;
{\cal R}(U)_\nu\cdot{\cal R}(U)^\nu -\nonumber
\\
& {\cal R}(U)_\mu & \cdot{\cal R}(U)^\nu\;{\cal R}(U)_\nu\cdot
{\cal R}(U)^\mu\biggr]
\end{eqnarray}
which, as we saw earlier, separates into a kinetic and a potential part: 
\begin{equation}
{\cal T}={f_\pi^2\over 2} {\cal R}_0\cdot{\cal R}_0 -
{1\over{2 e^2}}\biggl [{\cal R}_0\cdot{\cal R}_{\mathrm{i}}\;
{\cal R}_0\cdot{\cal R}_{\mathrm{i}} -
{\cal R}_0\cdot{\cal R}_0\;{\cal R}_{\mathrm{i}}\cdot
{\cal R}_{\mathrm{i}}\biggr ]
\end{equation}
\begin{equation}
{\cal V}=-{f_\pi^2\over 2} {\cal R}_{\mathrm{i}}\cdot{\cal R}_{\mathrm{i}} -
{1\over{4 e^2}}\biggl [{\cal R}_{\mathrm{i}}\cdot{\cal R}_{\mathrm{i}}\;
{\cal R}_{\mathrm{j}}\cdot{\cal R}_{\mathrm{j}} -
{\cal R}_{\mathrm{i}}\cdot{\cal R}_{\mathrm{j}}
\;{\cal R}_{\mathrm{j}}\cdot{\cal R}_{\mathrm{i}}\biggr ].
\end{equation}
The potential part has been extensively studied in the literature. It 
produces the following leading order term in the separation $d$ between the 
\break Skyrmions\cite{Skyrme,Jackson-Jackson-Pasquier,Isler}:
\begin{equation}
V = 2 M + 4\pi f_\pi^2 \kappa^2 {(1-\cos\theta) (3(\hat n\cdot\hat
d)^2-1)\over d^3}\label{eq:potsk}
\end{equation}
where $\hat d = \vec d/d$. We are interested by the kinetic part of the 
Lagrangian which is less understood. 

As noted earlier, to accurately describe the motion of the pair of Skyrmions 
we need the 
metric on the unstable manifold of the baryon number 2 sector, as well as the 
potential defined on it. Since both are not well known one must resort to 
approximating the pair of Skyrmions by a parametrization or ansatz. We 
chose the simplest parametrization for the system, the product ansatz, which 
maintains both solitons rigidly in the Skyrmion configuration at all time. 
Of course, 
since the particles deform when they come close to each other, our 
parametrization is only valid when the Skyrmions are far from each other. 
This is compatible with the low energy assumption we made and which is 
necessary to ensure that the degrees of freedom of the system can indeed be 
truncated to a finite (small) number. Then one takes
\begin{equation}
U(\vec x) = U_1^\dagger\; U_2
= A^\dagger U(\vec x - \vec R_1) A\;B^\dagger U(\vec x -\vec R_2) B.
\label{eq:ap}
\end{equation}
$A$ and $B$ are time dependent $\mathrm{SU}(2)$ matrices representing the 
orientation of the Skyrmions and $\vec R_1$ and $\vec R_2$ are their 
respective positions. A symmetrized 
product ansatz is possible, which respects the symmetry under exchange of the 
two particles\cite{RiskaMPA}, however such an elaboration does not affect the 
first 
order term in the mutual interaction. 

Replacing $U$ by $U_1^\dagger\; U_2$ in the kinetic part of 
the Skyrme Lagrangian $(\ref{eq:t1f})$ and working out the computations 
using the identity
\begin{equation}
{\cal R}^{\mathrm{a}}_\mu(U_1 U_2) = {\cal R}^{\mathrm{a}}_\mu(U_1) 
+ D_{\mathrm{ab}}(U_1)\;{\cal R}^{\mathrm{b}}_\mu(U_2)
\end{equation}
where $D_{\mathrm{ab}}(U)$ is
${1\over 2}\;\tr[\tau^{\mathrm{a}} U \tau^{\mathrm{b}} U^\dagger]$,
allows us to isolate the interaction term ${\cal T}_{\mathrm{int}}$ from 
contributions describing free individual particles.
The kinetic part of the Lagrangian then writes
\begin{equation}
{\cal T} = {\cal T}_1 + {\cal T}_2 + {\cal T}_{\mathrm{int}}
\end{equation} 
where 
\begin{eqnarray}
{\cal T}_{\mathrm{int}} = f_{\pi}^2\; & {\cal R}_0^1& \cdot D\cdot 
{\cal R}_0^2\nonumber
\\
+{1\over 2 e^2}
\Biggl[ &\bigl(&{\cal R}_0^1\cdot {\cal R}_0^1\; 
{\cal R}_{\mathrm{i}}^2\cdot {\cal R}_{\mathrm{i}}^2
+1\leftrightarrow 2\bigr)
+2\bigl({\cal R}_0^1\cdot{\cal R}_0^1+1\leftrightarrow 2\bigr) 
{\cal R}_{\mathrm{i}}^1\cdot D\cdot{\cal R}_{\mathrm{i}}^2\nonumber
\\
& + & 2\bigl( {\cal R}_{\mathrm{i}}^1\cdot {\cal R}_{\mathrm{i}}^2
+1\leftrightarrow 2\bigr) 
{\cal R}_0^1\cdot D\cdot{\cal R}_0^2
+4\;{\cal R}_0^1\cdot D\cdot{\cal R}_0^2\;
{\cal R}_{\mathrm{i}}^1\cdot D\cdot{\cal R}_{\mathrm{i}}^2\nonumber
\\
& - & \bigl({\cal R}_0^1\cdot D \cdot {\cal R}_{\mathrm{i}}^2\bigr)^2
 -\bigl({\cal R}_{\mathrm{i}}^1\cdot D\cdot {\cal R}_0^2\bigr)^2
-2\;{\cal R}_0^1\cdot {\cal R}_{\mathrm{i}}^1\;{\cal R}_0^2\cdot 
{\cal R}_{\mathrm{i}}^2\nonumber
\\
& - & 2 {\cal R}_0^1\cdot D \cdot{\cal R}_{\mathrm{i}}^2\;
{\cal R}_{\mathrm{i}}^1\cdot D\cdot{\cal R}_0^2\nonumber
\\
& &\qquad- 2\bigl({\cal R}_0^1\cdot
{\cal R}_i^1+1\leftrightarrow 2\bigr)
\bigl( {\cal R}_0^1\cdot D\cdot {\cal R}_{\mathrm{i}}^2 +
       {\cal R}_{\mathrm{i}}^1\cdot D\cdot {\cal R}_0^2\bigr)
\Biggr]\label{eq:tint}
\end{eqnarray}
where the dot implies contraction over isospin, ${\cal R}_\mu^1\equiv 
{\cal R}_\mu^{\mathrm{a}}(U_1)$, ${\cal R}_\mu^2\equiv 
{\cal R}_\mu^{\mathrm{a}}(U_2)$ and $D \equiv D_{\mathrm{ab}}(U_1)$ and
${\cal T}_1$ and ${\cal T}_2$ are similar to $(\ref{eq:t1f})$ for the fields 
$U_1$ and $U_2$ respectively.  

To obtain the full Lagrangian of the system we need to integrate the 
Lagrangian density over all space:
\begin{equation}
L_{\mathrm{int}} = \int\; {\d}^3 \vec x\; {\cal L}_{\mathrm{int}}.
\end{equation}
Since the Skyrmion profile function $f(r)$ is only known numerically, even 
though it falls to zero like $1/r^2$ for large $r$, it is not possible to  
compute the integral analytically. An expansion in inverse powers of the 
separation 
$d = ||\vec R_1-\vec R_2||$ between the Skyrmions is then used ($d$ was 
assumed to be large from the start in order for the product ansatz to be 
accurate). 

To do this we first write $U_1$ and $U_2$ as 
functions 
of the matrices $A$ and $B$ and the position vectors $\vec R_1$ and $\vec 
R_2$ using $(\ref{eq:ap})$ and the following identity:
\begin{eqnarray}
{\cal R}^{\mathrm{a}}_0(U_1)=\biggl(\delta^{\mathrm{ab}}-
D_{\mathrm{ab}}\bigl(AU(\vec x&-&\vec R_1)
A^\dagger\bigr) \biggr) \;{\cal R}^{\mathrm{a}}_0(A) -\nonumber
\\
& D_{\mathrm{ab}} & (A)\;\dot{R^{\mathrm{i}}_1}\;
{\cal R}^{\mathrm{b}}_{\mathrm{i}}\bigr(U(\vec x-\vec R_1)\bigl)
\label{eq:l0u1}
\end{eqnarray}
and correspondingly for $U_2$, $B$ and $\vec R_2$. We are interested in the 
leading contributions in inverse powers of $d$ of these tensors. Since the 
Skyrmion profile function $f(r)$ behaves like $\kappa/r^2$ at large 
separation, 
we get after a short calculation for large values of $|\vec x - \vec R_1|$:
\begin{eqnarray}
{\cal D}_{\mathrm{ab}}(U_1) &\equiv& \delta^{\mathrm{ab}}-
D_{\mathrm{ab}}\bigl(A\,U(\vec x-\vec R_1) \,A^\dagger\bigr) 
\\
&=& -2 \kappa {\epsilon^{\mathrm{abc}} D_{\mathrm{cd}}(A) 
\hat r_1^{\mathrm{d}}
\over |\vec x-\vec R_1|^2} + O\Biggl({1\over |\vec x - \vec R_1|^3}\Biggr)
\end{eqnarray}
and 
\begin{equation}
{\cal R}^{\mathrm{a}}_{\mathrm{i}}(A) = 
{\kappa\over |\vec x-\vec R_1|^3} \bigl( \delta^{\mathrm{ia}} - 
3 \hat r_1^{\mathrm{i}} \hat r_1^{\mathrm{a}} \bigr)
+ O\Biggl({1\over |\vec x - \vec R_1|^4}\Biggr)
\end{equation}
where $\hat r_1 = (\vec x-\vec R_1)/|\vec x - \vec R_1|$,
and correspondingly for $U_2$, $B$ and $\vec R_2$. To go further we will use 
a standard method for this type of calculation (for a very complete 
description of this method, its subtleties and its application to the 
computation of the Skyrmion-Skyrmion potential in the product ansatz see 
\cite{Isler}). 

\begin{figure}
\centering
\mbox{\epsfig{figure= 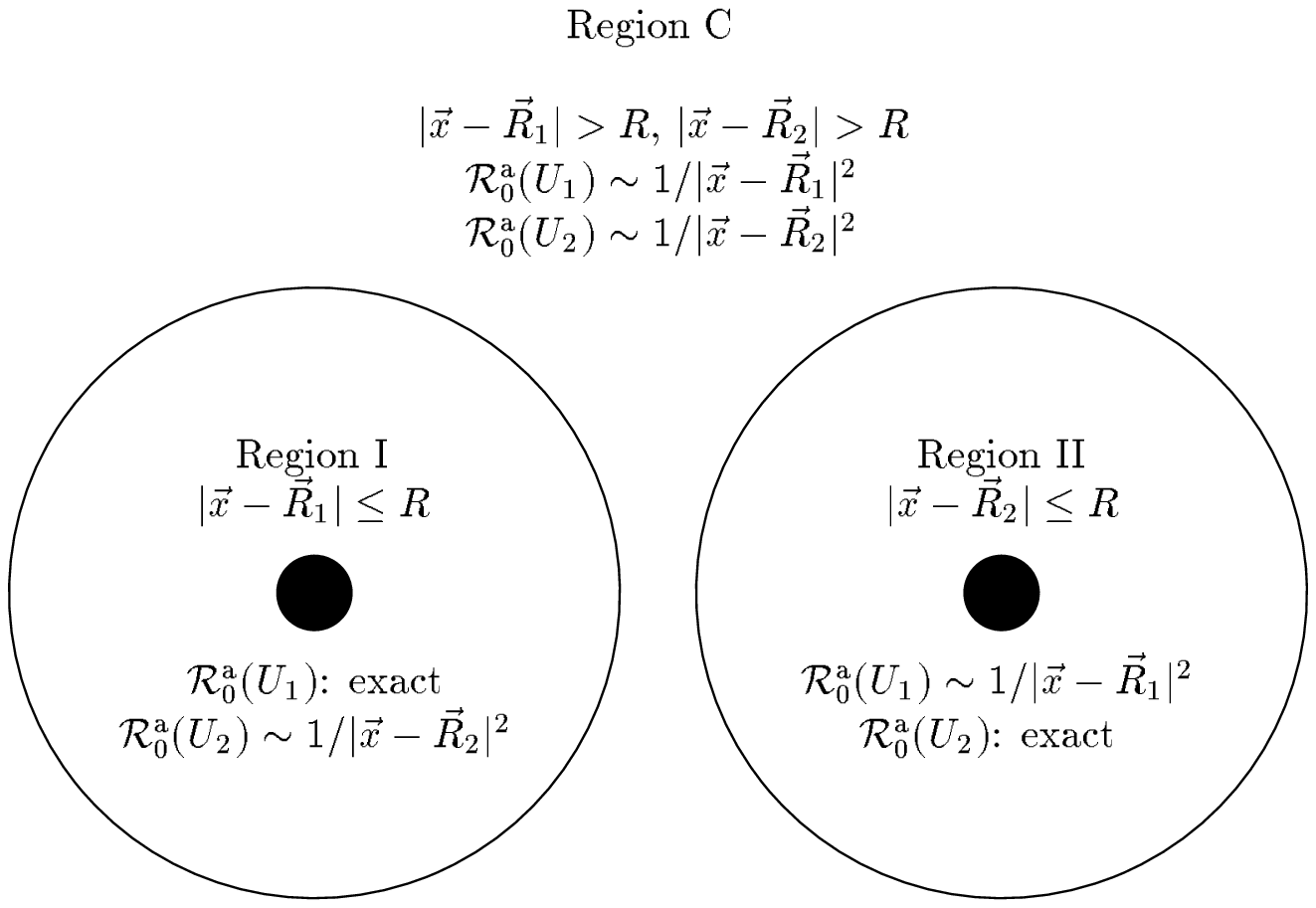,width=5.0in}}
\caption{Diagram of the three regions I, II and C used to divide space in 
order to compute the Lagrangian of a pair of Skyrmions as an expansion in 
inverse powers of the separation between the particles}\label{fig:fig28}
\end{figure}

In this method, we divide space in three regions $\mathrm{I}$, $\mathrm{II}$ 
and $\mathrm{C}$. $\mathrm{I}$ and $\mathrm{II}$ are regions of space close 
to the Skyrmions 1 and 2 respectively (see figure $\ref{fig:fig28}$). 
We define the notion of closeness by 
stating that these are regions where the profile function $f(r)$ 
is not well approximated by its 
asymptotic expression. For simplicity, and taking advantage of the spherical 
symmetry of the Skyrmion, we will take those regions to be spheres of some 
chosen radius $R$. Region $\mathrm{C}$ is the complementary region where both 
Skyrmion profiles fall off like $1/(\vec R_{\mathrm{i}}-\vec r)^2$ 
($i=1,\;2$). We then 
evaluate the integral over each region separately using the fact that in 
regions $\mathrm{I}$ and $\mathrm{II}$, one can use the ``exact'' expression 
for the Skyrmion profile, while in region $\mathrm{C}$ both Skyrmion profiles 
behave according to their asymptotic forms.

Let us start with the contributions from the non-linear $\sigma$ model, \ie 
the term quadratic in time derivatives. By using $(\ref{eq:lsk1f})$ and 
$(\ref{eq:l0u1})$ we easily find that:
\begin{eqnarray}
&&\!\!\!\!\!\!\!\!\!\!{\cal R}^0_{\mathrm{a}}(U_1)\;
D_{\mathrm{ab}}(U_1)\;
{\cal R}^0_{\mathrm{c}}(U_2)=\nonumber
\\
&&{\cal R}_0^{\mathrm{c}}(A) {\cal D}_{\mathrm{ca}}(U_1^\dagger) 
\;D_{\mathrm{ab}}(U_1)\; {\cal D}_{\mathrm{bd}}(U_2) {\cal R}_0^{\mathrm{d}}(B)
\nonumber
\\
&-&{\cal R}_0^{\mathrm{c}}(A) {\cal D}_{\mathrm{ca}}(U_1^\dagger)
\;D_{\mathrm{ab}}(U_1)\; D_{\mathrm{bd}}(B) \dot R_2^{\mathrm{j}} 
{\cal R}_{\mathrm{j}}^{\mathrm{d}}(U(\vec x-\vec R_2))
\nonumber
\\
&-&{\cal R}_{\mathrm{i}}^{\mathrm{c}}(U(\vec x-\vec R_1))
\dot R_1^{\mathrm{i}} D_{\mathrm{ca}}(A^\dagger) \;D_{\mathrm{ab}}(U_1)\;
{\cal D}_{\mathrm{bd}}(U_2) {\cal R}_0^{\mathrm{d}}(B)
\nonumber
\\
&+&{\cal R}_{\mathrm{i}}^{\mathrm{c}}(U(\vec x-\vec R_1))
\dot R_1^{\mathrm{i}} D_{\mathrm{ca}}(A^\dagger) \;D_{\mathrm{ab}}(U_1)\;
D_{\mathrm{bd}}(B) \dot R_2^{\mathrm{j}} 
{\cal R}_{\mathrm{j}}^{\mathrm{d}}(U(\vec x -\vec R_2))
\end{eqnarray}
When integrating this expression over region $\mathrm{I}$ we can use the 
asymptotic expansion for the tensors of Skyrmion 2, giving 
\begin{eqnarray}
\int_{\mathrm{I}}&& {\d}^3\,\vec x\; {\cal R}^0_{\mathrm{a}}(U_1)\,
D_{\mathrm{ab}}(U_1)\,
{\cal R}^0_{\mathrm{b}} (U_2) =\nonumber
\\ 
& & - 2 \kappa \int_{|\vec x- \vec R_1|\le R}\!\!\!{\d}^3\,\vec x\; 
{ {\cal R}_0^{\mathrm{a}}(A) {\cal D}_{\mathrm{ab}}(U_1^\dagger) 
D_{\mathrm{bc}}(U_1) \epsilon^{\mathrm{cde}} D_{\mathrm{ef}}(B) \hat 
r_2^{\mathrm{f}} {\cal R}_0^{\mathrm{d}}(B) \over |\vec x - \vec R_2|^2}
+\nonumber
\\
& & + 2 \kappa \int_{|\vec x- \vec R_1|\le R}\!\!\!\!\!{\d}^3\,\vec x\; 
{ {\cal R}_{\mathrm{i}}^{\mathrm{a}}(U(\vec x-\vec R_1)) \dot R^{\mathrm{i}}_1
D_{\mathrm{ab}}(A^\dagger) D_{\mathrm{bc}}(U_1) \epsilon^{\mathrm{cde}}
D_{\mathrm{ef}}(B) \hat r_2^{\mathrm{f}} {\cal R}_0^{\mathrm{d}}(B) \over 
|\vec x - \vec R_2|^2} \nonumber
\\
& & \qquad\quad+ O\biggl({1\over |\vec x - \vec R_2|^3}\biggr).
\end{eqnarray}
By translating the integration variable by $\vec R_1$, and defining a new 
integration variable $\vec y = \vec x - \vec R_1$, we easily see that this 
integral contributes at most terms of order $1/d^2$. 
Integration over region $\mathrm{II}$ yields the same result by 
symmetry. Integrating over the complementary region $\mathrm{C}$ is 
different since tensors from both Skyrmions take their asymptotic form and 
we find to leading order:
\begin{eqnarray}
\!\!\!\!\!\!&&\int_{\mathrm{C}} {\d}^3\, \vec x\;{\cal R}^0_{\mathrm{a}}(U_1)\;
D_{\mathrm{ab}}(U_1)\;
{\cal R}^0_{\mathrm{b}} (U_2) = \nonumber
\\
& & 4 \kappa^2 \int_{|\vec x - \vec R_1|\ge R,\; |\vec x - \vec R_2|\ge R} 
{\d}^3\,\vec x\;
{ 
\epsilon^{\mathrm{aeb}} D_{\mathrm{ef}}(A) \hat r_1^{\mathrm{f}} 
{\cal R}_0^{\mathrm{b}}(A) \epsilon^{\mathrm{adg}} {\cal R}_0^{\mathrm{d}}(B) 
D_{\mathrm{gh}}(B) \hat r_2^{\mathrm{h}}
\over |\vec x - \vec R_1|^2 |\vec x - \vec R_2|^2
}  + \ldots
\\
= & & 4 \kappa^2 \int_{\mathrm{I+II+C}} 
{\d}^3\,\vec y 
{
{\cal R}_0^{\mathrm{a}}(A) {\cal R}_0^{\mathrm{b}}(B)
[ D_{\mathrm{ac}}(B) \hat y^{\mathrm{c}} D_{\mathrm{bd}}(A) 
\widehat{(\vec y-\vec d)}^{\mathrm{d}}
- \widehat{(\vec y-\vec d)}^{\mathrm{c}} D_{\mathrm{cd}}(A^\dagger B) 
\hat y^{\mathrm{d}} \delta^{\mathrm{ab}}]
\over y^2 |\vec y - \vec d|^2
}\nonumber
\\
& & \qquad\qquad\qquad + \dots\label{eq:exp}
\\
= & & {4 \kappa^2 \over d} \int {\d}^3\,\vec z 
{
{\cal R}_0^{\mathrm{a}}(A) {\cal R}_0^{\mathrm{b}}(B)
[ D_{\mathrm{ac}}(B) \hat z^{\mathrm{c}} D_{\mathrm{bd}}(A) 
\widehat{(\vec z - \hat d)}^{\mathrm{d}} -
\widehat{(\vec z -\hat d)}^{\mathrm{c}} D{\mathrm{cd}}(A^\dagger B) \hat 
z^{\mathrm{d}} \delta^{\mathrm{ab}}]
\over z^2 |\vec z-\hat d|^2 }\nonumber
\\
&&\qquad\qquad\qquad\qquad+ O\biggl({1\over d^2}\biggr)\label{eq:unsurd}
\end{eqnarray}
where $\vec y = \vec x - \vec R_2$ and $\vec z=\vec y/d$.
To obtain $(\ref{eq:exp})$, we used the fact that expanding the integration 
bounds from 
$\mathrm{C}$ to the whole space only adds higher order contributions to the 
integral, but does not alter the value of the leading order terms (see 
\cite{Isler}
for details). The last line $(\ref{eq:unsurd})$ is obtained by shifting the 
integration variable by $R_1$ and absorbing 3 factors of $d$ in the measure. 
This is possible since we are integrating over all space. 

The fact that the main contribution comes from 
the faraway region might at first sight seem a bit strange, but it should be 
kept in mind that 
this same region makes a contribution in the analogous computation
of the 
potential $(\ref{eq:potsk})$, which is as important as the contribution from
the regions close to the Skyrmions 
(see \cite{Isler} for more details on that point). 

Similar computations are feasible for the Skyrme term, but because of its 
structure (quartic in derivatives), it only contributes at higher order. 
The leading contribution from the kinetic part of the Lagrangian is then of 
order $1/d$ while the potential part is 
of order $1/d^3$ rendering the latter, in principle, negligible compared to 
the contribution from the kinetic energy. Finally we obtain  
the following expansion of the Lagrangian for a pair of Skyrmions
far from each other in the product ansatz, using the expression 
$(\ref{eq:l1sk})$ for single Skyrmion Lagrangians:
\begin{eqnarray}
L = -2 &M& + {1\over 4} M \dot{\vec d^{\;2}} + 2 \Lambda 
\bigl({\cal R}^{\mathrm{a}}(A)\,
{\cal R}^{\mathrm{a}}(A) + {\cal R}^{\mathrm{a}}(B)\,
{\cal R}^{\mathrm{a}}(B)\bigr)\nonumber
\\
&+& {\Delta\over d} \epsilon^{\mathrm{iac}}\epsilon^{\mathrm{jbd}}\;
{\cal L}^{\mathrm{a}}(A)\,{\cal L}^{\mathrm{d}}(B)\;
\bigl(\delta^{\mathrm{ij}}-\hat{d}^{\mathrm{i}} 
\hat{d}^{\mathrm{j}}\bigr)\, D_{\mathrm{ab}}(A^\dagger B) +O(1/d^2)
\label{eq:l2sk}
\end{eqnarray}
where $\hat d = \vec d/d$, $\Delta=2 \pi \kappa^2 f_\pi^2$. 
${\cal L}^{\mathrm{a}} (A)\equiv {\cal L}^{\mathrm{a}}_0(A)$,
${\cal R}^{\mathrm{a}} (A)\equiv {\cal R}^{\mathrm{a}}_0(A)$ and we have used 
the relation ${\cal L}^{\mathrm{a}}(A)=D_{\mathrm{ab}}(A)\;
{\cal R}^{\mathrm{b}}(A)$. 

By keeping only the leading order terms in $1/d$, we neglect any contribution 
from the potential part of the original Lagrangian, and the motion is then 
completely specified by the geodesics of the metric induced by the kinetic 
term onto the unstable manifold of the baryon number two sector, 
as specified in subsection 4.3. 

The term in $1/d$ in $(\ref{eq:l2sk})$, though absent from the literature 
of the 
80's, was also independently obtained by Schroers\cite{Schroers}.  He found a 
leading contribution which behaves as $1/d$ and even calculated sub-leading 
spin-orbit coupling terms. The only other comparable calculation to our 
knowledge has been done by Walhout and Wambach\cite{Walhout} for the case of 
massive pions. The limit as $m_\pi\rightarrow 0$ of their
expression, however, does not leave a term which behaves as $1/d$ and hence
does not reproduce our result. We believe
that this contribution should come also from their evaluation of the integral
giving the induced kinetic energy in the faraway region (region 
$\mathrm{C}$) and then
recovering our result as $m_\pi\rightarrow 0$. Such a contribution would also
be proportional to $(e^{-m_\pi d})^2$, what they call ``two pion exchange''.
We also add that this $1/d$ term is of leading order in an expansion in inverse
separation with respect to a scale determined by $f_\pi$ and $e$ which has 
nothing to do with the length scale set by the pion mass.

The Skyrmion-Skyrmion Lagrangian we obtained describes dynamics in a 
12-dimensional 
moduli space via equations of motion which are highly non-linear 
and quite complex. The Lagrangian possesses several symmetries, and 
associated conserved quantities such as total isospin which is connected to 
invariance under left isorotation
\begin{equation}
A\rightarrow CA \qquad {\mathrm{and}}\qquad B\rightarrow CB
\end{equation}
and the total angular momentum related to invariance under the 
following operation:
\begin{equation}
A \rightarrow AC\qquad{\mathrm{and}}\qquad B \rightarrow BC\qquad{\mathrm{and}}
\qquad d^{\mathrm{a}} \rightarrow D_{\mathrm{ab}}(C^\dagger) d^{\mathrm{b}}
\end{equation}
(where $C$ is a constant $SU(2)$ matrix). However they are not much help in 
simplifying 
the equations of motion or even solving them. In order to go further, and to 
keep numerical analysis to a minimum, we use the perturbation method of 
Lagrange\cite{Goldstein} familiar in celestial mechanics to compute 
approximations to the equations of motion. We will describe how this 
perturbative scheme works.

We are dealing here with a system described by a Lagrangian of purely kinetic 
nature which we will note:
\begin{eqnarray}
L &=& T_1 + T_2 + T_{\mathrm{I}}
\nonumber
\\
&\equiv& T_0 + T_{\mathrm{I}}
\end{eqnarray}
where $T_0 = T_1+T_2$ is the free Lagrangian, $T_{\mathrm{I}}$ is the 
interaction part and
$T_{\mathrm{I}}\ll T_0$ because of some small factor in $T_{\mathrm{I}}$ 
($1/d$ in this case). 
Let $q_{\mathrm{i}}^0$ denote the free coordinates of the system, which we
take to completely describe its state. In our case $q_{\mathrm{i}}^0$ will 
represent 
$d^{\mathrm{i}}$, $A$ or $B$. The free canonically conjugate variables 
are defined as
\begin{equation}
p^0_{\mathrm{i}} = {\partial\over \partial {\dot q}^0_{\mathrm{i}}}
T_0(q^0_{\mathrm{i}},{\dot q}^0_{\mathrm{i}}).
\end{equation}
Here $\dot q_{\mathrm{i}}^0$ are taken to be $\dot d^{\mathrm{i}}$, 
${\cal R}^{\mathrm{i}}(A)$, ${\cal R}^{\mathrm{i}}(B)$, 
${\cal L}^{\mathrm{i}}(A)$ and ${\cal L}^{\mathrm{i}}(B)$. In the free system, 
which is described by $T_0$, these quantities are conserved. Indeed, without 
the interaction term, the Lagrangian describes a pair of free, spherically 
symmetrical tops moving and spinning at constant velocity. 

Adding $T_{\mathrm{I}}$ to $T_0$ complicates things and removes those 
conservation laws. Even though the system is still described by the same 
coordinates 
$q_{\mathrm{i}}= q_{\mathrm{i}}^0$, the canonically conjugate variables 
are changed; we find
\begin{eqnarray}
p_{\mathrm{i}} &=& {\partial\over \partial {\dot q}_{\mathrm{i}}}\Bigl (
T_0(q_{\mathrm{i}},{\dot q}_{\mathrm{i}}) + 
T_{\mathrm{I}}(q_{\mathrm{i}},{\dot q}_{\mathrm{i}})\Bigr)
\nonumber
\\
&=& p_{\mathrm{i}}(q^0_{\mathrm{j}},
p^0_{\mathrm{j}}) = p^0_{\mathrm{i}} + \Delta p_{\mathrm{i}}
\end{eqnarray}
where $\Delta p_{\mathrm{i}}$ has a well defined expansion in $1/d$. The aim 
of the method is to find an accurate expansion in $1/d$ of the equations of 
motion of the system, enabling one to only keep the dominant terms. This is 
done using Poisson brackets. 
Let us denote by $C^{\mathrm{k}}(q^0,p^0)$ quantitites which are 
conserved in the free system. In our case, they will be 
$\dot d_{\mathrm{i}}$, ${\cal R}^{\mathrm{i}}(A)$, 
${\cal R}^{\mathrm{i}}(B)$, ${\cal L}^{\mathrm{i}}(A)$ and 
${\cal L}^{\mathrm{i}}(B)$ but they could also represent more complicated 
functions of these quantities. $C^{\mathrm{k}}(q,p)$ 
will no longer necessarily be conserved and they can be  
separated in a free and an interacting part as follows:
\begin{eqnarray}
C^{\mathrm{k}}(q_{\mathrm{i}},p_{\mathrm{i}}) & = & 
C^{\mathrm{k}}(q_{\mathrm{i}},p_{\mathrm{i}}^0 + \Delta p_{\mathrm{i}})
\nonumber
\\
& \simeq & C^{\mathrm{k}}(q_{\mathrm{i}}^0,p_{\mathrm{i}}^0) +
{\partial\over \partial p_{\mathrm{i}}^0}
C^{\mathrm{k}}(q_{\mathrm{i}}^0,p_{\mathrm{i}}^0) \Delta p_{\mathrm{i}}.
\end{eqnarray}
The interacting part can be written in a $1/d$ expansion. The 
time derivative of $C^{\mathrm{k}}$ is given by the Poisson bracket of 
$C^{\mathrm{k}}$ with the Hamiltonian of the system:
\begin{equation}
{\d \over \d t} C^{\mathrm{k}} = \{ C^{\mathrm{k}},H\}\label{eq:ddtck}
\end{equation}
where $H = H_0 + H_{\mathrm{I}} = \sum_{\mathrm{i}} q_{\mathrm{i}} 
p_{\mathrm{i}} - (T_0 + T_{\mathrm{I}}).$ $H_{\mathrm{I}}$ being suppressed by
a factor of $1/d$ relative to $H_0$, writing $H$ as $H_0+H_{\mathrm{I}}$ 
also provides an expansion in this small parameter. Finally the Poisson 
bracket itself can be written as an expansion in $1/d$. Indeed, since:
\begin{equation}
{\partial\over\partial q_i} = {\partial\over\partial q^0_i}\quad \hbox{and}
\quad
{\partial\over\partial p_i}=\sum_j \Bigl({\partial p^0_j\over \partial
p_i} \Bigr) {\partial\over\partial p^0_j}
={\partial\over\partial p^0_i} - \sum_j
\Bigl({\partial\Delta p_j\over\partial p_i} \Bigr) {\partial\over\partial p^0_j}
\end{equation}
then
\begin{eqnarray}
&\{& A(q,p),B(q,p)\} = \sum_{\mathrm{i}}
{\partial A(q,p)\over\partial q_{\mathrm{i}}}
{\partial B(q,p)\over\partial p_{\mathrm{i}}} -
{\partial A(q,p)\over\partial p_{\mathrm{i}}}
{\partial B(q,p)\over\partial q_{\mathrm{i}}}\nonumber
\\
&=& \sum_{\mathrm{i}} \Biggl [
{\partial A\over\partial q^0_{\mathrm{i}}}
{\partial B\over\partial p^0_{\mathrm{i}}} -
{\partial A\over\partial p^0_{\mathrm{i}}}
{\partial B\over\partial q^0_{\mathrm{i}}}
-\sum_{\mathrm{j}} \Bigl (
{\partial A\over\partial q^0_{\mathrm{i}}}
{\partial B\over\partial p^0_{\mathrm{i}}} -
{\partial A\over\partial p^0_{\mathrm{i}}}
{\partial B\over\partial q^0_{\mathrm{i}}}
\Bigr )
\; \Bigl( {\partial \Delta p_{\mathrm{j}}\over\partial p_{\mathrm{i}}} \Bigr )
\Biggr ]\Biggr|_{(q^0,p^0+\Delta p)}\nonumber
\\
&=& \{ A,B\}_0 - \sum_{{\mathrm{i}},{\mathrm{j}}} \Bigl (
{\partial A\over\partial q^0_{\mathrm{i}}}
{\partial B\over\partial p^0_{\mathrm{j}}} -
{\partial A\over\partial p^0_{\mathrm{i}}}
{\partial B\over\partial q^0_{\mathrm{i}}} \Bigr )
\Bigl ({\partial \Delta p_{\mathrm{j}}\over\partial p_{\mathrm{i}}} \Bigr )
\nonumber
\\
&\equiv& \{ A,B\}_0 + \{A,B\}_{\mathrm{I}}
\end{eqnarray}
Replacing these expressions in $(\ref{eq:ddtck})$, we find the following 
expansion in $1/d$ of the time derivative of $C^{\mathrm{k}}$:
\begin{eqnarray}
{\d\over \d t} C^{\mathrm{k}} &=& \{C^k(q^0,p^0),H_0(q^0,p^0)\}_{\mathrm{I}}
+ \{C^{\mathrm{k}}(q^0,p^0),H_{\mathrm{I}}(q^0,p^0)\}_0+\nonumber
\\
&& \{ {\partial\over\partial p^0_{\mathrm{i}}} 
C^{\mathrm{k}}(q^0,p^0)\Delta p_{\mathrm{i}},H_0(q^0,p^0)\}_0 +\nonumber
\\
&&\{ C^{\mathrm{k}}(q^0,p^0),{\partial\over\partial p^0_{\mathrm{i}}}
H_0(q^0,p^0)\Delta p_{\mathrm{i}}\}_0+\cdots\label{eq:ddtckapp}
\end{eqnarray}
since $\{C^{\mathrm{k}},H_{\mathrm{I}}\}_{\mathrm{I}}$ is automatically of 
higher order in $1/d$ and $\{C^{\mathrm{k}},H_0\}_0=0$ exactly. The only 
asumption made here is that $d$ is large. 

For the case $C^{\mathrm{k}}=\dot d^{\mathrm{k}}$, since $H_{\mathrm{I}}$ 
does not depend on $\dot d^{\mathrm{k}}$, 
the last three terms in $(\ref{eq:ddtckapp})$ vanish, thus
\begin{eqnarray}
{\d\over \d t}\dot d^{\mathrm{k}} & = & \{\dot d^{\mathrm{k}},
H_{\mathrm{I}}\}_0
\nonumber
\\
& = & -{2\Delta\over M d^2}\biggr[\delta^{\mathrm{ij}}
\hat d^{\mathrm{k}} + \delta^{\mathrm{jk}}\hat
d^{\mathrm{i}} +\delta^{\mathrm{ik}}\hat d^{\mathrm{j}} - 3\hat 
d^{\mathrm{i}} \hat d^{\mathrm{j}} \hat d^{\mathrm{k}}\biggl]\times\nonumber 
\\
& & \quad\qquad\qquad\qquad\epsilon^{\mathrm{iac}} \epsilon^{\mathrm{jbd}} 
{\cal L}^{\mathrm{c}}(A) 
{\cal L}^{\mathrm{d}}(B) D_{\mathrm{ab}}(A^\dagger B)\label{eq:ddot}
\end{eqnarray}
using the free Poisson brackets of the free system:
\begin{eqnarray}
\{ d^{\mathrm{i}},\Pi^{\mathrm{j}}\} & = & \delta^{\mathrm{ij}}
\\
\{{\cal R}^{\mathrm{a}}(A),{\cal R}^{\mathrm{b}}(A)\} & = & 
-{1\over 2\Lambda} \epsilon^{\mathrm{abc}} {\cal R}^{\mathrm{c}}(A)
\\
\{{\cal L}^{\mathrm{a}}(A),{\cal L}^{\mathrm{b}}(A)\} & = & {1\over 2\Lambda}
\epsilon^{\mathrm{abc}} {\cal L}^{\mathrm{c}}(A)
\\
\{{\cal R}^{\mathrm{a}}(A),{\cal L}^{\mathrm{b}}(A)\} & = & 0
\\
\{{\cal R}^{\mathrm{a}}(A),D_{\mathrm{bc}}(A)\} & = & -{1\over 2\Lambda}
\epsilon^{\mathrm{abd}} D_{\mathrm{dc}}(A)
\\
\{{\cal L}^{\mathrm{a}}(A),D_{\mathrm{bc}}(A)\} & = & {1\over 2\Lambda} 
\epsilon^{\mathrm{acd}}D_{\mathrm{db}}(A)
\end{eqnarray}
where $\Pi^{\mathrm{i}}$ are the conjugate momenta to $d^{\mathrm{i}}$. 
(Because of the symmetric
nature of the free Hamiltonian, the same brackets are true if we replace $A$ 
by $B$ everywhere. Furthermore 
all the mixed brackets between $A$ and $B$ are zero.)

In the case where $C^{\mathrm{k}}$ is ${\cal R}^{\mathrm{k}}(A)$, 
${\cal L}^{\mathrm{k}}(A)$, ${\cal R}^{\mathrm{k}}(B)$
or ${\cal L}^{\mathrm{k}}(B)$, we get
\begin{eqnarray}
{\d\over \d t}{\cal R}^{\mathrm{k}}(A) & = &  {\Delta\over 2 M d} 
\epsilon^{\mathrm{iac}} \epsilon^{\mathrm{jbd}}
{\cal L}^{\mathrm{c}}(A) {\cal L}^{\mathrm{d}}(B) 
\bigl(\delta^{\mathrm{ij}}-\hat d^{\mathrm{i}}\hat d^{\mathrm{j}}\bigr)
\epsilon^{\mathrm{kef}} D_{\mathrm{fa}}(A) D_{\mathrm{eb}}(B)\nonumber
\\
& &\qquad\qquad\qquad + \cdots\label{eq:ladot}
\\
{\d\over \d t}{\cal R}^{\mathrm{k}}(B) & = &  {\Delta\over 2 M d} 
\epsilon^{\mathrm{iac}} \epsilon^{\mathrm{jbd}}
{\cal L}^{\mathrm{c}}(A) {\cal L}^{\mathrm{d}}(B) 
\bigl(\delta^{\mathrm{ij}}-\hat d^{\mathrm{i}}\hat d^{\mathrm{j}}\bigr)
\epsilon^{\mathrm{kef}} D_{\mathrm{ae}}(A^\dagger)D_{\mathrm{fb}}(B)\nonumber
\\
& & \qquad\qquad\qquad + \cdots\label{eq:lbdot}
\\
{\d\over \d t}{\cal L}^{\mathrm{k}}(A) & = & - {\Delta\over 2 M d}
\epsilon^{\mathrm{iac}} \epsilon^{\mathrm{jbd}}
{\cal L}^{\mathrm{d}}(B)\bigl(\delta^{\mathrm{ij}}-
\hat d^{\mathrm{i}}\hat d^{\mathrm{j}}\bigr)\times\nonumber
\\
& & \qquad\biggl[\epsilon^{\mathrm{kcf}} {\cal L}^{\mathrm{f}}(A) 
D_{\mathrm{ab}}(A^\dagger B) +
\epsilon^{\mathrm{kaf}} D_{\mathrm{fb}}(A^\dagger B) 
{\cal L}^{\mathrm{c}}(A)\biggr] + \cdots\label{eq:radot}
\\
{\d\over \d t}{\cal L}^{\mathrm{k}}(B) & = & - {\Delta\over 2 M d}
\epsilon^{\mathrm{iac}} \epsilon^{\mathrm{jbd}}
{\cal L}^{\mathrm{c}}(A)
\bigl(\delta^{\mathrm{ij}}-\hat d^{\mathrm{i}}\hat d^{\mathrm{j}}\bigr)\times
\nonumber
\\
& &\qquad \biggl[\epsilon^{\mathrm{kdf}} 
{\cal L}^{\mathrm{f}}(B) D_{\mathrm{ab}}(A^\dagger B) +
\epsilon^{\mathrm{kbf}} D_{\mathrm{af}}(A^\dagger B) 
{\cal L}^{\mathrm{d}}(B)\biggr] + \cdots\label{eq:rbdot}
\end{eqnarray}
where $\{C^{\mathrm{k}},H_{\mathrm{I}}\}_0$ is exhibited and the dots 
represent the remaining very complicated terms which are non zero and 
actually are not negligible, being of the same order in $1/d$ as 
$\{C^{\mathrm{k}},H_{\mathrm{I}}\}_0$.

Our approximation is reliable as long as the separation
$d$ between the particles is large enough for the conjugate momenta to stay
close to their free values. As we have already worked with the undeformed
product ansatz approximation, and neglected the potential, which are both valid
for large $d$, we feel confident that we have not lost any meaningful 
information by making this further approximation. If $d$ is kept large we 
should then find geodesics similar (qualitatively at least) to those given by 
the exact equations of motion. 

The system of equations $(\ref{eq:ddot})$, 
$(\ref{eq:ladot})$, $(\ref{eq:lbdot})$, $(\ref{eq:radot})$ and 
$(\ref{eq:rbdot})$ is still quite complicated and we will treat it with one 
final approximation method, namely the method of variation of constants.
This perturbation scheme consists of replacing the variables in the right hand
side of these ``conservation equations'' by their free trajectories.
This gives rise to a ``variation'' of the previously conserved
``constants''. The procedure can be iterated indefinitely to give higher order
corrections.
One should however always maintain consistency with the first (Lagrange)
approximation. The range of validity of this further approximation
is rather hard to define, but it is clear that only slowly varying trajectories
in phase space with large $d$ can be considered. The method of
``variation of constants'' is only useful with respect
to the equation for $\vec d$, where it gives the scattering trajectory. The
change in the spin or the isospin governed by equations $(\ref{eq:ladot})$, 
$(\ref{eq:lbdot})$, $(\ref{eq:radot})$ and $(\ref{eq:rbdot})$ cannot
be treated with this approximation method because of the long range nature
of the interaction. The results give an infinite change in these angular
momenta, which is not reliable. We will get back to this problem in the last 
subsection of this article.

%% file: section432.tex
We now use the semi-classical Bohr-Sommerfeld quantization rules applied to 
the classical motion of the free Skyrmion to construct quantum states 
corresponding to nucleons. We take such an unusual route because we need 
quantum states which are described by specific classical trajectories and
consquently suitable for the approximation of variation of 
constants. In contrast, the ordinary quantization method does not fit into 
our  
scheme since it provides nucleonic states as quantum wave functions. It is 
nevertheless interesting to consider it in the light of our previous 
developments and we will briefly describe it in the following paragraph. 

Adkins \etal \cite{Adkins} quantized the Skyrmion and constructed  
spin and isospin 1/2 states. They considered the Lagrangian for a 
single spinning Skyrmion $(\ref{eq:l1sk})$ which can be written as
\begin{eqnarray}
L & = & - M +  \Lambda \,\tr\bigl[ \dot A^\dagger \dot A\bigr]
\nonumber
\\
& = & - M +  2 \Lambda \,\sum_{\mathrm{i}=0}^{3} \dot a_{\mathrm{i}}^2
\label{eq:l1ska}
\end{eqnarray}
where $a_{\mathrm{i}}$ parametrizes $A=a_0+{\mathrm{i}}\vec a\cdot 
\vec\tau$ with $a_0^2 + {\vec a}^2=1$. The time dependent $\mathrm{SU}(2)$ 
matrix $A$ defines the rotational
characteristics of the Skyrmion. By doing so they make the low energy 
hypothesis, describing the system approximately using only its zero modes 
(all radial oscillations, deformations, etc. are neglected). This is exactly 
like the BPS case: geodesic motion on the minimum energy manifold. 
One must solve\cite{Mantongfc} the Schr\"odinger equation on the manifold of 
low energy 
dynamics (in this case, static solutions), finding the wave functions for 
each state of the system in a highly non-Cartesian (\ie curved) moduli space. 
The one Skyrmion 
system is a very simple and elegant illustration of this method.

By considering the truncation of the system $(\ref{eq:l1ska})$, the 
configuration of the system is just represented by a point moving on the 
3-sphere of the $\mathrm{SU}(2)$ group defined by the familiar constraint:
\begin{equation}
AA^\dagger = 1 = a_0^2 + {\vec a}^2.
\end{equation}
This is a straightforward generalization of the system studied in the first 
example of subsection 4.2.1 where the 
system was a particle following great circles on a 2-sphere. By computing 
$\vec\pi_{\mathrm{a}}$ (the variables canonically conjugate to the $\vec a$) 
\begin{equation}
\pi_{\mathrm{a}}^{\mathrm{i}} = {\partial L\over \partial \dot a_{\mathrm{i}}},
\end{equation}
substituting in the Lagrangian $(\ref{eq:l1ska})$, 
and performing the usual Legendre transformation, we obtain 
the familiar expression for the Hamiltonian of a spinning rigid body of mass 
$M$ and moment of inertia $\Lambda$:
\begin{equation}
H = M + {1\over 2 \Lambda} {\vec\pi_{\mathrm{a}}}^2. 
\end{equation}
The usual canonical quantization procedure gives:
\begin{equation}
H = M - {1\over 2 \Lambda} \sum_{\mathrm{i}=0}^3 {\partial^2 
\over\partial a_{\mathrm{i}}^2}.
\end{equation}
which is the Laplacian over the 3-sphere of $\mathrm{SU}(2)$: the 
Schr\"odinger equation we have to solve on the low-energy manifold 
comes naturally. Solutions to this equation are well known: symmetrical, 
traceless polynomials in $a_{\mathrm{i}}$. Statistical considerations 
impose the 
order of the polynomial to be odd for the Skyrmion to be a fermion, and even 
for a boson. Adkins \etal\cite{Adkins} 
of course chose the former. They found the 
following states four spin/isospin states:
\begin{eqnarray}
&& |p\,\uparrow\rangle ={1\over \pi} (a_1 + {\mathrm{i}} a_2)
\\
&& |p\,\downarrow\rangle = -{\mathrm{i}\over \pi} (a_0 - {\mathrm{i}} a_3)
\\
&& |n\,\uparrow\rangle = {\mathrm{i}\over \pi} (a_0 + {\mathrm{i}} a_3) 
\\
&& |n\,\downarrow\rangle = - {1\over \pi} (a_1 - {\mathrm{i}} a_2) 
\label{eq:anw}
\end{eqnarray}
These wave functions are only suitable for a completely quantum mechanical 
treatment of the Skyrmion-Skyrmion system. As stated earlier we follow the 
semi-classical route and will obtain these same states 
via the Bohr-Sommerfeld method\cite{Bohr}. 

We also start from the classical Lagrangian of equation
$(\ref{eq:l1ska})$. Using the expression of the matrix $A$ as a function of 
the usual Euler angles:
\begin{equation}
A = e^{-\mathrm{i}\alpha \tau_3/2}\,
e^{-\mathrm{i}\beta \tau_2/2}\,
e^{-\mathrm{i}\gamma \tau_3/2}
\label{eq:aeuler}
\end{equation}
where $\alpha,\gamma\in [0,2\pi]$ and $\beta\in [0,\pi]$, or for the $a_\mu$ 
\begin{eqnarray}
a_0 & = & \cos({\beta\over 2})\cos({\alpha+\gamma\over 2})
\\
a_1 & = & \sin({\beta\over 2})\sin({\alpha-\gamma\over 2}) 
\\
a_2 & = & - \sin({\beta\over 2}) \cos ({\alpha-\gamma\over 2})
\\
a_3 & = & - \cos({\beta\over 2}) \sin({\alpha+\gamma\over 2}),
\end{eqnarray}
we find in another form the Lagrangian for a rotating rigid body:
\begin{equation}
L = - M + {1\over 2}\Lambda\,\bigl[ \dot \alpha^2+ \dot \beta^2 +\dot \gamma^2
+ 2 \dot \alpha \dot \gamma \cos \beta\bigr].
\label{eq:lskeuler}
\end{equation}
The angles $\alpha$, $\beta$ and $\gamma$ obey the following equations of 
motion:
\begin{eqnarray}
\ddot\alpha + \ddot\gamma\cos\beta-\dot\gamma\dot\beta\sin\beta=0\nonumber
\\
\ddot\beta+ \dot\alpha\dot\gamma\sin\beta=0\label{eq:eqeuler}
\\
\ddot\gamma + \ddot\alpha\cos\beta-\dot\alpha\dot\beta\sin\beta=0\nonumber
\end{eqnarray}
and refer, in connection with the rigid body system, to the motion 
of the body fixed axes relative to the laboratory fixed
axes. In our case however, they have a different interpretation. 
Indeed, contrary to an ordinary rigid body which 
has only one conserved vector 
quantity in the laboratory reference frame, (namely 
the total angular momentum) 
the quantized Skyrmion will have 
two such conserved quantities: the spin and the isospin. The Euler angles will 
then specify those two quantities. In fact we can say that the angle $\beta$ 
roughly will
fix the relative orientation of these two vector quantities, while the time 
derivatives
of $\alpha$ and $\gamma$ will fix their magnitudes. For simplicity 
without loss of generality we will 
choose the axis of polarization of spin and isospin parallel to the axis 3 of 
isospace. 
Following the convention used by Adkins \etal \cite{Adkins}, we will chose
the following expressions for the spin and isospin generators:
\begin{eqnarray}
I_3 & = & - 2 \Lambda{\cal R}^3_0(A)\equiv - 2 \Lambda{\cal R}^3
\\
S_3 & = & 2 \Lambda{\cal L}^3_0(A)\equiv 2 \Lambda{\cal L}^3.
\end{eqnarray}
Following our choice of quantization axes, we take $\beta$ to be either 0 or 
$\pi$ and look for solutions to the system of equations $(\ref{eq:eqeuler})$ 
which are now given by:
\begin{eqnarray}
\ddot\alpha \pm \ddot\gamma=0\label{eq:eqeulerb1}
\\
\ddot\gamma \pm \ddot\alpha=0\label{eq:eqeulerb2}
\end{eqnarray}
The variables $\alpha$ and $\gamma$ actually decouple now and the spin and 
isospin generators are given by (if $\beta$ is constant)
\begin{eqnarray}
I_3 & = & \Lambda (\dot\alpha + \cos\beta \dot\gamma)\nonumber
\\
&=& \Lambda (\dot\gamma\pm\dot\alpha)\label{eq:i3a}
\\
S_3 & = & - \Lambda (\dot\alpha\cos\beta + \dot\gamma)\nonumber
\\
&=& - \Lambda (\dot\gamma\pm\dot\alpha),\label{eq:s3a}
\end{eqnarray} 
depending on whether $\beta=0$ (``+'' sign) or $\pi$ (``$-$'' sign). 
It is then easy to choose $\beta$ so that spin and isospin are parallel or 
antiparallel, and $\alpha$ and $\gamma$ so they are positive or negative. 
There only remains to apply 
the quantization rules of Bohr-Sommerfeld to the system. In the original
problem of the hydrogen atom, this method was used to compute the allowed radii
of the electron orbits. In the case of the Skyrmion, 
the radius of the orbits is already defined since the system moves at 
constant angular velocity along trajectories which are the great circles of 
$\mathrm{SU}(2)$, therefore having radius 1. The quantization condition 
will fix the angular velocity so that the spin and isospin have the 
right value, namely 1/2 for nucleons, 3/2 for the first nucleon resonance and
so on. The Bohr-Sommerfeld quantization condition, {\em derived\/} 
by path integral methods, is\cite{Rajaraman}
\begin{equation}
W = \sum_{\mathrm{i}} J_{\mathrm{i}} = (n+\xi) h, \qquad n= 0,\pm 1,\pm 2,
\cdots 
\end{equation}
where the $J_{\mathrm{i}}$ are the action angle variables and
$\xi$ is a correction factor arising from the functional integral over Gaussian
fluctuations about the classical trajectory, which we will neglect. 
The action-angle variables $J_{\mathrm{i}}$ are defined by
\begin{equation}
J_{\mathrm{i}} = \oint p_{\mathrm{i}} \,dq_{\mathrm{i}}
\end{equation}
where $p_{\mathrm{i}}$ is the momentum conjugate to the coordinate 
$q_{\mathrm{i}}$, and the integral is taken along a closed path followed by 
the system during one period in the plane $(q_{\mathrm{i}},p_{\mathrm{i}})$ 
of phase space. In our case $q_{\mathrm{i}}$ represents the angles $\alpha$, 
$\beta$ and $\gamma$. Due to the cyclic nature of the angles $\alpha$ and 
$\gamma$, $J_\alpha$ and $J_\gamma$ are readily computed:
\begin{equation}
J_\alpha = \oint p_\alpha \d\alpha= 2\pi p_\alpha=
2 \pi\Lambda\bigl[ \dot \alpha + \dot \gamma \cos \beta\bigr]\label{eq:ja}
\end{equation}
\begin{equation}
J_\gamma = \oint p_\gamma \d_\gamma= 2\pi p_\gamma=
2\pi\Lambda\bigl[ \dot \gamma + \dot \alpha \cos \beta\bigr]\label{eq:jg}.
\end{equation}
and by fixing $\beta$ to either 0 or $\pi$, $J_\beta$ is 0. Then we find,
applying the quantization rules,
\begin{equation}
J_\alpha+ J_\gamma = n h, \qquad n= 0,\pm 1,\pm 2,\cdots. \label{eq:condqu}
\end{equation}
The spin and isospin generators can also be expressed as functions of the  
action angle variables using $(\ref{eq:i3a})$, $(\ref{eq:s3a})$, 
$(\ref{eq:ja})$ and $(\ref{eq:jg})$:
\begin{eqnarray}
I_3 &=& {J_\alpha\over 2 \pi}
\\
S_3 &=& -{J_\gamma\over 2 \pi}.
\end{eqnarray} 
By symmetry, the Skyrmion has equal magnitude of spin and isospin, $|I_3| = 
|J_3|$, so for $n=1$ in $(\ref{eq:condqu})$ we have 
$|J_\alpha|=|J_\gamma|=1/2$. 

In the case where $\beta=0$, then using $(\ref{eq:i3a})$ and
$(\ref{eq:s3a})$
\begin{eqnarray}
I_3 & = & \Lambda(\dot\alpha+\dot\gamma)\equiv 2 \Lambda \omega=\pm{1/2}
\\
S_3 & = &-\Lambda(\dot\alpha+\dot\gamma)\equiv -2 \Lambda \omega=\mp{1/2}
\end{eqnarray}
with $(\alpha+\gamma)/2=\phi(t) = \omega t + \phi_0$, a solution of
$(\ref{eq:eqeulerb1})$ and $(\ref{eq:eqeulerb2})$. This type of 
angular motion produces Skyrmions with spin and isospin
antiparallel, and the states $|p\downarrow>$ and $|n\uparrow>$.
The proton state corresponds to isospin $+1/2$ along the 3 axis in isospace
while the neutron corresponds to isospin $-1/2$. Replacing this solution in 
equation $(\ref{eq:aeuler})$ gives
\begin{eqnarray}
A &=& \cos\phi(t) - \mathrm{i} \sin\phi(t) \tau^3
\\
&=& e^{-\mathrm{i} \phi(t)\tau^3/2}
\end{eqnarray}
where $\omega>0$ corresponds to the state $|p\downarrow>$ and $\omega<0$ to
$|n\uparrow>$. 

In the case where $\beta=\pi$,  
\begin{eqnarray}
I_3 & = & \Lambda(\dot\alpha-\dot\gamma)\equiv 2 \Lambda \omega=\pm 1/2
\\
S_3 & = & -\Lambda(\dot\gamma-\dot\alpha)\equiv 2 \Lambda \omega=\pm 1/2
\end{eqnarray}
with $(\alpha-\gamma)/2=\psi(t) = \omega t + \psi_0$ and the corresponding 
matrix
\begin{eqnarray}
A & = & \mathrm{i}\bigl[ \sin\psi(t) \tau^1-\cos\psi(t)\tau^2\bigr]
\\
& = & -\mathrm{i} e^{-\mathrm{i} \psi(t)\tau^3/2}\,\tau^2\, 
e^{\mathrm{i} \psi(t)\tau^3/2}
\end{eqnarray}
represents the state $|p\uparrow>$ with $\omega>0$ and $|n\downarrow>$ with
$\omega<0$. We can see the similarity of the forms for the matrices $A$
corresponding to nucleon states and the wave functions obtained in the directly
quantum version of Adkins {\it et al} $(\ref{eq:anw})$. Finally there 
remains to compute the magnitude of $\omega$ so as to have spin 
and isospin 1/2. This is done by solving the following equation:
\begin{equation}
|I_3| = {1\over 2} = 2\Lambda |\omega|\Longrightarrow
\omega = {1\over 4 \Lambda}\sim 50\mathrm{-}100\;\mathrm{MeV},
\end{equation}
using the value of $\Lambda$ from reference\cite{Walhout}.
The uncertainty is caused by the values of $f_\pi$ and $e$ (via the moment of
inertia  $\Lambda$) which are subject to variations depending on which 
observables are chosen to be best reproduced by the model.

%% file: section433.tex
We are now ready to compute nucleon-nucleon scattering trajectories.
As mentioned earlier, we 
use the approximation method of ``variation of constants''
and simply replace into the right hand side of
equations $(\ref{eq:ddot})$, $(\ref{eq:ladot})$, $(\ref{eq:lbdot})$
$(\ref{eq:radot})$ 
and $(\ref{eq:rbdot})$ the semi-classically quantized
trajectories 
found in the previous section for spin and isospin, as well as taking $\vec
d(t)$ corresponding to its free trajectory, a straight line 
at constant velocity.
To calculate the change induced in the previously constant quantities we 
integrate the equations from $t=-\infty$ to $t=+\infty$ over one free 
trajectory. This will enable us to obtain scattering angles for the 
trajectories.

As we mentioned earlier, this computation scheme does not work for the time 
evolution of the spins and isospins. Equations $(\ref{eq:ladot})$,
$(\ref{eq:lbdot})$, $(\ref{eq:radot})$ and $(\ref{eq:rbdot})$ all have on
the right hand side a factor $1/d$ which behaves like $1/t$ for large values 
of $t$ since $\vec d(t) = \vec v\; t + \vec \gamma$. Then
\begin{equation}
{\d \over \d t} {\cal L}^{\mathrm{k}} \sim {1\over d} \sim {1\over t}
\end{equation}
and similarly for ${\cal R}^{\mathrm{k}}$. Then ${\cal R}$ and ${\cal L}$ 
change by an infinite amount
between $t=-\infty$ and $t=+\infty$ (because the right hand 
side of the equations integrate to a $\log$ which varies very slowly). This 
means that our approximation scheme is too crude or that our 
treatment is not valid for zero pion mass. Indeed, if $m_\pi\not=0$ then 
the usual Yukawa cut-off factors 
$e^{-m_\pi r}$ will arise on the right hand side of the equations of motion,
and will render the changes in ${\cal L}$ and ${\cal R}$
finite.
There does not seem to be any easy solution to the problem so we will not 
discuss
spin/isospin changes further. We will just say that results are (roughly) 
compatible with the exchange of charge carrying (pions) and spin carrying 
(vector mesons) as intermediate particles.

By contrast, equation $(\ref{eq:ddot})$ shows that 
${\d}/{\d} t\; \dot d^{\mathrm{i}}$ behaves like $1/d^2$, and the 
approximation 
method works well. We present below the results for the scattering of 
nucleons for some particular cases of the initial polarizations, using our 
semi-classical formalism, namely scattering of particles whose spin and 
isospin are polarized along the $z$ axis
\begin{eqnarray}
&&A(t), B(t) = \cos\phi(t) - \mathrm{i}\sin\phi(t) \tau^3\nonumber
\;{\mathrm{for\;the\;states}}\;|p\downarrow>\;{\mathrm{and}}\;|n\uparrow>
\\
&&\qquad\;\,\, {\mathrm{or }}\;\; \mathrm{i} [ \sin\psi(t) \tau^1 - 
\cos\psi(t) \tau^2]
\;{\mathrm{for}}\;{\mathrm{the}}\;{\mathrm{states}}\;|p\uparrow>\;
{\mathrm{and}}\;|n\downarrow>
\end{eqnarray}
and the relative motion initially given by
\begin{equation}
\vec d(t) = \vec v\; t + \vec\gamma
\end{equation} 
where $\vec\gamma$ is the impact parameter vector if $\vec v$ and $\vec\gamma$
are chosen orthogonal and the time of closest approach is at $t=0$. 
These give simple two-dimensional orbits which can be obtained analytically.
The tensorial nature of the
interaction implies that the forces depend on the angle between the axis of
separation and the spin polarization. If we choose the spin polarization 
along an
axis tilted with respect to the normal to the initial scattering plane we get
complicated, three dimensional scattering trajectories.

It is important to observe that in our formalism an additional 
parameter arises which describes the initial state of two incoming, polarized 
nucleons. This parameter, along with
the impact parameter, the initial velocity and the direction of polarization, 
actually selects the particular
scattering trajectory followed by the nucleons. The parameter describes the
relative orientation of the Skyrmions at a fixed (initial) time. It plays in 
fact a role
similar to a hidden variable. An incoming pair of physical nucleons, in our
formalism, has a fixed value for this parameter, which is only ``measured'' 
after 
the scattering takes place. In a physical experiment consisting of incoming
beams of nucleons giving rise to collisions or scattering of pairs of nucleons,
the value of this parameter will be uniformly distributed. A similar parameter
arises in the case of the scattering of BPS monopoles. This parameter enters 
the computations via $\delta$ and $\epsilon$ which are defined by
\begin{eqnarray}
\delta&=&\phi_0^{\mathrm{A}}-\phi_0^{\mathrm{B}}
\\
\epsilon&=&\phi_0^{\mathrm{A}}+\phi_0^{\mathrm{B}} 
\end{eqnarray}
in self evident notation.

There is an immediate separation of the scattering into 
two cases, depending on
whether $D_{\mathrm{ab}}(A^\dagger B)$ is time independent or not. When it 
depends on time for large values of the ratio $(\omega\gamma/v)$ there is an 
exponential suppression
of the scattering where $v$ is the relative velocity and $\gamma$ the impact
parameter. This is quite evident: for slowly translating Skyrmions, the
prescribed rotations imposed by selecting semi-classically quantized nucleon
states have the effect of averaging the interaction to zero. Interactions 
which depend on the relative orientation of rapidly spinning bodies are
common in the classical or semi-classical treatment of soliton systems. A
similar numerical example is presented by Piette \etal\cite{Piette} in the 
case of ``baby-Skyrmions'' (Skyrmions in a two dimensional space), where
classical trajectories and scattering of spinning solitons is studied. They
too exhibit an ``oscillatory'' interaction which nonetheless generates a net
force between the particles and non-trivial scattering. 
 
\paragraph{First case: $D_{\mathrm{ab}}(A^\dagger B)$ time independent}
We first present the expression for time variations of the previously
constant relative momentum $\vec p = (M/2) \dot{\vec d}$ for the scattering 
of protons or of neutrons with various spin polarisations. 

\begin{equation}
\!\!\!\!\!\!\!\!\!\!\!\!(i)\qquad
\begin{array}{l}
p\uparrow p\uparrow
\\
n\downarrow n\downarrow
\\
p\downarrow p\downarrow
\\
n\uparrow n\uparrow
\end{array}
\qquad {\d\over \d t} p^{\mathrm{k}} = - {\Delta\omega^2\over d^2} 
\cos({2\delta}) \hat d^{\mathrm{k}}\label{eq:firstcase}
\end{equation}

\begin{equation}
\!\!\!\!\!\!\!\!\!\!\!\!(ii)\qquad p\uparrow p\downarrow
\qquad 
\begin{array}{l}
{\d\over \d t} p^{\mathrm{k}} = - {\Delta\omega^2\over d^2}\bigl[
\hat d^{\mathrm{k}} + 4 r^{\mathrm{k}} \hat r\cdot\hat d - 6 \hat 
d^{\mathrm{k}} (\hat r\cdot\hat d)^2 \bigr]
\\
\qquad\qquad\qquad\qquad\hat r^{\mathrm{k}} = (-\sin(\delta),\cos(\delta),0)
\end{array}
\end{equation}

\begin{equation}
\!\!\!\!\!\!\!\!\!\!\!\!(iii)\qquad\uparrow n\downarrow
\qquad
\begin{array}{l}
{\d\over \d t} p^{\mathrm{k}} = - {\Delta\omega^2\over d^2}\bigl[
\hat d^{\mathrm{k}} + 4 r^{\mathrm{k}} \hat r\cdot\hat d - 
6 \hat d^{\mathrm{k}} (\hat r\cdot\hat d)^2 \bigr]
\\
\qquad\qquad\qquad\qquad\hat r^{\mathrm{k}} = (\sin(\delta),-\cos(\delta),0)
\end{array}
\end{equation}
The right hand 
sides can be interpreted, at this level of our approximation, as coming from 
a spin-spin channel and a tensor channel interaction. We stress that this is 
only a correspondence: the true effect of the kinetic term is to supply a 
non-trivial connection in the geodesic equations on the low energy 
sub-manifold and not to modify the potential.

To find the actual change in $\vec p$ and
hence the scattering angle, we integrate these equations from $t=-\infty$ to
$t=+\infty$. For the cases of scattering of protons or neutrons on each other
respectively we find 
that the scattering angle depends on the variable $\delta$
which corresponds to the phase lag between the rotation of $A(t)$ and $B(t)$. 
Straightforward integration gives
\begin{eqnarray}
p^{\mathrm{k}}(t) & = & -\Delta\,\omega^2 \cos(2\delta) 
\int_{-\infty}^t {v^{\mathrm{k}} t + 
\gamma^{\mathrm{k}} \over (v^2 t^2 + \gamma^2)^{3/2}}{\d} t + 
{M\over 2} v^{\mathrm{k}}
\\
& = & -\Delta\,\omega^2 \cos(2\delta)\Biggl[ 
-{v^{\mathrm{k}} \over v^2 (v^2 t^2 +
\gamma^2)^{1/2}} + {\gamma^{\mathrm{k}} t\over \gamma^2 
(v^2 t^2 + \gamma^2)^{1/2}}
\Biggr]^t_{-\infty}\nonumber
\\
&&\qquad\qquad\qquad\qquad + {M\over 2} v^{\mathrm{k}}.
\end{eqnarray}
This yields
\begin{equation}
p^{\mathrm{k}}(+\infty) = -2 \Delta\,\omega^2 \cos 2\delta 
{\gamma^{\mathrm{k}}\over \gamma^2 v} +
{M\over 2} v^{\mathrm{k}}
\end{equation}
from which we calculate the cosine of the scattering angle
\begin{eqnarray}
\cos\theta & = & {\vec p(+\infty)\cdot\vec p(-\infty)\over 
|\vec p(+\infty)| |\vec p(-\infty)|} 
\\
& = & {M\gamma v^2\over 4 \Delta} {1\over\bigl( {M^2\gamma^2 v^4\over 
16 \Delta^2} + \omega^4 \cos^2 2\delta \bigr)^{1/2}}\label{eq:anglediff}.
\end{eqnarray} 

\paragraph{Second case: $D_{\mathrm{ab}}(A^\dagger B)$ time dependent}
For the cases of collisions between protons and neutrons, the expression for
$p^k(+\infty)$ is more complicated, because of the time dependence of
$D_{\mathrm{ab}}(A^\dagger B)$. The expressions for the scattering of $p$ on 
$n$ each contain a
time dependent $A^\dagger B$. When integrated these yield an exponentially
suppressed variation in the dimensionless group $(\omega\gamma/v)$
\begin{equation}
\sim e^{-({\omega\gamma\over v})}.
\end{equation}
Thus in the limit $v\to 0$ we get negligible scattering in these cases.
These exponential suppression factors appear in the solutions via $G$ functions
which can be expressed from Bessel functions by 
\begin{equation}
G_{\mathrm{n,m}}(x,y) = {{\d}^{\mathrm{n}}\over {\d} x^{\mathrm{n}}} 
x^{\mathrm{m}} K_{\mathrm{m}}(x,y)
\end{equation}
where $K_{\mathrm{m}}(x)$ is the Bessel function of the second kind. We get
\begin{equation}
\!\!\!\!\!\!\!\!\!\!\!\!(i)\qquad
\begin{array} {l}
p\uparrow n\downarrow
\\
p\downarrow n\uparrow
\end{array}
\qquad {\d\over \d t} p^{\mathrm{k}} = {\Delta\omega^2\over d^2} 
\cos(4\omega t+2\epsilon) \hat d^{\mathrm{k}}
\end{equation}
Which gives after integration
\begin{equation}
\begin{array} {l}
p_{\mathrm{x}}(+\infty) = {2 \Delta\omega^2\over M \gamma v} \sin{2\epsilon}\;
G_{1,1}(4\omega,\gamma/v) + {M v\over 2}
\\
p_{\mathrm{y}}(+\infty) = {2 \Delta\omega^2\over M v^2} \cos{2\epsilon}\;
G_{0,1}(4\omega,\gamma/v).
\end{array}
\end{equation}

\begin{equation}
\!\!\!\!\!\!\!\!\!\!\!\!(ii)\qquad
p\uparrow n\uparrow
\qquad
\begin{array}{l}
{\d\over \d t} p^{\mathrm{k}} = {\Delta\omega^2\over d^2}\bigl[
\hat d^{\mathrm{k}} + 4 r^{\mathrm{k}} \hat r\cdot\hat d - 
6 \hat d^{\mathrm{k}} (\hat r\cdot\hat d)^2 \bigr]
\\
\qquad\qquad\qquad\hat r^{\mathrm{k}} = (-\sin(2\omega t + \epsilon),
\cos(2\omega t + \epsilon),0)
\end{array}
\end{equation}
similarly integrates to
\begin{equation}
\begin{array} {l}
p_{\mathrm{x}}(+\infty) = -{2 \Delta\omega^2\over 3 } \sin{2\epsilon} \bigl [ 
{8\gamma\over v^3}
G_{0,2}(4\omega,\gamma/v) + {20\over v^2} G_{1,2}(4\omega,\gamma/v) 
\\
\qquad\qquad\qquad\qquad+{16\over \gamma v} 
G_{2,2}(4\omega,\gamma/v) + 
{4\over \gamma^2} G_{3,2}(4\omega,\gamma/v)\Bigr] + {M v\over 2}
\\
p_{\mathrm{y}}(+\infty) = -{2 \Delta\omega^2\over 3 } \cos{2\epsilon} \bigl 
[{4\gamma\over
v^3} G_{0,2}(4\omega,\gamma/v) + {16\over v^2} G_{1,2}(4\omega,\gamma/v) 
\\
\qquad \qquad \qquad\qquad\qquad\qquad+{20\over \gamma v} 
G_{2,2}(4\omega,\gamma/v) + 
{8\over \gamma^2} G_{3,2}(4\omega,\gamma/v)\Bigr]
\end{array}
\end{equation}
and
\begin{equation}
\!\!\!\!\!\!\!\!\!\!\!\!(iii)\qquad p\downarrow n\downarrow
\qquad
\begin{array}{l}
{\d\over \d t} p^{\mathrm{k}} = {\Delta\omega^2\over d^2}\bigl[
\hat d^{\mathrm{k}} + 4 r^{\mathrm{k}} \hat r\cdot\hat d - 
6 \hat d^{\mathrm{k}} (\hat r\cdot\hat d)^2 \bigr]
\\
\qquad\qquad\qquad\hat r^{\mathrm{k}} = (-\sin(2\omega t + \epsilon),
-\cos(2\omega t + \epsilon),0)
\end{array}
\end{equation}
to
\begin{equation}
\begin{array} {l}
p_{\mathrm{x}}(+\infty) = -{2 \Delta\omega^2\over 3 } \sin{2\epsilon} \bigl [
{8\gamma\over v^3} G_{0,2}(4\omega,\gamma/v) - {20\over v^2}
G_{1,2}(4\omega,\gamma/v)  
\\
\qquad \qquad \qquad\qquad+{16\over \gamma v} 
G_{2,2}(4\omega,\gamma/v) -
{4\over \gamma^2} G_{3,2}(4\omega,\gamma/v)\Bigr] + {M v\over 2}
\\
p_{\mathrm{y}}(+\infty) = -{2 \Delta\omega^2\over 3 } \cos{2\epsilon} \bigl [
-{4\gamma\over v^3} G_{0,2}(4\omega,\gamma/v) + {16 \over v^2}
G_{1,2}(4\omega,\gamma/v)  
\\
\qquad \qquad \qquad\qquad\qquad\qquad- {20\over \gamma v} 
G_{2,2}(4\omega,\gamma/v) +
{8\over \gamma^2} G_{3,2}(4\omega,\gamma/v)\Bigr].
\end{array}
\end{equation}
This is to our knowledge the first analytical calculation of 
nucleon-nucleon scattering from
essentially first principles, without recourse to {\it ad hoc} models or
potentials. To calculate the classical scattering 
cross-section we need to compute the scattering for all different polarizations
relative to the initial scattering plane. This would comprise a different
project which would probably be best achieved by numerical methods. Therefore
we are unable at this point to make a direct comparison with experiment. 

Let us now make a few remarks on our results.
In the limit that the initial velocity vanishes, 
for fixed $\omega$ and $\gamma$, we recover 90$^\circ$ scattering. This is, 
however, not surprising as it is a property also shared by the Coulomb and 
many other 
interactions treated within our approximation. 90$^\circ$ scattering is hardly 
remarkable except at zero impact parameter, where of course, it is impossible 
to avoid the region of close proximity of the nucleons and it is important 
that the configurations pass through the minimal, toroidal configuration. 

We have made several approximations in our treatment,
which deserve some discussion.
First we want consider the method of variation of 
constants. To check its accuracy, we observe that 
equation $(\ref{eq:firstcase})$ is a  
simple Kepler problem (for this particular equation). 
We solve it directly to find 
the exact value of the scattering angle and then compare wiith the result 
obtained by variation of constants. 
The exact scattering angle $\Theta$ is:
\begin{equation}
\cos\Theta = 1 - {2\over 1 + 
{ \gamma^2 M^2 v^4\over 4 \Delta^2 \omega^4 \cos^2{2\delta}}
}.\label{eq:anglediffK}
\end{equation} 
Defining a parameter $\chi$
\begin{equation}
\chi = {\cos^2{2\delta} \Delta^2 \omega^4\over \gamma^2 M^2 v^4},
\end{equation}
and choosing $v$ and $\gamma$ so that $\chi$ is much smaller than 1
we get from equations $(\ref{eq:anglediff})$ and $(\ref{eq:anglediffK})$:
\begin{eqnarray}
\cos\theta & = & {1\over \sqrt{1 + 16 \chi}} \simeq 1 - 8 \chi + O(\chi^2)
\\
\cos\Theta & = & {1 - 4 \chi\over 1 + 4 \chi} \simeq 1 - 8 \chi + O(\chi^2).
\end{eqnarray}
With the previously defined values and definitions we get 
\begin{equation}
\chi \sim 10^{-10}\;{\mathrm{MeV}}^{-2}\;\times {1\over \gamma^2 v^4} \ll 1.
\end{equation}
This imposes very loose restrictions to $v$ and $\gamma$. For 
these conditions, 
the approximation seems to work well and respect the other assumptions.

The second main approximation is the expansion in inverse powers of $d$,  
neglecting all terms beyond the dominant
contribution (from the kinetic term). We have 
from the start restricted ourselves to large separations
between the Skyrmions. In this regime, the Skyrmions are well described to
leading order by the product ansatz and the manifold of collective
coordinates
is parametrized by the variables of the product ansatz. The induced metric,
calculated to leading order behaves as $1/d$ while the induced potential
behaves as $1/d^3$. In principle there can be a region where the contribution
of the induced metric dominates and we can neglect the potential. We
find that the metric induces an interaction which can be interpreted, within
our approximation method, as a spin-spin and a tensor interaction. 
Unfortunately
it seems that the domination by the metric term is not physically realized. The
induced kinetic term is multiplied by essentially the frequencies of angular
rotation of the Skyrmions: 
\begin{eqnarray}
T_{\mathrm{int}} &=& {2\pi \kappa^2 f_\pi^2\over d}
\epsilon^{\mathrm{iac}}\epsilon^{\mathrm{jbd}}\;
{\cal L}^{\mathrm{a}}(A)\,{\cal L}^{\mathrm{d}}(B)\;
\bigl(\delta^{\mathrm{ij}}-\hat{d}^{\mathrm{i}} 
\hat{d}^{\mathrm{j}}\bigr)\, D_{\mathrm{ab}}(A^\dagger B) +O(1/d^2)\nonumber
\\
&\sim& {2\pi \kappa^2 f_\pi^2\over d} \omega^2
\end{eqnarray}
while the potential term has two extra powers of the
separation in the denominator: 
\begin{eqnarray}
V_{\mathrm{int}} &=& 
4\pi f_\pi^2 \kappa^2 {(1-\cos\theta) (3(\hat n\cdot\hat
d)^2-1)\over d^3}\nonumber
\\
&\sim& {4\pi f_\pi^2 \kappa^2 \over d^3}.
\end{eqnarray}
So 
\begin{eqnarray}
H_{\mathrm{int}} &=& T_{\mathrm{int}} + V_{\mathrm{int}}\nonumber
\\
&\simeq& {2 \pi \kappa^2 f_\pi^2 \omega^2\over d} \biggl(1
+ {2\over d^2 \omega^2}\biggr)
\end{eqnarray}
implies that $T_{\mathrm{int}}\gg V_{\mathrm{int}}$ if 
\begin{equation}
d \gg {\sqrt{2}\over \omega}.
\end{equation} 
Thus for the kinetic term to dominate, the
frequencies should be much larger than the separation.
This corresponds to a region of validity for a separation of about 3 fm and
greater. However there is much latitude
available since the values of $f_\pi$ and $e$ which go into determining
$\omega$ are fixed only by choosing two experimental inputs. $f_\pi$, $e$ can
vary as much as 10-30\% thus we do not feel overly concerned with exceeding the
regime of validity. Our approximation would of course be better justified for
the case of $\Delta\!-\!\Delta$ scattering where $\omega\sim 300$ MeV
corresponding
to a separation of 1 fm. In any case, we do not believe that it is
physically reasonable to consider the scattering of nucleons with the metric
term alone and we expect a contribution from the potential term which is of
the same order of magnitude.  We do not expect, however, any great,
qualitative modification of the scattering upon inclusion of the potential
term, it is of similar strength but actually contributes only in the tensor
channel for the case of massless pions.

We have shown how 
to formulate the nucleon states within the semi-classical
approximation. We have treated the scattering and computed the scattering
angles in a systematic perturbative approximation. Future work should include
consideration of a non-zero pion mass, which leads to a central channel
interaction, a better control of the perturbative method, a departure from the
product ansatz, and a proper treatment of the region of close proximity, to
test the validity of our formalism in the phenomenology of low energy
nucleon-nucleon scattering and of the static quantum states in this sector.